\documentclass[12pt,a4paper]{article}

\usepackage{a4,amsmath,amssymb,ifpdf}
\usepackage[british]{babel}
\ifpdf
   \usepackage{graphicx}
   \usepackage[pdftex]{hyperref}
\else
   \usepackage[dvips]{hyperref}
   \usepackage[arrow,line,dvips]{xy}
\fi

\newcommand{\di}{\genfrac{}{}{0pt}{}}
\def\under#1{\kern.4pt\underline{\kern-.4pt{}#1\kern-.4pt}\kern.4pt}

\renewcommand{\title}[1]{\vspace{10mm}\noindent{\Large{\bf #1}}\vspace{8mm}}
\newcommand{\authors}[1]{\noindent{\large #1}\vspace{3mm}}
\newcommand{\address}[1]{{\itshape #1\vspace{2mm}}}

\newtheorem{Theorem}{Theorem}
\newtheorem{Proposition}[Theorem]{Proposition}
\newtheorem{Assumption}[Theorem]{Assumption}
\newtheorem{Lemma}[Theorem]{Lemma}
\newtheorem{Corollary}[Theorem]{Corollary}

\makeatletter
\def\section{\@startsection{section}{1}{\z@}{-3.25ex plus -1ex minus
    -.2ex}{1.5ex plus .2ex}{\normalfont\large\bfseries}}
\def\subsection{\@startsection{subsection}{1}{\z@}{-3.25ex plus -1ex
    minus -.2ex}{1.5ex plus .2ex}{\normalfont\itshape}
}

\renewcommand*\l@section[2]{%
  \ifnum \c@tocdepth >\z@
    \addpenalty\@secpenalty
    \addvspace{0.6em \@plus\p@}%
    \setlength\@tempdima{1.5em}%
    \begingroup
      \parindent \z@ \rightskip \@pnumwidth
      \parfillskip -\@pnumwidth
      \leavevmode \bfseries
      \advance\leftskip\@tempdima
      \hskip -\leftskip
      #1\nobreak\hfil \nobreak\hb@xt@\@pnumwidth{\hss #2}\par
    \endgroup
  \fi}

\renewenvironment{thebibliography}[1]
         {\section*{References}\addcontentsline{toc}{section}{References}%
          \frenchspacing\small
          \begin{list}{[\arabic{enumi}]}
         {\usecounter{enumi}\parsep=2pt\@plus\p@\topsep 0pt
         \settowidth{\labelwidth}{[#1]}
         \leftmargin=\labelwidth\advance\leftmargin\labelsep
         \rightmargin=0pt\itemsep=0pt\sloppy}}{\end{list}}

\@addtoreset{equation}{section}
\@addtoreset{Theorem}{section}
\makeatother

\allowdisplaybreaks[4]

\begin{document}

\begin{titlepage}

\begin{center}

\title{Self-dual noncommutative $\phi^4$-theory \\[1mm]
in four dimensions is a non-perturbatively solvable \\[3mm] 
and non-trivial 
quantum field theory }

\authors{Harald {\sc Grosse}$^1$ and Raimar {\sc Wulkenhaar}$^2$}

\address{$^{1}$\,Fakult\"at f\"ur Physik, Universit\"at Wien\\
Boltzmanngasse 5, A-1090 Wien, Austria}

\address{$^{2}$\,Mathematisches Institut der Westf\"alischen
  Wilhelms-Universit\"at\\
Einsteinstra\ss{}e 62, D-48149 M\"unster, Germany}

\footnotetext[1]{harald.grosse@univie.ac.at}
\footnotetext[2]{raimar@math.uni-muenster.de}

\vskip 1.5cm

\textbf{Abstract} \vskip 3mm
\begin{minipage}{13cm}%

  We study quartic matrix models with partition function
  $\mathcal{Z}[E,J]=\int dM \exp(\mathrm{trace}(JM- EM^2
    -\frac{\lambda}{4} M^4))$. The integral is over the space of
  Hermitean $\mathcal{N}\times \mathcal{N}$-matrices, the external
  matrix $E$ encodes the dynamics, $\lambda > 0$ is a scalar coupling
  constant and the matrix $J$ is used to generate correlation
  functions. For $E$ not a multiple of the identity matrix, we prove a
  universal algebraic recursion formula which gives all higher correlation
  functions in terms of the $2$-point function and the distinct
  eigenvalues of $E$. The 2-point function itself satisfies a closed
  non-linear equation which must be solved case by case for given $E$.
  These results imply that if the $2$-point function of a quartic
  matrix model is renormalisable by mass and wavefunction
  renormalisation, then the entire model is renormalisable and has
  vanishing $\beta$-function.

  As main application we prove that Euclidean $\phi^4$-quantum field
  theory on four-dimensional Moyal space with harmonic propagation,
  taken at its self-duality point and in the infinite volume limit, is
  exactly solvable and non-trivial. This model is a quartic matrix
  model, where $E$ has for $\mathcal{N}\to\infty$ the same spectrum as
  the Laplace operator in 4 dimensions. Using the theory of singular
  integral equations of Carleman type we compute (for $\mathcal{N}\to
  \infty$ and after renormalisation of $E,\lambda$) the free energy
  density $(1/\text{volume}) \log(\mathcal{Z}[E,J]/\mathcal{Z}[E,0])$
  exactly in terms of the solution of a non-linear integral equation.
  Existence of a solution is proved via the Schauder fixed point
  theorem. 

The derivation of the non-linear integral equation relies
  on an assumption which we verified numerically for coupling
  constants $0 < \lambda \leq \frac{1}{\pi}$.

\end{minipage}
\end{center}

\bigskip

\noindent
{\bf Keywords:} quantum field theory in 4 dimensions; 
noncommutative geometry; exactly solvable models; matrix models; 
Schwinger-Dyson equations; singular integral equations

\end{titlepage}

\tableofcontents

\section{Introduction}

A rigorous construction of quantum field theories in \emph{four
  dimensions} was not very successful so far. In this paper we show
that for $\phi^4$-theory on four-dimensional Moyal space with harmonic
propagation, taken at critical frequency and in the infinite volume
limit, much more is true: \emph{The model is exactly solvable}.

We know that this is a toy model in so far as classical locality and
Poincar\'e symmetry are not realised and the Minkowskian continuation
of that Euclidean model needs to be investigated. On the other hand,
the model 
\begin{itemize} \itemsep 0pt \parskip 0pt  
\item[--] carries an action of an infinite-dimensional symmetry group,

\item[--] is invariant under a duality transformation between position
and momentum space \cite{Langmann:2002cc}, 

\item[--] is almost scale invariant \cite{Disertori:2006nq},

\item[--] is known
to have a realisation as a matrix model (with non-constant kinetic
term),

\item[--] has perturbatively an infinite
number of divergent but renormalisable Feynman graphs
\cite{Grosse:2004yu}.  
\end{itemize}
Each renormalised Feynman graph has subleading
logarithmic terms which make the perturbatively renormalised
correlation functions divergent at large energy. 
The model and its solution touch many aspects of quantum field
theory which we recall in the next subsections. The reader in a hurry
may jump to Sec.~\ref{subsec:outline}.

\subsection{Perturbative, axiomatic and algebraic quantum 
field theory}

Starting with the Lamb shift in the 1940s and culminating in the
experimental tests of the Standard Model, perturbatively renormalised
quantum field theory is an enormous phenomenological success. However,
this success lacks a mathematical understanding. The perturbation
series is at best an asymptotic expansion which cannot converge at
physical coupling constants such as the electron charge $e^2\approx 
\frac{1}{137}$. In addition there are physical effects such as
confinement which seem out of reach for perturbation theory. Therefore,
the development of a mathematical foundation of quantum field theory that
permits non-trivial examples is one of the most urgent challenges in
mathematical physics. 

In the early 1950's, G\aa{}rding and Wightman gave a rigorous
mathematical foundation to relativistic quantum field theory by
casting the unquestionable physical principles (locality, covariance,
stability, unitarity) into a set of axioms.  These ideas were
published years later \cite{Wightman:1956zz, Wightman:1964??,
  Streater:1964??}. The difficulty to provide non-trivial examples to
these axioms motivated the development of equivalent or more general
frameworks such as Algebraic quantum field theory and
Constructive/Euclidean quantum field theory.  Algebraic quantum field
theory shifts the focus from the field operators to the Haag-Kastler net of
algebras assigned to open regions in space-time \cite{Haag:1963dh}.
Fields merely provide coordinates on the algebra. Over the years this
point of view turned out to be very fruitful \cite{Haag:1992hx}.  One
important advantage over the axiomatic setup is the natural
possibility to describe quantum field theory on curved space-time
\cite{Brunetti:2009pn}.

\subsection{Euclidean quantum field theory}

Of central importance for us is the Euclidean approach.  Wightman
functions admit an analytic continuation in time. At purely imaginary
time they become the Schwinger functions \cite{Schwinger:1959zz} of a
Euclidean quantum field theory. Symanzik emphasised the powerful
Euclidean-covariant functional integral representation
\cite{Symanzik:1964zz}, which yields a Feynman-Kac formula of the heat
kernel \cite{Kac:1949??}. In this way the Schwinger functions become the
moments of a statistical physics probability distribution. 
This tight connection between Euclidean quantum field theory and
statistical physics led to a fruitful exchange 
of concepts and methods, most importantly that of the 
renormalisation group \cite{Wilson:1973jj}.

It is sometimes possible to rigorously prove the existence of a
Euclidean quantum field theory or of a statistical physics model
without knowing or using that this model derives from a true
relativistic quantum field theory. This is, for instance, the case for
the model constructed in this paper. Sufficient conditions on the
Euclidean model which guarantee the Wightman axioms were first given
by Nelson \cite{Nelson:1971??}. These conditions based on Markov
fields turned out to be too strong or inconvenient. Shortly later,
Osterwalder and Schrader established a set of axioms
\cite{Osterwalder:1973dx,Osterwalder:1974tc} by which the Euclidean
quantum field theory is equivalent to a Wightman theory. The most
decisive axiom is reflection positivity which yields existence of the
Hilbert space and a positive energy Hamiltonian. The Euclidean
approach together with the Osterwalder-Schrader axioms turned out to
be the key to construct relativistic quantum field theories in
dimension less than four. Two successful methods were developed: The
correlation inequality method \cite{Ginibre:1970??, Guerra:1973gd}
relies on positivity and monotonicity of Schwinger functions and is
suitable for bosonic theories. The phase space expansion method 
\cite{Glimm:1974??, Glimm:1987ng, Rivasseau:1991ub} 
works both for bosonic and fermionic theories. It uses lattice partitions 
and iterated cluster and Mayer expansions to control the utraviolet
limit of the theory. It can also typically establish that the 
Schwinger functions built constructively are the Borel sums of their 
ordinary perturbative series. The cluster expansion \cite{Glimm:1974??} is 
also used to prove a mass gap in the spectrum of the Hamiltonian.

\subsection{Solvable models}

Under solvable models we understand models in quantum field theory or
statistical physics where all correlation functions can be exactly
evaluated in terms of ``known'' functions, at least in principle.
With a few three-dimensional exceptions, solvable models were
established only in one or two dimensions.

The first field-theoretical example was the Thirring model
\cite{Thirring:1958in} which describes a quartic self-interaction of a
Dirac field in 1+1 dimensions. In the massless case, Johnson
\cite{Johnson:1961cs} found the exact expression for the 2- and
4-point functions. This work was extended by Hagen \cite{Hagen:1967??}
and Klaiber \cite{Klaiber:1967jz} to the explicit solution of any
correlation function. The massive case is more complicated; we mention
the construction of the S-matrix by Korepin \cite{Korepin:1979qq}
using the Bethe ansatz \cite{Bethe:1931hc}. Another famous model that
is exactly solvable is the Schwinger model \cite{Schwinger:1962tp}, or
QED in 2 dimensions.

Many 2-dimensional models of statistical physics are known to be
solvable. The first one was the 2-dimensional square-lattice critical
Ising model \cite{Ising:1925em}, which after important work by Kramers
and Wannier \cite{Kramers:1941kn} was solved by Onsager
\cite{Onsager:1943jn} using the transfer matrix method. 
The list of more involved exactly solvable models contains the
6-vertex model (or ice model) solved by Lieb \cite{Lieb:1967zz}, the
8-vertex model solved by Baxter \cite{Baxter:1971cr} and the hard
hexagon model also solved by Baxter \cite{Baxter:1980??}. The quantum
inverse scattering method \cite{Faddeev:1995??} and Yang-Baxter
equations \cite{Jimbo:1990??}  are important tools for these
achievements.

The Ising model turned out to be the prototype of a discrete series of
solvable two-dimensional field theories: the Minimal Models in
conformal field theory \cite{Belavin:1984vu, Friedan:1983xq}. Minimal
Models correspond to highest-weight representations of the Virasoro
algebra at central charge $c=1-\frac{6}{m(m+1)}$. Some of these models give
non-trivial realisations of the Osterwalder-Schrader or Wightman
axioms.  A solvable model of different type in conformal quantum field
theory is the Wess-Zumino-(Novikov-)Witten model \cite{Wess:1971yu,
  Witten:1983ar} whose solutions are realised by affine Kac-Moody
algebras.

\subsection{Constructive methods} 

Constructing a model means to prove for a specific candidate
Hamiltonian or Lagrangian the axioms of Wightman, Haag-Kastler or
Osterwalder-Schrader. Thereby the Wightman or Schwinger functions,
although not computed/solved explicitly, are shown to have the
required properties. Constructing a model is less than solving it.
However, most of the solved models are disguised free fields whereas the
constructed models are true interacting field theories, unfortunately
only in dimension $<4$.  For a historical review (that we used already
in the section on Euclidean quantum field theory) we refer to
\cite{Jaffe:2000ub}.

The first successfully constructed model is the $\phi^4$-model in 2
dimensions, $\phi^4_2$ for short, constructed in a series of articles
by Glimm-Jaffe \cite{Glimm:1968kh}. Whereas this was directly achieved
in the Wightman or Haag-Kastler setup, the Euclidean approach was
decisive for the generalisation to the $P(\phi)_2$ model
\cite{Glimm:1974??, Guerra:1973gd}, i.e.\ a two-dimensional scalar
field with polynomial interaction. The construction of $\phi^4_3$
\cite{Glimm:1973kp, Feldman:1976im} turned out to be much harder.
There is a stability problem which was overcome by renormalisation
group ideas. Another problem not present in two dimensions is the
inequivalence of representations of the canonical commutation
relations for the interacting and the free field.

Perturbatively, these models $\phi^4_2$, $\phi^4_3$ and $P(\phi)_2$
are all super-renormalisable. It is therefore not surprising that a
construction of the $\phi^4_4$-model (which is perturbatively just
renormalisable) faces far more problems. In $4+\epsilon$ dimensions,
the $\phi^4$-model does not exist. As shown by Aizenman and Fr\"ohlich
\cite{Aizenman:1981du, Frohlich:1982tw}, the model is trivial , i.e.\
it only exists if the coupling constant vanishes. Related to
triviality is the appearance of Landau poles \cite{Landau:1954??},
first discovered for quantum electrodynamics (QED$_4$): The
$\beta$-function which describes the running of the coupling constant
when changing the scale could develop a singularity at finite momentum
cut-off. This means that the theory is only consistent below that
cut-off, or has to be trivial if the cut-off is removed.
Perturbatively, the $\beta$-function of $\phi^4_4$-theory develops a
Landau pole, which together with the triviality of
$\phi^4_{4+\epsilon}$ strongly suggests (although a proof is still
missing) that $\phi^4_4$ cannot be constructed.

There is a variant of the Euclidean $\lambda\phi^4_4$-model, the
wrong-sign model $\lambda<0$, which can be constructed if restricted
to planar graphs \cite{'tHooft:1982cx,Rivasseau:1983jj}.  Because of
the wrong sign and the resulting instability, this Euclidean model
does not give rise to a model of relativistic quantum field theory. We
mention this model because it shares some aspects with the model we
study and solve in this paper. Our model also restricts to the planar
sector, but as result of the thermodynamic limit and not by hand. Our
model is stable but violates locality in the traditional sense. It is
unknown so far whether or not our model satisfies a non-local variant
of the Osterwalder-Schrader axioms.
  
Recall that QED$_4$ is ruled out for the same Landau pole problem
\cite{Landau:1954??}. It is therefore important that non-abelian
Yang-Mills theory is asymptotically free \cite{Gross:1973id,
  Politzer:1973fx} and as such does not have a Landau pole (in
perturbation theory). This means that Yang-Mills theory is a candidate
for a constructive quantum field theory in four dimensions.
Unfortunately, Yang-Mills theory is too difficult so that its
construction is an open problem \cite{Jaffe:2000ne}.  A toy model of
asymptotic freedom is the Gross-Neveu model \cite{Gross:1974jv}, a
generalisation of the Thirring model to $N$-component fermions. In two
dimensions this model (GN$_2$ for short) is perturbatively
renormalisable but not super-renormalisable.  The construction of
(GN$_2$) was achieved in \cite{Gawedzki:1985ez, Feldman:1985ar} and
constitutes the first example of a constructed just renormalisable
quantum field theory.

\subsection{Noncommutative geometry}

Our present fundamental physics rests on two pillars: Quantum field
theory and General relativity. One of the main questions in this area
concerns the matching of these two concepts. In perturbative quantum
field theory, the standard model which correctly describes all
experimentally observed particles is renormalisable, whereas general
relativity which correctly describes all observed gravity effects is
not. The renormalisation group tells us that non-renormalisable
interactions are scaled away. To put it differently, the existence of
gravity means that space-time cannot be viewed as a differentiable
manifold over all length scales. From the numerical value of the
gravitational coupling constant the renormalisation group estimates
the fundamental scale at which the manifold structure breaks down: it
is the Planck scale $10^{-35}\,\mathrm{m}$. The most active approaches
of quantum gravity all agree on the assumption that space-time is
fundamentally different at the Planck scale; there is only
disagreement about the structure which replaces it.

The approach to quantum gravity which is relevant for this paper is
noncommutative geometry \cite{Connes:1994yd}. Noncommutative geometry
has its roots in the mathematical description of quantum mechanics. It
has been vastly developed over the years and achieved spectacular
success in mathematics. Its relevance to physics comes from the fact
that both Yang-Mills theory (as a classical field theory) and general
relativity are unquestionably of geometrical origin. Noncommutative
geometry achieved, on the level of classical field theories, a true
unification of Yang-Mills theory, and even of the whole standard
model, with general relativity \cite{Connes:1996gi,
  Chamseddine:1996zu}: General relativity put on a noncommutative
space which is the product of a manifold with a discrete space is
nothing but Yang-Mills theory. The astonishing picture which results
is that the breakdown of the manifold structure of space-time already
occurs at the much larger length scale given by the Compton length of
the $W$ and $Z$ bosons of the standard model.  This breakdown is mild;
what at larger distances is a point of space-time becomes a couple of
points at scales shorter than the standard model scale. But once
accustomed to this idea it is natural to assume that passing to
shorter and shorter distances there could be a cascade of phase
transitions in which the manifold structure of space-time fades out
more and more.

Quantum field theory as defined by the Wightman, Haag-Kastler or
Osterwalder-Schrader axioms relies on the manifold structure of
space-time. In giving up the manifold one has to adapt the axioms. A
reasonable replacement is not yet available. One of the reasons is
that whereas noncommutative geometry has a clear concept of compact
manifolds with Euclidean signature \cite{Connes:2008vs}, it is unknown
how to describe non-compact manifolds with Lorentzian signature.
Therefore, the focus has been on Euclidean quantum field theories on
noncommutative manifolds. Because of their functional integral
realisation, such models are easy to define: It suffices to specify a
parametrisation of the fields and an action functional for them.
Scalar fields, for instance, can classically be viewed as sections of
a vector bundle over the manifold. In noncommutative geometry, such
sections form a projective module over an algebra, with the algebra
itself being the simplest example. The traditional pointlike
interaction of scalar fields is then translated into the product in a
noncommutative algebra.

The simplest examples are given by deformations. This means that one
takes the same continuous or smooth functions on the manifold as
parametrisation but equipped with a deformed product $\star$. Rieffel
showed \cite{Rieffel:1993} that if the manifold carries an action of
$\mathbb{R}^d$ by translations, as it is the case for Euclidean space,
it is possible to turn this group action into a deformed associative
product. Doplicher-Fredenhagen-Roberts studied quantum space-time
arising from space-time uncertainty relations at the Planck scale
\cite{Doplicher:1994tu}, Filk derived the Feynman rules for deformed
Euclidean space \cite{Filk:1996dm}, and in 1999 many authors showed
perturbative one-loop renormalisability of a number of models.
Shortly later Minwalla, van Raamsdonk and Seiberg discovered a severe
problem at higher loop order, the so-called ultraviolet/infrared
mixing \cite{Minwalla:1999px}.

A possible way to cure this problem for the $\phi^4_4$-model has been
found by us in previous work \cite{Grosse:2004yu}. It leads to an
action functional which has $4$ relevant/marginal operators
$Z,\mu^2,\lambda,\Omega$ (instead of $3$):
\begin{align}
S=\int_{\mathbb{R}^4}  dx\Big(
\frac{Z}{2} \phi (-\Delta+ \Omega^2 \|2\Theta^{-1}x\|^2 +\mu^2) \phi
+ \frac{Z^2\lambda}{4} \phi\star \phi\star \phi \star \phi \Big)(x)\;.
\label{GW0}
\end{align}
Here $\Theta$ is the deformation matrix which defines the $\star$-product.
We have been able to show that the resulting model is renormalisable
up to all orders in perturbation theory. In addition, a new fixed point
appears at $\Omega=1$. At one-loop order the theory flows into this
fixed point, leaving the ratio $\frac{\lambda}{\Omega^2}$ constant
\cite{Grosse:2004by, Grosse:2004ik}: In contrast to the usual
$\phi^4_4$-theory, the running coupling constant is one-loop bounded.
The group around Rivasseau then showed that for the critical case
$\Omega=1$ the $\beta$-function vanishes up to three loop order
\cite{Disertori:2006uy}, and finally succeeded in proving $\beta=0$ up
to all orders in perturbation theory \cite{Disertori:2006nq}. In this
way, the Landau pole problem is tamed.

Vanishing of the $\beta$-function is often a consequence of
integrability. It was therefore conjectured that the model (\ref{GW0})
at critical frequency $\Omega=1$ has a chance of construction. The
construction was searched in two directions. There is the loop vertex
expansion developed by Rivasseau \cite{Rivasseau:2007fr} which
combines the Hubbard-Stratonovich transform with the
Brydges-Kennedy-Abdesselam-Rivasseau forest formula
\cite{Brydges:1987??, Abdesselam:1994ap}. This method already proved
useful in traditional constructive quantum field theory
\cite{Magnen:2007uy, Rivasseau:2011df} and has been applied in
\cite{Wang:2012gn} to the two-dimensional version of (\ref{GW0}).  The
second approach was proposed by us in \cite{Grosse:2009pa}. By
extending the Ward identities and Schwinger-Dyson equations used by
Disertori-Gurau-Magnen-Rivasseau in \cite{Disertori:2006nq}, we
derived a self-consistent non-linear integral equation for the
renormalised 2-point function of the model (\ref{GW0}). This is a
non-perturbative equation, which we solved perturbatively up to third
order in the coupling constant.

\subsection{Matrix models}

The large-$\mathcal{N}$ limit of $SU(\mathcal{N})$ Yang-Mills theory
led to matrix models with dominant planar diagrams
\cite{'tHooft:1973jz}. Since that observation, matrix models play an
important r\^ole in mathematical physics: Triangulations of
two-dimensional manifolds as simple models of quantum gravity coupled
to matter led to many studies of matrix models \cite{Di
  Francesco:1993nw}.  The IKKT \cite{Ishibashi:1996xs} and BFSS
\cite{Banks:1996vh} models are important as reductions of string
theory models. Of great interest is the observation 
\cite{Brezin:1990rb, Douglas:1989ve, Gross:1989vs} 
that in the double scaling limit of matrix models a new phase
transition occurs.

The partition function of a one-matrix model as toy model for 2D
gravity is given by $\mathcal{Z}=\int dM \; \exp(-\sum_n \alpha_n
\mathrm{tr}(M^n))$, where the integral is over ($\mathcal{N}\times
\mathcal{N}$)-Hermitean matrices and the $\alpha_n$ are scalar
coefficients which may depend on $\mathcal{N}$. In the double scaling
limit $\alpha_n=\mathcal{N} t_n$ and $\mathcal{N}\to \infty$, the
partition function becomes a series in $(t_n)$ which can be expressed
in terms of the $\tau$-function for the Korteweg-de Vries (KdV) hierarchy
(see \cite{Di Francesco:1993nw} for a review). There is another
approach to topological gravity in which the partition function is a
series in $(t_n)$ with coefficients given by intersection numbers of
complex curves.  Witten conjectured \cite{Witten:1990hr} that the
partition functions of the two approaches coincide. This conjecture
was proved by Kontsevich \cite{Kontsevich:1992ti} who achieved the
computation of the intersection numbers in terms of weighted sums over
ribbon graphs (fat Feynman graphs), which he proved to be generated
from the Airy function matrix model (Kontsevich model)
\begin{equation}
\mathcal{Z}[E]=
\frac{\displaystyle \int dM \;\exp \big(-\tfrac{1}{2}\mathrm{tr}(EM^2)
+\tfrac{\mathrm{i}}{6} \mathrm{tr}(M^3)\big)}{
\displaystyle \int dM \;\exp \big(-\tfrac{1}{2}\mathrm{tr}(EM^2)
\big)}\;,
\label{Kontsevich}
\end{equation}
where $\mathrm{i}^2=-1$ and $E$ is a positive Hermitean matrix 
related to the series $(t_n)$ by $t_n=(2n-1)!!
\mathrm{tr}(E^{-(2n-1)})$. The large-$\mathcal{N}$ limit of 
(\ref{Kontsevich}) gives the KdV evolution equation, thus proving
Witten's conjecture. 

The Rieffel deformation of Euclidean space can be described by a
matrix product \cite{GraciaBondia:1987kw}.  Langmann, Szabo and
Zarembo showed \cite{Langmann:2003if} that the model for a complex
scalar field $\Phi$ with interaction $(\overline{\Phi}\star \Phi)^2$
in an external magnetic field $B=\Theta^{-1}$ which is dual
\cite{Langmann:2002cc} to the deformation matrix gives rise to a
matrix model with partition function $\displaystyle \int dM dM^\dag \;
\exp\big(-\mathrm{tr}(E M^\dag M - \lambda M^\dag M M^\dag
M)\big)$.  They showed that in the large-$\mathcal{N}$ limit 
the model is exactly solvable but trivial.
In joint work with Steinacker \cite{Grosse:2005ig, Grosse:2006tc}, one
of us used the relation of the Langmann-Szabo self-dual noncommutative 
$\phi^{\star 3 }$-model to the Kontsevich model to perform a non-perturbative
renormalisation and to extract the running of the asymptotically free
$\phi^{\star 3 }$-model in 6 dimensions.

In this paper we relate the Langmann-Szabo self-dual $\phi^{\star
  4}$-model (which in contrast to $\phi^ {\star 3}$ is stable) in 4
dimensions to a matrix model with partition function
\begin{equation}
\mathcal{Z}[E,J,\lambda]=
\frac{\displaystyle
\int dM \;\exp \big(\mathrm{tr}(JM)
-\mathrm{tr}(EM^2)
-\tfrac{\lambda}{4} \mathrm{tr}(M^4)\big)}{
\displaystyle
\int dM \;\exp \big(-\mathrm{tr}(EM^2)
-\tfrac{\lambda}{4} \mathrm{tr}(M^4)\big)}
\;.
\label{ZEJ}
\end{equation}
The matrix $E$ is fixed by identification with the noncommutative
field theory, but our key results hold for general $E$.  The
external source $J$ is used to generate correlation functions.  We
prove a universal algebraic recursion formula for all higher
correlation functions in terms of the two-point function and the
eigenvalues of $E$. The formula implies that if the two-point function is
renormalisable, then all higher correlation functions are
renormalisable, too, with vanishing $\beta$-function.  

For the specific case of the $\phi^{\star 4}$-model in 4 dimensions
we identify a scaling limit $\mathcal{N}{\to} \infty$ in which we can
also prove existence of the two-point function. Thanks to the
recursion formulae, this provides the exact solution of the model,
which in contrast to the Langmann-Szabo-Zarembo model is non-trivial.

It is
tempting to speculate that there might be relations to mathematical
structures which are the analogues of KdV hierarchy and intersection
theory of complex curves to which the Kontsevich model is related.

\subsection{Outline of the paper}

\label{subsec:outline}

This paper is a far-reaching generalisation of our previous work
\cite{Grosse:2009pa}. We first define in Sec.~\ref{sec:matrixmodel} a
general framework of field-theoretical matrix models for a compact
operator $M$ on a Hilbert space\footnote{We write $\phi$ instead of
  $M$ in Sec.~\ref{sec:matrixmodel}. For this outline we write $M$ to
  be consistent with (\ref{ZEJ}).}. Whereas the interaction is, as
usual, the trace of a polynomial in $M$, the kinetic term is
$\mathrm{tr}(EM^2)$ for a selfadjoint unbounded operator $E$ with
compact resolvent. The action of the group of unitaries on Hilbert
space induces an infinite number of Ward identities. In
Theorem~\ref{Thm:Ward} we turn this Ward identity into a formula for
the second derivative of the partition function. This is the most
crucial step which goes far beyond previous glimpses
\cite{Disertori:2006nq, Grosse:2009pa} of that formula.  It is then
straightforward to algebraically derive in Sec.~\ref{sec:SD} the
Schwinger-Dyson equation of \emph{any} correlation function (i.e.\
including non-planar functions and with several boundary components)
of the matrix model, and not only the planar 2- and 4-point functions
as in \cite{Grosse:2009pa}.

In the infinite volume limit, the non-planar sector with genus $g>0$
is suppressed, but there remain non-trivial sectors which have the
topology of a sphere with $B$ punctures/marked points/boundary
components. Accordingly, the theory is given by the set of
$(N_1{+}\dots{+}N_B)$-point functions, where the $i^{\mathrm{th}}$
boundary component carries $N_i$ external sources. For $B=1$ we prove
that \emph{the $2$-point function $G_{|ab|}$ satisfies a non-linear equation for
that function alone}. The matrix indices $a,b$ label the eigenvalues
$\{E_a\}$ of the external matrix $E$ in (\ref{ZEJ}), which may occur
with multiplicities. For a real theory with $M=M^*$ we prove in
Sec.~\ref{sec:recursion} an algebraic recursion formula which computes
any $(N\geq 4)$-point function $G_{|b_0\dots b_{N-1}|}$ in terms of
$G_{|ab|}$ and the eigenvalues of the external matrix $E$.  The
solution for the $4$-point function is remarkably simple,
$G_{b_0b_1b_2b_3}=(-\lambda) \frac{G_{b_0b_1}G_{b_2b_3}-
  G_{b_0b_3}G_{b_2b_1}}{(E_{b_0}-E_{b_2})(E_{b_1}-E_{b_3})}$.

The $N$-point function $G_{b_0\dots b_{N-1}}$ has a similar structure,
namely a sum of fractions with products of two-point functions in the
numerator and differences of eigenvalues of $E$ in the
denominator. This structure can be visualised in a graphical manner,
see Sec.~\ref{sec:graphics}. We put the indices $b_0,\dots,b_{N-1}$ in
cyclic order on a circle and symbolise a factor $G_{b_ib_j}$ by a
chord between $b_i,b_j$ and a factor $\frac{1}{E_{b_i}-E_{b_j}}$ by an
arrow from $b_i$ to $b_j$. This produces the non-crossing chord diagrams
counted by the Catalan numbers, decorated by two oriented trees, one connceting
all even vertices and one connecting all odd vertices, subject to the
condition that any of the trees intersects the chords only in the
vertices. The solution for $N\in \{4,6,8\}$ is given explicitly.

In Appendix \ref{appendix:B=2} we extend these results to 
$(N_1{+}N_2)$-point functions. The Schwinger-Dyson equations for the
$(1{+}1)$- and $(2{+}2)$-point functions are linear in the top degree
and depend otherwise only on the $N$-point functions already
known. Given these basic functions, any other $(N_1{+}N_2)$-point
function with one $N_i\geq 3$ is again purely algebraic. This pattern
continues to arbitrary $(N_1{+}\dots{+}N_B)$-point functions: The
basic functions with all $N_i\leq 2$ satisfy linear equations to be
solved case by case, whereas the higher functions with one $N_i\geq 3$
are universal and purely algebraic. 

The main consequence of these universal algebraic recursion formlulae
is that \emph{if the $2$-point function is renormalisable by
  wavefunction and mass renormalisation $E\mapsto
  Z(E+\frac{\mu^2-\mu_{bare}^2}{2})$, $\lambda \mapsto Z^2\lambda$,
  and if this renders the basic functions with $N_i\leq 2$ finite,
  then the whole quartic matrix model with action
  $\mathrm{tr}(EM^2+\frac{\lambda}{4}M^4)$ is non-perturbatively
  renormalisable and has vanishing $\beta$-function}
(Theorem~\ref{Thm:beta=0}). To the best of our knowledge such a
statement for the quartic matrix model was not known before; it came
in this generality completely unexpected for us.

So far this applies to any field-theoretical matrix model with
cut-off. One example is the model (\ref{GW0}) of a four-dimensional
noncommutative quantum field theory which we recall in
Sec.~\ref{sec:GW}. By results of Gracia-Bond\'{\i}a and V\'arilly
\cite{GraciaBondia:1987kw}, the model has a matrix realisation.  As
shown by us and together with Gayral in \cite{Grosse:2007jy,
  Gayral:2011vu}, the model has, as a noncommutative geometry, a
finite volume of diameter $\sqrt{\frac{\theta}{\Omega}}$, where
$\theta$ is the deformation parameter which determines the
$\star$-product. Since $\Omega=1$ is fixed, going to the thermodynamic
(infinite volume) limit amounts to the limit $\theta\to \infty$. The
limit $(\Omega=0,\;\theta\to \infty)$ was already emphasised by
Minwalla, Van Raamsdonk and Seiberg in \cite{Minwalla:1999px} and then
further studied by Becchi, Giusto and Imbimbo in
\cite{Becchi:2002kj,Becchi:2003dg} in the renormalisation group
framework.  Becchi et al found for graphs in momentum space, and we
confirm this for whole functions at $\Omega=1$, that the non-planar
sector is scaled away but there remains a non-trivial topology in form
of spheres with several holes (they called them `swiss cheese' in
\cite{Becchi:2002kj}).

Moreover, and this is the key to all further progress, the infinite
volume limit $\theta\to \infty$ turns the Schwinger-Dyson equation for
the basic 2-point function into an \emph{integral equation}
(Sec.~\ref{sec:Integral}). Performing the wavefunction renormalisation
in order to prepare for the continuum limit $\Lambda\to \infty$, this
integral equation for the \mbox{2-point} function $G_{ab}$ at
continuous ``matrix indices'' $a,b \in [0,\Lambda^2]$ is cubic in
$G_{ab}$ and has a singular integral kernel.  In our previous work
\cite{Grosse:2009pa} we have already arrived at this point.  The focus
there was put on a perturbative solution up to third order in the
coupling constant which revealed certain number-theoretic properties.
In this paper we arrange the equation in different manner. In this way
we recognise that the equation for $G_{ab}$ is linear in the function
with $a,b\neq 0$ and non-linear only in its boundary value $G_{a0}$.
Even more, the resulting linear equation at given boundary $G_{a0}$ is
the well-studied \emph{Carleman type singular integral equation}
\cite{Carleman, Tricomi, Muskhelishvili}.  Using the known solution of
that particular Riemann-Hilbert problem, we solve exactly the equation
for $G_{ab}$ in terms of $G_{a0}$. But because the 2-point function is
symmetric, and $G_{0b}=G_{b0}$ is solved in terms of $G_{a0}$, this
gives rise to a consistency equation $G=TG$
\begin{align}
G_{b0} 
= \frac{1}{1+b}
\exp\Bigg(- \lambda
\int_0^b \!\! dt \int_0^\infty \!\!\! dp\;
\frac{(G_{p0})^2}{\big(\lambda\pi p G_{p0}\big)^2 
+\big( 1 + tG_{p0} +\lambda\pi p
{\mathcal{H}_p}^{\!\!\!\!\infty}[G_{\bullet 0}]\big)^2} 
\Bigg) .
\label{G-b0-intro}
\end{align}
Here, ${\mathcal{H}_a}^{\!\!\!\!\infty}$ is the \emph{Hilbert transform}
$\displaystyle {\mathcal{H}_a}^{\!\!\!\!\infty}[f(\bullet)]=\frac{1}{\pi}
\lim_{\Lambda \to \infty}\lim_{\epsilon\to 0}
\Big(\int_0^{a-\epsilon}+\int_{a+\epsilon}^{\Lambda^2}\Big) dp
\;\frac{f(p)}{p-a}$.

The big progress over \cite{Grosse:2009pa} is that the non-linear
operator $T$ is, although still complicated, an improvement operator:
We prove in Sec.~\ref{sec:Master} that for $\lambda>0$ the operator
$T$ satisfies the assumptions of the Schauder fixed point theorem
which therefore guarantees that (\ref{G-b0-intro}) has a solution
$(b\mapsto G_{b0})\in \mathcal{C}^1_0(\mathbb{R}_+)$ which is
monotonously decreasing, positive, and vanishing together with its
derivative at infinity.

To be precise, the consistency equation (\ref{G-b0-intro}) relies on
an assumption which we do not check. In an accompanying paper
\cite{GW-numerical} we study (\ref{G-b0-intro}) numerically and
validate this assumption for coupling constants $0< \lambda \leq
\lambda^*=\frac{1}{\pi}$. For
$\lambda>\lambda^*$ we find that the required symmetry $G_{ab}=G_{ba}$
does not hold for the numerical solution of (\ref{G-b0-intro}). We
trace this back to the non-trivial solution \cite{Tricomi} of the
homogeneous Carleman equation which we have put to zero in the
derivation of (\ref{G-b0-intro}). The additional term is related to a
winding number \cite{Muskhelishvili} for the boundary value of an
analytic function.  It is very well possible that at $\lambda^*$ one
enters another sheet of the logarithm which would provide a
continuation of the solution to $\lambda> \frac{1}{\pi}$. Moreover, 
the  non-trivial solution of the homogeneous Carleman equation also
affects the uniqueness proof. All this is left for
future investigation.

In sec.~\ref{sec:effective} we determine the effective coupling
constant $\lambda_\textit{eff}=-G_{0000}$ explicitly in
terms of $\lambda$ and the fixed point solution $G_{b0}$. In agreement
with the general Theorem~\ref{Thm:beta=0} that all renormalisable
quartic matrix models have vanishing $\beta$-function, 
$\lambda_\textit{eff}$ differs only by a finite factor from the bare
coupling $\lambda$. This confirms at non-perturbative level the
perturbative result of Disertori, Gurau, Magnen and Rivasseau
\cite{Disertori:2006nq} that the self-dual noncommutative
$\phi^4_4$-model has vanishing $\beta$-function.

Our solution for the 2-point function and the universal algebraic
solutions for higher correlation functions can be seen as 
summation of infinitely many renormalised Feynman graphs. We check in
appendix~\ref{app:perturbative} that the perturbative expansion (at
next-to-lowest order) of our exact results coincides with the Feynman
graph computation.

We stress again that this is an exact solution of a Euclidean quantum
field theory. But thanks to the explicit knowledge of all correlation
functions it should be possible in not too distant future to decide
whether or not the model satisfies reflection positivity. An
affirmative answer is not impossible in view of the positivity and
monotonicity properties of the 2-point function. We are convinced
that in perturbation theory there is no chance to settle this question. 
We feel that getting thus far with this model was only
possible because there is a deep mathematical structure behind which
is still to be discovered.  We hope that the novel methods we developed
and the fascinating results we obtained might be relevant also for 
realistic quantum field theories in four dimensions.

\section{Ward identities in matrix models}

\label{sec:matrixmodel}

A classical scalar field on $\mathbb{R}^d$ is (e.g.) a continuous
function $\phi \in \mathcal{C}_0(\mathbb{R}^d)$ vanishing at $\infty$.
The space of such functions forms a commutative $C^*$-algebra and
hence is realised as a bounded linear operator on Hilbert space. One
natural translation of existence of the mass term $\int_{\mathbb{R}^d}
dx \;\frac{m^2}{2}\phi^2(x)$ to \emph{noncommutative operator
  algebras} is to require that $\phi^2$ is a trace-class operator,
which is the case if $\phi$ is a Hilbert-Schmidt compact operator on
Hilbert space.

\subsection{Field-theoretical matrix models}

A matrix for us is a compact operator on Hilbert space $H$.  For the
Hilbert space $H=\mathbb{C}^n$ this is an ordinary $n\times n$-matrix,
for $H=\ell^2(\mathbb{N})$ an infinite matrix and for
$H=L^2(\mathbb{R}^n)$ an integral kernel operator. Accordingly, we let
$I$ be a set of indices, which can be finite, countable or continuous.

We let $\mathcal{A}_I$ be the
algebra of matrices $\phi=(\phi_{ab})_{a,b \in I}$ associated with
$I$, i.e.\ the vector space of maps $I\times I\ni (a,b)\mapsto
\phi_{ab} \in \mathbb{C}$ equipped with the product
$(\phi\psi)_{ab}=\sum_{c\in I} \phi_{ac}\psi_{cb}$. 
For continuous $I$ the space $\mathcal{A}_I$ is
the space of integral kernels $\phi_{ab}\equiv \phi(a,b)$, and the
above sum is a short-hand notation for the integral $\sum_{c\in I}
\phi_{ac}\psi_{cb} \equiv \int_I d\mu(c)\; \phi(a,c)\psi(c,b)$ with
respect to a measure $\mu$ on $I$. It might be convenient to also
allow for weight factors in discrete sums, $(\phi\psi)_{ab}=\sum_{c\in I} \mu_c
\phi_{ac}\psi_{cb}$.

Recall that for continuous $I$ a matrix $(\phi_{ab})$ defines a
compact operator on \mbox{$H=L^2(I,\mu)$} (in fact a Hilbert-Schmidt operator)
if $(\phi_{ab}) \in L^2(I\times I,\mu\times \mu)$. The case of
infinite discrete $I$ is then a special case for the Dirac measure.
For $\psi=(\psi_{ab})$ in the trace-class ideal of compact operators we let
$\mathrm{tr}(\psi):=\sum_a \psi_{aa}$ be the trace ($\mathrm{tr}(\psi)
:=\sum_a \mu_a\psi_{aa}$ in presence of a measure), and
$\phi^*_{ab}=\overline{\phi_{ba}}$ defines the adjoint matrix of $\phi$.

We consider Euclidean quantum field theory for a single real scalar
field $\phi=\phi^* \in \mathcal{A}_I$ living on a space of volume $V$. At a
later point we will pass to densities in the infinite volume limit
$V\to \infty$ so that we must carefully keep track of volume
factors. The field theory is defined by an action functional 
\begin{align}
S[\phi]=V\; \mathrm{tr}(E\phi^2+P[\phi])\;.
\label{action}
\end{align}
Here, $P[\phi]$ is a polynomial in $\phi$ with
coefficients in $\mathbb{R}$, and
$S_{int}[\phi]=V\,\mathrm{tr}(P[\phi])$ would be a standard matrix
model action. Being interested in a situation closer
to field theory, we need an analogue of the kinetic term
$\int_{\mathbb{R}^d} dx\; \frac{1}{2} \phi(x) (-\Delta \phi)(x)$
involving the Laplacian.  We therefore let $E$ to be an
\emph{unbounded selfadjoint operator on Hilbert space with compact
  resolvent}. This means that the resolvent $(E-\mathrm{i})^{-1}$ is a
compact operator on $H$, i.e.\ a matrix, so that by little abuse of
notation we can view $E$ as a matrix, too. To make sense of
$V\,\mathrm{tr}(E\phi^2)$ we require $\phi$ to belong to an appropriate
subspace of the compact operators such that $E\phi^2$ is
trace-class. Note that the Kontsevich model \cite{Kontsevich:1992ti}
with action (\ref{Kontsevich}) fits into this setting, with the
resolvent of $E$ being trace-class.

The action (\ref{action}) gives rise to the partition function
\begin{align}
\mathcal{Z}[J]=\int \mathcal{D}[\phi]\;\exp(-S[\phi]
+V\,\mathrm{tr}(\phi J))\;,
\label{partition}
\end{align}
where $\mathcal{D}[\phi]=\prod_{a,b\in I}d\phi_{ab}$ is the extension
of the Lebesgue measure defined on finite-rank operators to the space
of selfadjoint matrices $\phi \in \mathcal{A}_I$ for which $E\phi^2$
is trace-class.  By $J\in\mathcal{A}_I$ we denote the external source.
The partition function gives rise to connected correlation functions
by
\begin{align}
\langle \varphi_{a_1b_1}\dots \varphi_{a_nb_n}\rangle_c := 
\frac{\partial^n \mathcal{W}[J]}{\partial J_{b_1a_1}\dots 
\partial J_{b_na_n}}\;,\qquad 
\mathcal{W}[J] := \frac{1}{V} \log \mathcal{Z}[J]\;.
\end{align}
The functional $\mathcal{W}[J]$ is the free energy density which
should have a limit for $V\to \infty$. Unless $I$ is finite, the
resulting index sums may diverge and require a renormalisation.

\subsection{Ward identity}

The action of the group $\mathcal{U}(H)$ of unitaries on Hilbert space
induces an adjoint action of $U\in \mathcal{U}(H)$ on bounded operators, 
$\mathcal{A}_I \ni \phi \mapsto U\phi U^*\in \mathcal{A}_I$. 
We let $\phi^U:= U\phi U^*$. Since the
adjoint action preserves the space of selfadjoint matrices and
$\phi$ is a dummy integration variable in the partition function, one
has 
\begin{align}
\mathcal{Z}[J]=\int \mathcal{D}[\phi]\;\exp(-S[\phi]
+V\,\mathrm{tr}(\phi J))
=\int \mathcal{D}[\phi^U]\;\exp(-S[\phi^U]+V\,\mathrm{tr}(\phi^U J))\;.
\end{align}
For finite matrices $A=A^*,B\in M_n(\mathbb{C})$, the Jacobian of the
transformation $A\mapsto BAB^*$ is $\det \frac{\partial
  BAB^*}{\partial A}= (\det B \det B^*)^n$. In particular,
$\det \frac{\partial  UAU^*}{\partial A}=1$ for $U\in
U(n)$. Therefore, the transformation formula for $\phi^U[\phi]$ yields
\begin{align}
0 = \int \mathcal{D}[\phi]\;\Big\{
\exp(-S[\phi]+V\,\mathrm{tr}(\phi J))
-\exp(-S[U\phi U^*]+V\,\mathrm{tr}(U\phi U^* J))\Big\}
\label{Ward-U}
\end{align}
for any unitary $U$. Note that the integrand itself does not vanish
because $V\,\mathrm{tr}(E\phi^2)$ and $V\,\mathrm{tr}(\phi J)$ are not
invariant under $\phi\mapsto \phi^U$. Linearisation of (\ref{Ward-U})
about $U=\mathrm{id}$ gives for the action 
$S[\phi]=V\,\mathrm{tr}(E\phi^2+P[\phi])$ the \emph{system of Ward identities}
\begin{align}
0 = \int \mathcal{D}[\phi]\;\Big(E\phi^2-\phi^2 E -J\phi+\phi J\Big)
\exp\big(-S[\phi]+V\,\mathrm{tr}(\phi J)\big)\;.
\label{Ward-E}
\end{align}
Note that (\ref{Ward-E}) is a matrix equation, which means that for an
infinite index set $I$ there is an infinite number of (scalar) Ward
identities. By adjoint action with $U$ we can always achieve that the
external matrix $E$ is diagonal, $E_{mn}=E_m \delta_{mn}$. In this
case the system takes the form
\begin{align}
0 = \int \mathcal{D}[\phi]\;\sum_{n\in I} \Big((E_p-E_a)
\phi_{pn}\phi_{na}
-J_{pn}\phi_{na}+J_{na}\phi_{pn} \Big)
\exp\big(-S[\phi]+V\,\mathrm{tr}(\phi J)\big)\;.
\label{Ward-Emn}
\end{align}
Writing under the integral $\phi_{ab}=\frac{1}{V} \frac{\partial}{\partial
  J_{ba}}$, we finally obtain (this was first 
derived in \cite{Disertori:2006nq}):
\begin{Proposition}
\label{Proposition-Ward}
The partition function $\mathcal{Z}[J]$ of a real scalar matrix model
with action $S[\phi]=V\,\mathrm{tr}(E\phi^2+P[\phi])$, where $E_{mn}=E_m
\delta_{mn}$ is diagonal and $P[\phi]$ a polynomial in
$\phi$, satisfies the system of Ward identities
\begin{align}
0 = \sum_{n\in I} \Big(\frac{(E_p-E_a)}{V}
\frac{\partial^2 \mathcal{Z}}{\partial J_{an} \partial J_{np}}
+J_{na}\frac{\partial \mathcal{Z}}{\partial J_{np}}  
-J_{pn}\frac{\partial \mathcal{Z}}{\partial J_{an}} 
\Big)\;,\qquad a,p\in I\;.
\label{eq:Ward}
\\*[-3ex]
\tag*{\mbox{$\square$}}
\end{align}
\end{Proposition}
We shall make a technical
\begin{Assumption}
The map $\iota:I\to \mathbb{R}$ defined by $\iota(m)=E_m$
is injective.
\label{assumption}
\end{Assumption}
This assumption is in fact less restrictive than it seems.  For fixed
potential $\mathrm{tr}(P[\phi])$, the partition function is completely
determined by the spectrum $\sigma(E)$ of $E$ and the spectral
density. Since $E$ is selfadjoint and has compact resolvent,
$\sigma(E)$ is discrete and has finite-dimensional eigenspaces.
Moreover, by achieving $E$ to be diagonal we have already placed
ourselves in an eigenbasis of $E$. The requirements on $E$ thus imply
that $I$ is necessarily countable, and the chosen diagonalisation
yields a bijection between $I$ and the set
$\{(\varepsilon,n_\varepsilon)\;:~ \varepsilon \in \sigma(E)\;,~
n_\varepsilon \in \{1,2,\dots,\mathrm{dim}(\ker (E-\varepsilon)) \}$.  
This means that index summations over $m\in I$ of a function $f$ which
depends only on the spectrum of $E$ but not on the multiplicity 
can be partitioned into 
\[
\sum_{m\in I} f(m)=\sum_{\varepsilon\in \sigma(E)} 
\sum_{n_\epsilon=1}^{\mathrm{dim}\ker (E-\varepsilon)} 
f(\epsilon,n_\varepsilon)
=\sum_{\varepsilon\in \sigma(E)} 
\mathrm{dim}(\ker (E-\varepsilon)) \cdot 
f(\epsilon)\;.
\]
Now the new summation over $\sigma(E)\ni \varepsilon$ with measure
$\mathrm{dim}(\ker (E-\varepsilon))$ is automatically injective.  In
other words, the Assumption~\ref{assumption} already covers the
general case if we include an appropriate measure in the index
summation. In order to make use of Proposition~\ref{Proposition-Ward}
we need at least two indices $a\neq p$ with $E_a\neq E_p$. This means
that the only case which we really exclude is the standard framework
of matrix models $V\,\mathrm{tr}(E\phi^2)=V\mu^2\mathrm{tr}(\phi^2)$ for
which there are highly developed technologies \cite{Di
  Francesco:1993nw}. Note that this $E=\mu^2 \mathrm{id}_H$ does not
have compact resolvent on infinite-dimensional Hilbert spaces.

\subsection{Expansion into  boundary components}

Feynman graphs in matrix models are ribbon graphs drawn on a genus-$g$
Riemann surface with $B$ boundary components (or
punctures/holes/marked points/faces) to which the external legs are
attached \cite{Grosse:2003aj}. Diagonality of the external matrix
$E_{mn}=E_m \delta_{mn}$ guarantees that the matrix index is constant
on every face (i.e.\ single line of the ribbon). This means that the
right index $b$ of a source $J_{ab}$ coincides with the left index of
another source $J_{bc}$, or of the same source $J_{bb}$.  Accordingly,
there is a decomposition of the free energy density 
$\mathcal{W}[J]=\frac{1}{V}\log \mathcal{Z}[J]$ into products of
$J$-cycles $J_{p_1p_2}J_{p_2p_3}\cdots J_{p_{n-1}p_n}
J_{p_np_1}=:J_{P^n}$, where $P^n=p_1\dots p_n$ stands for a collection
of $n$ indices. The decomposition according to the longest cycle reads
\begin{align}
\mathcal{W}[J]=\mathcal{W}[0]
+\sum_{L=1}^\infty \sum_{\genfrac{}{}{0pt}{1}{n_1,\dots,n_L = 0}{n_L\geq 1}}^\infty
\Big(\prod_{j=1}^L \frac{1}{n_j!j^{n_j}}\Big)
\sum_{P^j_{i_j} \in I^j}
G_{|P_1^{1}|\dots| P_{n_1}^{1}|
\dots |P_1^{L}|\dots| P_{n_L}^{L}|}
\prod_{j=1}^L \prod_{i_j=1}^{n_j} J_{P_{i_j}^{j}}\;.
\label{WGJ}
\end{align}
The total number of $J$-cycles in a function $G_{|P_1^{1}|\dots|
  P_{n_1}^{1}| \dots |P_1^{L}|\dots| P_{n_L}^{L}|}$ with
$N=\sum_{j=1}^L j n_j$ external sources is its number
$B=n_1+\dots+n_L$ of boundary components. For convenience we also list
those terms of this decomposition which contain at most four sources:
{\allowdisplaybreaks[4]
\begin{align}
\mathcal{W}[J]&=\mathcal{W}[0]+ \sum_{p \in I} G_{|p|} J_{pp} 
+ \sum_{p,q \in I} \Big\{ \frac{1}{2} G_{|pq|} J_{pq} J_{qp} 
+ \frac{1}{2} G_{|p|q|} J_{pp} J_{qq} \Big\}
\nonumber
\\*
&+ \sum_{p,q,r \in I} \Big\{  \frac{1}{6} G_{|p|q|r|} J_{pp} J_{qq} J_{rr} 
+ \frac{1}{2} G_{|p|qr|} J_{pp} J_{qr} J_{rq} 
+ \frac{1}{3} G_{|pqr|} J_{pq} J_{qr} J_{rp} \Big\}
\nonumber
\\
&+ \sum_{p,q,r,s \in I} \Big\{
 \frac{1}{24} G_{|p|q|r|s|} J_{pp} J_{qq} J_{rr}J_{ss} 
+ \frac{1}{4} G_{|p|q|rs|} J_{pp} J_{qq} J_{rs} J_{sr} 
+ \frac{1}{8} G_{|pq|rs|} J_{pq} J_{qp} J_{rs} J_{sr} 
\nonumber
\\*
& \qquad\qquad\quad 
+ \frac{1}{3} G_{|p|qrs|} J_{pp} J_{qr} J_{rs}J_{sq}
+ \frac{1}{4} G_{|pqrs|} J_{pq} J_{qr} J_{rs} J_{sp} \Big\} 
+ \mathcal{O}(J^6)\;.
\label{WJ4}
\end{align}}

In the sequel it is important to understand that the decomposition
into cycles also persists for coinciding indices. For example,
$G_{|pp|}J_{pp}J_{pp}$ is topologically different from
$G_{|p|p|}J_{pp}J_{pp}$. We keep track of this distinction by a
continuity argument. For that we use an appropriate embedding
$I\subset \mathbb{R}^k$ and a corresponding extension of $E_{m}$ to
any $\mathcal{C}^1$-function (continuously differentiable) on
$\mathbb{R}^k$ with local extrema outside $I$. Perturbatively, the
coefficients $G_{|\dots|}$ are rational functions of $E_m$ so that
they become differentiable functions of the indices, too. Identifying
the connected functions by functional derivative of $\mathcal{W}$ with
respect to $J_{ab}$ it is then important to perform the derivative at
generic indices, e.g.\
\[
G_{|pq|}
=\frac{\partial^2 \mathcal{W}}{\partial J_{pq} \partial J_{qp}}
\Big|_{J=0,q\neq p}\;,\qquad 
G_{|p|q|}
=\frac{\partial^2 \mathcal{W}}{\partial J_{pp} \partial J_{qq}}
\Big|_{J=0,q\neq p}\;.
\]
The case of coinciding indices is then obtained by continuity, 
$G_{|pp|}=\lim_{q\to p} G_{|pq|}$ and 
$G_{|p|p|}=\lim_{q\to p} G_{|p|q|}$.

We are now going to prove the following Ward identity, which is the
key to solve our model.
\begin{Theorem}
\label{Thm:Ward}
For injective $m\mapsto E_m$, the partition function 
$\mathcal{Z}[J]$ of a real scalar matrix model
with action $S[\phi]=V\,\mathrm{tr}(E\phi^2+P[\phi])$, where $E_{mn}=E_m
\delta_{mn}$ is diagonal and $P[\phi]$ a polynomial in
$\phi$, satisfies the system of Ward identities
{\allowdisplaybreaks[4]
\begin{align}
\sum_{n\in I} 
\frac{\partial^2 \mathcal{Z}[J]}{\partial J_{an} \partial J_{np} }
&= \delta_{ap} (V^2 W_a^1[J]+V W_a^2[J])\mathcal{Z} 
+ \frac{V}{E_{p}-E_{a}}\sum_{n \in I}\Big(
J_{pn}\frac{\partial \mathcal{Z}[J]}{\partial J_{an}}
{-}J_{na}\frac{\partial\mathcal{Z}[J]}{\partial J_{np}}\Big)\;,
\raisetag{3ex}
\label{ZJJ}
\\
W_a^2[J]&:=
\sum_{L=0}^\infty \sum_{n_1,\dots,n_L = 0}^\infty
\Big(\prod_{j=1}^L \frac{1}{n_j!j^{n_j}}\Big) 
\sum_{P^j_{i_j} \in I^j}
\Big(\prod_{j =1}^L \prod_{i_j=1}^{n_j} J_{P_{i_j}^{j}}\Big) \times 
\nonumber
\\*[-1ex]
&\quad
\Big( G_{|a|a|P_1^{1}|\dots| P_{n_L}^{L}|}
+ 
\sum_{n\in I} G_{|P_1^{1}|\dots| P_{n_1}^{1}|an|P_1^{2}|
\dots| P_{n_L}^{L}|}
\nonumber
\\*[-1ex]
&\qquad
+ \sum_{k=3}^L \;
\sum_{n,q_1,\dots,q_{k-3}\in I}  \!\!\!
G_{|P_1^{1}|\dots| P_{n_{k-1}}^{k-1}|
nan q_1\dots q_{k-3}|P_1^{k}|\dots| P_{n_L}^{L}|}
J_{nq_1}J_{q_1q_2}\dots J_{q_{k-3}n}
\Big)\;,
\nonumber
\\
W_a^1[J]&:=
\sum_{L,L'=0}^\infty \;
\sum_{n_1,\dots,n_L,m_1,\dots,m_{L'} = 0}^\infty\;
\Big(\prod_{j=1}^L \frac{1}{n_j!j^{n_j}}\Big) 
\Big(\prod_{k=1}^{L'} \frac{1}{m_k!k^{m_k}}\Big) \times 
\nonumber
\\*[-1ex]
&\qquad \quad
\sum_{P^j_{i_j} \in I^j}
\sum_{Q^k_{l_k} \in I^k}
\Big(\prod_{j =1}^L \prod_{i_j=1}^{n_j} J_{P_{i_j}^{j}}\Big) 
\Big(\prod_{k =1}^{L'} \prod_{l_k=1}^{m_k} J_{Q_{l_k}^{k}}\Big) 
G_{|a|P_1^{1}|\dots| P_{n_L}^{L}|}
G_{|a|Q_1^{1}|\dots| Q_{m_{L'}}^{L'}|} \Big\}\;.
\nonumber
\end{align}}
\end{Theorem}
\emph{Proof.} 
In $\displaystyle \sum_{n\in I} \frac{\partial^2
  \mathcal{Z}[J]}{\partial J_{an} \partial J_{np} } = \displaystyle
\sum_{n\in I} \Big( V \frac{\partial^2 \mathcal{W}[J]}{\partial J_{an}
  \partial J_{np} } + V^2 \frac{\partial \mathcal{W}[J]}{\partial J_{an}}
\frac{\partial \mathcal{W}[J]}{\partial J_{np}} \Big)
\mathcal{Z}[J]$ we write 
\[
\frac{\partial^2 \mathcal{W}[J]}{\partial J_{an}  \partial J_{np} }
= \delta_{ap} W_a^{2}[J]+W^{2,reg}_{ap}[J]\;,
\qquad
\frac{\partial \mathcal{W}[J]}{\partial J_{an}}
\frac{\partial \mathcal{W}[J]}{  \partial J_{np} }
= \delta_{ap} W_a^1[J]+W^{1,reg}_{ap}[J]\;,
\]
where $\delta_{ap}W_a^i[J]$ contain all terms in which the derivatives
generate a factor $\delta_{ap}$, and $W^{i,reg}_{ap}[J]$ are the
remainders. Applying the functional derivatives to (\ref{WGJ}) we
derive the structure of $W_a^1,W_a^2$ as given in the Theorem. 
Namely, a $\delta_{ap}$ arises in 
$\frac{\partial^2 \mathcal{W}[J]}{\partial J_{an}  \partial J_{np} }$
if (in that order) 
\begin{itemize}\itemsep 0pt\parsep 0pt \topsep-1ex\partopsep-1ex
\item both $\frac{\partial}{\partial J_{an} }$
and $\frac{\partial}{\partial J_{np} }$ act on different
$J$-cycles each of length 1, or

\item both $\frac{\partial}{\partial J_{an} }$
and $\frac{\partial}{\partial J_{np} }$ act on the same 
$J$-cycle of length 2, or

\item $\frac{\partial}{\partial J_{np} }$ acting on a $J$-cycle of
  length $k\geq 3$ gives the chain $J_{pq_0}J_{q_0q_1}\dots
    J_{q_{k-3}n}$. The second derivative 
$\frac{\partial}{\partial J_{an} }$ then acts on $J_{pq_0}$ in that chain.
\end{itemize}
In $\frac{\partial \mathcal{W}[J]}{\partial J_{an}}
\frac{\partial \mathcal{W}[J]}{\partial J_{np}}$, the only possibility
to generate a $\delta_{ap}$ is to let
\begin{itemize}\itemsep 0pt \topsep-1ex
\item both $\frac{\partial}{\partial J_{an} }$
and $\frac{\partial}{\partial J_{np} }$ act on 
$J$-cycles each of length 1.
\end{itemize}
The combinatorial factors in (\ref{WGJ}) are reproduced after a shift.
It is straightforward to also write down the remainders
$W^{i,reg}_{ap}[J]$ in explicit form; but as we do not need this we
refrain from listing the result.

We consider the case $a\neq p$. On one hand this implies
$\delta_{ap}W_a^{i}=0$, on the other hand we may divide the expression
in Proposition~\ref{Proposition-Ward} by $\frac{E_p-E_a}{V} \neq 0$ to
obtain
\begin{align}
a\neq p ~~\Rightarrow~~ \sum_{n\in I} 
(V^2 W^{1,reg}_{ap}+V W^{2,reg}_{ap})\mathcal{Z} 
=\sum_{n\in I}
\frac{V}{E_p-E_a} \Big(
J_{pn}\frac{\partial \mathcal{Z}[J]}{\partial J_{an}}
{-}J_{na}\frac{\partial\mathcal{Z}[J]}{\partial J_{np}}\Big)\;.
\label{proof-1}
\end{align}
It is now important to realise that the identity (\ref{proof-1})
also extends to $p=a$. Namely, the partition function and its derivatives
such as $W^{1,reg}_{ap},W^{2,reg}_{ap}$ are completely determined by
the entries $E_m$ in the external matrix. Extending $m\mapsto E_m$ to
a differentiable function, the rhs of (\ref{proof-1}) is at least
continuous at $p\to a$ by l'H\^opital's rule and agrees with the lhs
for which there is no obstruction to go to $p=a$. This finishes the
proof. \hfill $\square$%

\section{Schwinger-Dyson equations}

\label{sec:SD}

\subsection{Strategy}

A standard procedure in quantum field theory consists in expressing
the fields $\phi$ in the interaction
$S_{int}[\phi]=V\,\mathrm{tr}(P[\phi])$ of the partition function
(\ref{partition}) as derivatives with respect to the sources,
\[
\mathcal{Z}[J]=\int \mathcal{D}[\phi]\;
\exp\Big(-S_{int}\Big[\frac{1}{V}\frac{\partial}{\partial J}\Big] \Big)
\exp\big(-V\,\mathrm{tr}(E\phi^2) +V\,\mathrm{tr}(\phi J)\big)\;.
\]
In this formulation the Gau\ss{}ian $\phi$-integration can formally 
be carried out and gives
\begin{align}
\mathcal{Z}[J]= C \exp\Big(-S_{int}
\Big[\frac{1}{V}\frac{\partial}{\partial J}\Big]
\Big) \exp\big(\tfrac{V}{2}\langle J,J\rangle_E\big)\;,\qquad
\langle J,J\rangle_E:=\sum_{m,n\in I} \frac{J_{mn} J_{nm}}{E_m+E_n}\;.
\label{partition-2}
\end{align}
The constant (possibly ill-defined) prefactor $C$ will be omitted.
The power-series expansion of $e^{-S_{int}[\frac{1}{V}\frac{\partial}{\partial J}]}$
gives rise to Feynman graphs, but since the expansion looses the
distinction between $e^{-S_{int}[\frac{1}{V}\frac{\partial}{\partial J}]}$ and
$e^{+S_{int}[\frac{1}{V}\frac{\partial}{\partial J}]}$, the expansion does not
converge. A better strategy is to keep $e^{-S_{int}[\frac{1}{V}
\frac{\partial}{\partial J}]}$ intact and instead let functional derivatives
$\phi_{pq}=\frac{1}{V}\frac{\partial}{\partial J_{qp}}$ act on $\mathcal{W}$ in
order to produce the connected functions $G$ in the expansion
(\ref{WGJ}). These ``external'' derivatives combine with the
``internal'' derivatives in $e^{-S_{int}[\frac{1}{V}\frac{\partial}{\partial J}]}$ to
certain identities called \emph{Schwinger-Dyson equations}. It turns
out that for each function $G$ the Ward identity of
Theorem~\ref{Thm:Ward} can be used at an intermediate step to generate
new relations.

We shall demonstrate this for the regular (i.e.\ $B=1$) two-point
function $G_{|ab|}$. As usual we impose $a\neq b$ so that
$0=(\frac{\partial \mathcal{Z}}{\partial J_{ab}})\big|_{J=0}$. We have
\begin{align}
G_{|ab|}&=\frac{1}{V} 
\frac{\partial^2 (\log \mathcal{Z})[J]}{\partial J_{ba} \partial
  J_{ab}}\Big|_{J=0}
=\frac{1}{V} \frac{1}{\mathcal{Z}[0]}\frac{\partial^2 \mathcal{Z}[J]}{\partial
  J_{ba}\partial J_{ab}} \Big|_{J=0}-
\frac{1}{V}\frac{1}{\mathcal{Z}[0]}\frac{\partial \mathcal{Z}[J]}{\partial
  J_{ba}}\Big|_{J=0}
\frac{1}{\mathcal{Z}[0]}\frac{\partial \mathcal{Z}[J]}{\partial
  J_{ab}}\Big|_{J=0}
\nonumber
\\
&= \frac{1}{V}\frac{1}{\mathcal{Z}[0]}
\Big\{ \frac{\partial}{\partial J_{ba}}
e^{-S_{int}[\frac{1}{V}\frac{\partial}{\partial J}]}
\frac{\partial}{\partial J_{ab}}e^{\frac{V}{2}\langle
  J,J\rangle_E}\Big\}_{J=0}
\nonumber
\\
&= \frac{1}{E_a+E_b}+\frac{1}{(E_a+E_b)\mathcal{Z}[0]} 
\Big\{ \Big(\phi_{ab}\frac{\partial(-S_{int})}{\partial
  \phi_{ab}}\Big)\Big[\frac{1}{V}\frac{\partial}{\partial J}\Big]\Big\}
\mathcal{Z}[J]\Big|_{J=0}\;.
\label{Gab-general}
\end{align}
For the most general interaction
$\displaystyle
S_{int}[\phi]=V\sum_{k=3}^K 
\sum_{p_1,\dots,p_k\in I} 
\frac{\lambda_k}{k} \phi_{p_1p_2}\cdots 
\phi_{p_{k-1}p_k} \phi_{p_kp_1}$ we have 
\begin{align}
\Big(\phi_{ab}\frac{\partial(-S_{int})}{\partial
  \phi_{ab}}\Big)\Big[\frac{1}{V}\frac{\partial}{\partial J}\Big]
=  \sum_{k=3}^K 
\sum_{n,p,p_1,\dots,p_{k-3}\in I} 
\frac{(-\lambda_k)}{V^{k-1}} \frac{\partial^{k-2}}{
\partial J_{pp_1}\dots \partial J_{p_{k-3}b}
\partial J_{ba}  } 
\frac{\partial^2}{\partial J_{an} \partial J_{np}}\;,
\label{Vphi-gen}
\end{align}
i.e.\ in any part we encounter the 
two-fold derivative $\frac{\partial^2}{\partial J_{an} \partial
  J_{np}}$ known from Theorem~\ref{Thm:Ward}. This appearance of 
$\frac{\partial^2}{\partial J_{an} \partial
  J_{np}}$ is a general feature of any connected function $G$.

\subsection{Schwinger-Dyson equations for a $\phi^4$-model:
$B=1$ cycle}

We now specify to the case
$S_{int}[\phi]=V\,\frac{\lambda_4}{4}\mathrm{tr}(\phi^4)$. We first evaluate
(\ref{Gab-general}) using (\ref{Vphi-gen}) and Theorem~\ref{Thm:Ward},
taking $a\neq b$ into account and the fact that $G_{|q|}=0$:
\begin{align}
G_{|ab|}&=\frac{1}{E_a+E_b}+\frac{(-\lambda_4)}{V^3 (E_a+E_b)\mathcal{Z}[0]} 
\Big\{\frac{\partial^2\big((V^2W_a^1[J]+V W_a^2[J])\mathcal{Z}[J]\big)}{
\partial J_{ab}\partial J_{ba}  }
\nonumber
\\
&\qquad\qquad\qquad+\sum_{p,n \in I} 
\frac{V}{E_p-E_a} 
\frac{\partial^2}{\partial J_{pb} \partial J_{ba}}
\Big(J_{pn}\frac{\partial \mathcal{Z}}{\partial J_{an}}-
J_{na}\frac{\partial \mathcal{Z}}{\partial J_{np}}\Big)\Big\}_{J=0}
\nonumber
\\
&=\frac{1}{E_a+E_b}+\frac{(-\lambda_4)}{(E_a+E_b)} 
\Big\{ \frac{1}{V} G_{|ab|}\Big(G_{|a|a|}+\sum_{n\in I}G_{|an|}\Big)
\nonumber
\\
&\qquad\qquad\qquad\qquad 
+ \frac{1}{V^2}\Big(G_{|a|a|ab|}+\sum_{n\in I} G_{|an|ab|}
+ G_{|aaab|}+G_{|baba|}\Big)
\nonumber
\\
&\qquad\qquad\qquad
+\frac{1}{V}\sum_{p \in I} \frac{G_{|ab|}- G_{|pb|}}{E_p-E_a} 
+\frac{1}{V} \frac{G_{|a|b|}- G_{|b|b|}}{E_b-E_a} 
\Big\}\;.
\label{G2-regular}
\end{align}
To obtain the last line one has to use 
$\frac{\partial \mathcal{Z}}{\partial J_{an}}
= V\mathcal{Z}[0]\big(G_{|an|} J_{na}+\delta_{an} \sum_{q\in I}
G_{|a|q|} J_{qq}+\mathcal{O}(J^2)\big)$
and $\frac{\partial \mathcal{Z}}{\partial J_{np}}
= V\mathcal{Z}[0]\big(G_{|pn|} J_{pn}+\delta_{np} \sum_{q\in I}
G_{|p|q|} J_{qq}+\mathcal{O}(J^2)\big)$.

In a similar manner we obtain for pairwise different indices $a,b_i$
the following equation for the regular (i.e.\ single-cycle) 
$(N\geq 4)$-point function: 
\begin{align*}
G_{|a b_1\dots b_{N-1}|}
&= \frac{(-\lambda_4)}{V^3(E_a+E_b)\mathcal{Z}[0]} 
\Big\{\frac{\partial^N\big((V^2 W_a^1[J]+V W_a^2[J])\mathcal{Z}[J]\big)}{
\partial J_{ab_1}\partial J_{b_1b_2} \dots \partial J_{b_{N-2}b_{N-1}} 
\partial J_{b_{N-1}a} }
\\*
&  
+\sum_{p,n \in I} 
\frac{V}{E_p-E_a} 
\frac{\partial^N}{\partial J_{pb_1} \partial J_{b_1b_2}\cdots 
\partial J_{b_{N-2}b_{N-1}} \partial J_{b_{N-1}a}}
\Big(J_{pn}\frac{\partial \mathcal{Z}}{\partial J_{an}}-
J_{na}\frac{\partial \mathcal{Z}}{\partial J_{np}}\Big)\Big\}_{J=0}\;.
\end{align*}
Since all indices $a,b_i$ are pairwise different, the first line gives
$V^2W_a^1[0]+VW_a^2[0]$ times the application of all derivatives to
$\mathcal{Z}[J]$ plus $\mathcal{Z}[0]$ times the application of all
derivatives to $(V^2W_a^1[J]+VW_a^2[J])$. In the first case there is
actually only a contribution from the application of all derivatives
to $\log \mathcal{Z}[J]=V\mathcal{W}[J]$, because the indices are
pairwise different. The second line is a sum of terms where one of the
derivatives acts on $J_{pn}$ minus the action of $J_{b_{N-1}a}$ on
$J_{na}$, with all other derivatives acting on $\frac{\partial
  \mathcal{Z}[J]}{\partial J_{an}}$ and $\frac{\partial
  \mathcal{Z}[J]}{\partial J_{np}}$, respectively. Depending on the
Kronecker-$\delta$ arising in this way, the derivatives acting on
$\mathcal{Z}[J]$ either form a single cycle or two cycles. For the
derivative $\frac{\partial^N\mathcal{Z}[J]}{\partial J_{pb_1}\partial
  J_{b_1b_2} \dots \partial J_{b_{N-2}b_{N-1}}\partial J_{b_{N-1}p}}$
special care is needed in the cases $p=b_k$. Either these are the
necessarily arising cases in derivatives which form a single cycle, or
the derivatives form two cycles $\frac{\partial^k\mathcal{Z}[J]}{
\partial J_{b_kb_1}\partial J_{b_1b_2}
\dots \partial J_{b_{k-1}b_{k}}}$ and 
$\frac{\partial^{N-k}\mathcal{Z}[J]}{
  \partial J_{b_kb_{k+1}} \dots \partial
  J_{b_{N-2}b_{N-1}}J_{b_{N-1}b_k}}$ with one common index $b_k$.  The
single-cycle cases reduce to its action on $\log\mathcal{Z}[J]$. For
the case of two cycles there are two possibilities: Either both act on
a two-cycle contribution of $\log\mathcal{Z}[J]$, or on one-cycle
contributions of $\frac{1}{2!}(\log\mathcal{Z}[J])^2$.  With these
remarks we find, noting that each $\log \mathcal{Z}$ contributes a
factor $V$,
\begin{align}
&G_{|a b_1\dots b_{N-1}|}
\nonumber
\\
&= \frac{(-\lambda_4)}{E_a+E_{b_1}}\Big\{
\frac{1}{V} \Big(G_{|a|a|}+\sum_{n\in I} G_{|an|}\Big) G_{|ab_1\dots b_{N-1}|}
\nonumber
\\
&
+ \frac{1}{V^2}\Big(G_{|a|a|ab_1\dots b_{N-1}|}
+ \sum_{n\in I} G_{|an|ab_1\dots b_{N-1}|}
+ G_{|aaab_1\dots b_{N-1}|}
+\sum_{k=1}^{N-1} 
G_{|b_1\dots b_kab_k\dots b_{N-1}a|}\Big)
\nonumber
\\
& - \frac{1}{V} \sum_{p\in I}
\frac{G_{|pb_1\dots b_{N-1}|}-G_{|ab_1\dots b_{N-1}|}}{E_p-E_a} 
-\frac{1}{V}\sum_{k=1}^{N-1}\frac{G_{|b_1\dots b_k|b_{k+1}\dots b_{N-1}b_k|}-
G_{|b_1\dots b_k|b_{k+1}\dots b_{N-1}a|}}{E_{b_k}-E_a}
\nonumber
\\
&
-\sum_{l=1}^{\frac{N-2}{2}} G_{|b_1\dots b_{2l}|}
\frac{G_{|b_{2l+1}\dots  b_{N-1}b_{2l}|}-G_{|b_{2l+1}\dots b_{N-1}a|}
}{E_{b_{2l}}-E_a}
\Big\}\;.
\label{GN-regular}
\end{align}

In Appendix \ref{appendix:B=2} we derive the Schwinger-Dyson
equations for functions with $B=2$ boundary components.

\subsection{An algebraic recursion formula for a real field theory}

\label{sec:recursion}

The generating functional $\mathcal{W}[J]$ for connected $N$-point
functions is real. For real fields $\phi=\phi^*$ also the source $J$
is real, i.e.\ $J=J^*$ or
$J_{\under{p}\under{q}}=\overline{J_{\under{q}\under{p}}}$.
Selfadjointness of $E$ implies $E_a=\overline{E_a}$. Therefore, the
expansion coefficients of $\mathcal{W}[J]$ in (\ref{WGJ}) are all
real, $G_{|P^1_1|\dots |P^L_{n_L}|}=\overline{G_{|P^1_1|\dots
    |P^L_{n_L}|}}$. This is clear in perturbation theory where the
$G_{|P^1_1|\dots |P^L_{n_L}|}$ are represented by ribbon graphs made
of propagators $\frac{1}{E_a+E_b}$ and vertices $(-\lambda_4)$. It
follows non-perturbatively from the reality of the equations
(\ref{G2-regular}), (\ref{GN-regular}) and their $(B\geq
2)$-analogues, possibly after a decoupling of these equations by genus
expansion introduced in section~\ref{sec:genusexpansion}.
These considerations imply for the $B=1$ sector
\begin{align*}
&\sum_{p_0,\dots,p_{N-1} \in I} \!\!\!\!\!
G_{|p_0\dots p_{N-1}|} J_{p_0p_1}
\cdots J_{p_{N-2}p_{N-1}} 
J_{p_{N-1}p_0} 
= \!\!\! \sum_{p_0,\dots,p_{N-1} \in I}\!\!\!\!\!
G_{|p_0\dots p_{N-1}|}
J_{p_{0}p_{N-1}} 
J_{p_{N-1}p_{N-2}} \cdots 
J_{p_{1}p_{0}}\;.
\end{align*}
Renaming the indices $p_k \leftrightarrow p_{N-k}$ for $k\in
\{1,2,\dots \frac{N}{2}\}$ we notice that $N$-point functions are not
only invariant under cyclic permutations but also \emph{invariant
  under reversal of the order of its indices} (or orientation
reversal), and this extends to arbitrary $B\geq 1$:
\begin{align}
G_{|p^1_0p^1_1\dots p^1_{N_1-1}|\dots 
|p^B_0p^B_1\dots p^B_{N_B-1}|}
=
G_{|p^1_0p^1_{N_1-1}\dots p^1_{1}|\dots 
|p^B_0p^B_{N_B-1}\dots p^B_{1}|}\;.
\label{GN-reversal}
\end{align}
The equation (\ref{GN-reversal}) is an empty condition if all
$N_i\leq 2$. But as soon as $N_i > 2$ for at least one $1\leq i\leq
B$ it allows us to completely solve $G_{|p^1_0p^1_1\dots p^1_{N_1-1}|\dots 
|p^B_0p^B_1\dots p^B_{N_B-1}|}$ in terms of the functions having all 
$N_i\leq 2$. Here we demonstrate this for $B=1$. The case
$B=2$ and an outlook to $B>2$ is given in Appendix~\ref{appendix:B=2}.

Starting point is (\ref{GN-regular}) which we multiply by
$(E_a+E_{b_1})$. From the resulting equation we subtract the equation
obtained by the same procedure but with renamed indices
$b_k\leftrightarrow b_{N-k}$ for $1\leq k\leq \frac{N}{2}$. From
(\ref{GN-reversal}) we conclude $(E_a+E_{b_1})G_{|ab_1\dots b_{N-1}|}-
(E_a+E_{b_{N-1}})G_{|ab_{N-1}\dots b_{1}|}=
(E_{b_1}-E_{b_{N-1}})G_{|ab_1\dots b_{N-1}|}$. On the rhs of
(\ref{GN-regular}) all terms not having $E_{b_k}$ in the denominator
cancel by (\ref{GN-reversal}), leaving
\begin{align*}
(E_{b_1}-E_{b_{N-1}})G_{|ab_1\dots b_{N-1}|}
&= \frac{\lambda_4}{V}
\sum_{k=1}^{N-1}
\Big(
\frac{G_{|b_1\dots b_k|b_{k+1}\dots b_{N-1}b_k|}-
G_{|b_1\dots b_k|b_{k+1}\dots b_{N-1}a|}}{E_{b_k}-E_a}
\nonumber
\\
& \qquad\qquad 
-\frac{G_{|b_{N-1}\dots b_{N-k}|b_{N-k-1}\dots b_{1}b_{N-k}|}-
G_{|b_{N-1}\dots b_{N-k}|b_{N-k-1}\dots b_{1}a|}}{E_{b_{N-k}}-E_a}
\Big)
\nonumber
\\
&
+\lambda_4 \sum_{l=1}^{\frac{N-2}{2}} \Big(
G_{|b_1\dots b_{2l}|}
\frac{G_{|b_{2l+1}\dots  b_{N-1}b_{2l}|}-G_{|b_{2l+1}\dots b_{N-1}a|}
}{E_{b_{2l}}-E_a}
\nonumber
\\
& \qquad\qquad 
-
G_{|b_{N-1}\dots b_{N-2l}|}
\frac{G_{|b_{N-2l-1}\dots  b_{1}b_{N-2l}|}-G_{|b_{N-2l-1}\dots b_{1}a|}
}{E_{b_{N-2l}}-E_a}
\Big)
\;.
\end{align*}
Rearranging the sums to common denominators 
$E_{b_k}-E_a$ and taking (\ref{GN-reversal}) into account, we conclude:
\begin{Theorem}
\label{Thm:GN-algebraic}
  The (unrenormalised) connected $(N \geq 4)$-point function at $B=1$
  of a real scalar matrix model with action
  $S=V\mathrm{tr}\,(E\phi^2+\frac{\lambda_4}{4} \phi^4)$ is for
  injective $m\mapsto E_m$ given by the algebraic recursion formula
\begin{align}
G_{|b_0b_1\dots b_{N-1}|}
&= (-\lambda_4)
\sum_{l=1}^{\frac{N-2}{2}} 
\frac{G_{|b_0 b_1 \dots b_{2l-1}|} G_{|b_{2l}b_{2l+1}\dots b_{N-1}|} 
- G_{|b_{2l} b_1 \dots b_{2l-1}|} G_{|b_0 b_{2l+1}\dots b_{N-1}|} 
}{(E_{b_0}-E_{b_{2l}})(E_{b_1}-E_{b_{N-1}})}
\nonumber
\\
& + \frac{(-\lambda_4)}{V} 
\sum_{k=1}^{N-1} 
\frac{G_{|b_0 b_1 \dots b_{k-1}|b_{k}b_{k+1}\dots b_{N-1}|} 
- G_{|b_{k} b_1 \dots b_{k-1}|b_0 b_{k+1}\dots b_{N-1}|} 
}{(E_{b_0}-E_{b_{k}})(E_{b_1}-E_{b_{N-1}})}\;.
\label{GN-algebraic-E}
\\[-3ex]
\tag*{\mbox{$\square$}}
\end{align}
\end{Theorem}

The equation (\ref{G2-regular}) for $G_{|ab|}$ is not algebraic. It involves
  sums over the index set $I$ which in general will diverge for
  infinite $I$. One would try to absorb these divergences in a field
  redefinition $\phi \mapsto \sqrt{Z}\phi$ in the initial action 
$S=V\mathrm{tr}\,(E\phi^2+\frac{\lambda_4}{4} \phi^4)$ and a mass
renormalisation of the smallest eigenvalue of $E$. This amounts to
replace $E_a \mapsto Z(E_a+\frac{\mu^2}{2}-\frac{\mu_{bare}^2}{2})$
and $\lambda_4 \mapsto Z^2\lambda_4$. Remarkably, these replacements
leave (\ref{GN-algebraic-E}) invariant. Anticipating a similar
recursion formula for $B=2$ we can formulate this
observation as follows:
\begin{Theorem}
\label{Thm:beta=0}
  Given a real scalar matrix model with action
  $S=V\,\mathrm{tr}(E\phi^2+\frac{\lambda_4}{4} \phi^4)$ and $m\mapsto
  E_m$ injective. If the basic functions
  $G_{|p_1|\dots|p_{2k}|p_{2k+1}p_{2k+2}|\dots|p_{N-1}p_{N}|}$ with at
  most two sources per cycle are turned finite by 
  renormalisation $E_a \mapsto
  Z(E_a+\frac{\mu^2}{2}-\frac{\mu_{bare}^2}{2})$ and $\lambda_4
  \mapsto Z^2\lambda_4$, then all ($N{\geq} 4$)-point functions
  $G_{|b_0b_1\dots b_{N-1}|}$ are finite (at least for pairwise
  different indices) without further need of a renormalisation of
  $\lambda_4$. This means that all such $\phi^4$-matrix models have
  vanishing $\beta$-function. 

  All higher correlation functions are given by algebraic recursion
  formulae in terms of renormalised basic functions
  $G_{|p_1|\dots|p_{2k}|p_{2k+1}p_{2k+2}|\dots|p_{N-1}p_{N}|}$.
  \hfill $\square$%
\end{Theorem}

There remains some worry about $G_{|b_0b_1\dots b_{N-1}|}$ at
coinciding indices. In perturbation theory these are rational
functions of $E_{b_i}$ so that the $E_{b_i}-E_{b_j}$ in the
denominator of (\ref{GN-algebraic-E}) cancel. In a thermodynamic limit
$V\to \infty$ and $\frac{1}{V}\sum_{i\in I}$ finite one can hope that
the basic functions
$G_{|p_1|\dots|p_{2k}|p_{2k+1}p_{2k+2}|\dots|p_{N-1}p_{N}|}$ become
smooth functions of continuous indices so that the limit to coinciding
indices can exist.

Observe that the limit $V\to \infty$ eliminates the second line of
(\ref{GN-algebraic-E}). We show in section~\ref{sec:genusexpansion}
that this agrees with the restriction to the planar sector. In
section~\ref{sec:graphics} we give a graphical solution of the planar
part of (\ref{GN-algebraic-E}). This involves non-crossing chord
diagrams counted by the Catalan numbers.

We recall that Disertori, Gurau, Magnen and Rivasseau proved in
\cite{Disertori:2006nq} that the $\beta$-function of self-dual
noncommutative $\phi^4_4$-theory vanishes to all orders in
perturbation theory. We have generalised the strategy of
\cite{Disertori:2006nq} to a very general class of $\phi^4$-matrix
models and proved by Theorem~\ref{Thm:beta=0} that \emph{all these models
have a non-perturbatively vanishing $\beta$-function, provided that
the two-point function is renormalisable}.

\subsection{The genus expansion}

\label{sec:genusexpansion}

Our strategy will be to perform the infinite volume limit $V\to
\infty$ in such a manner that the the densitised summation
$\frac{1}{V}\sum_{p\in I}$ has a limit. This is achieved by relating a
cut-off in the index sum over $p\in I$ to the volume. This limit
scales away all terms in (\ref{G2-regular}), (\ref{GN-regular}) and
(\ref{GN-algebraic-E}) with more inverse powers of $V$ than the number
of index summations. We argue in this subsection that the
contributions which are scaled away in this limit are precisely the
non-planar parts of genus $g\geq 1$. These arguments are not necessary
to understand the rest of the paper so that the reader may jump to
(\ref{SD-G}) and (\ref{SD-GN}) which summarise the scaling limit of
(\ref{G2-regular}) and (\ref{GN-algebraic-E}). In fact the direct
scaling limit $(V\to \infty, \frac{1}{V}\sum_{p\in I}\,\text{finite})$
avoids the infinite sum over all genera $\geq 1$ which can at best 
be expected to be Borel summable\footnote{We thank the referee for 
this remark.\label{referee}}.

\smallskip

Perturbatively, the connected functions 
$G_{|P^1_1|\dots|P^L_{n_L}|}$ are expanded into ribbon
graphs, the matrix analogue of Feynman graphs. 
A ribbon graph represents a two-dimensional simplicial complex and as
such encodes the genus $g$ of a Riemann surface on which the ribbon
graph can be drawn. We thus decompose the connected functions
according to the genus, 
$G_{|P^1_1|\dots|P^L_{n_L}|}=\sum_{g=0}^\infty
G^{(g)}_{|P^1_1|\dots|P^L_{n_L}|}$. We are not going to discuss 
the convergence of this series. We shall see in 
Proposition~\ref{prop:genus} that this expansion can be viewed as formal 
power series in $V^{-2g}$. Understanding the resummation of this series 
for large enough $V$ would be an interesting problem of constructive 
physics$^{\scriptsize\ref{referee}}$. 

The Schwinger-Dyson equations of subsection~\ref{sec:SD} describe how a
given connected function $G_{|a|P^1_1|\dots|P^L_{n_L}|}$ or
$G_{|abp_1\dots p_{k-2}|P^{k+1}_1|\dots|P^L_{n_L}|}$ is represented as
a contraction of the special (planar) vertex $\big(\frac{\partial
  (-S_{int})}{\partial \phi_{aa}}\big)$ or $\big(\phi_{ap_{k-2}}\dots
\phi_{p_1b} \frac{\partial (-S_{int})}{\partial \phi_{ab}}\big)$, which
carries the external source $J_{aa}$ or $J_{ab}$, to a smaller
subgraph. This contraction may increase the genus.  We have
investigated a similar problem in \cite{Grosse:2003aj} and use these
results to identify the topology of the various contributions to the
Schwinger-Dyson equations. We can summarise the outcome of 
\cite{Grosse:2003aj} as follows:
\begin{Proposition}
\label{Prop:genus}
\begin{enumerate}
\item[i)] Attaching one leg of a (planar) vertex to a connected subgraph
  does not change the genus. If this attachment is to a $J$-cycle of
  length $1$, then the other legs of the distinguished vertex form a
  cycle of its own. If the attachment is to a $J$-cycle of length $\geq
  2$, then the other legs insert in cyclic order into the connected chain.

\item[ii)] Joining two legs of different cycles of a connected graph
  increases the genus by $1$. At the same time, these two cycles are
  connected to a single one, unless both are of length 1 in which case
  both cycles disappear.

\item[iii)] Joining two legs of the same cycle of a connected graph does not
  change the genus. The cycle disappears if before the contraction it
  was of length 2. If the original cycle was of length $k\geq 3$ and
  neighboured legs are joint, then the remaining legs form a single
  cycle of length $k-2$. If the original cycle was of length $k\geq 4$
  and legs are joint which are not neighbours, then the joining splits
  the original cycle into two, one consisting of the `inner' legs
  relative to the contraction, the other one of the `outer' legs.
\hfill $\square$
\end{enumerate}
\end{Proposition}

We now identify the topology of individual parts of Ward identity and
Schwinger-Dyson equations. We start with the two parts $W^1_a[J]$,
$W^2_a[J]$ in Theorem \ref{Thm:Ward}: $W^2_a[J]$ arises by insertion
of $(\phi^2)_{pa}$ into a connected function of $\mathcal{W}$, and
$W^1_a[J]$ arises by joining a pair of disjoint connected functions
from $\mathcal{W}\cdot \mathcal{W}$ via the vertex $(\phi^2)_{pa}$. In
the application to Schwinger-Dyson equation later on it will be
important that $(\phi^2)_{pa}$ is part of a larger vertex 
$(\phi^2)_{pa}(\phi^{\otimes r})_{(b_i)}$ similar to 
(\ref{Vphi-gen}). The $r$ legs of $(\phi^{\otimes r})_{(b_i)}$
have to be taken into account in the cycle structure. We find:
\begin{enumerate}
\item $G_{|a|a|P_1^1|\dots| P^L_{n_L}|}$\quad
arises by contracting two cycles of the same graph, each of
length $1$, via $(\phi^2)_{pa}$. We are in the situation of
Proposition~\ref{Prop:genus}.ii) so that the genus \emph{first 
increases by $1$}. But if $(\phi^2)_{pa}$ is actually part of a larger
vertex $(\phi^2)_{pa}(\phi^r)_{ab}$, then the remainder $(\phi^r)_{ab}$
forms a cycle of its own although notationally not visible in 
$G_{|a|a|P_1^1|\dots| P^L_{n_L}|}$. If these derivatives in the cycle 
$(\phi^r)_{ab}$ contract to $h$ of the cycles 
$|P_1^1|,\dots,| P^L_{n_L}|$, then the genus \emph{further increases by $h$}. 
The total increase of the genus is thus by $h+1$.

\item $G_{P_1^1|\dots| P^1_{n_1}|an|P^2_1|\dots|P^L_{n_L}|}$\quad 
arises by joining both legs of a cycle of length $2$ via
$(\phi^2)_{pa}$. According to Proposition~\ref{Prop:genus}.iii), the
genus \emph{first is unchanged}. 
The original $J$-cycle disappears in this way, but
the remaining part $(\phi^{\otimes r})_{(b_i)}$ 
of the vertex forms again a cycle of its own. If these derivatives in  
$(\phi^{\otimes r})_{(b_i)}$ contract to $h$ of the cycles 
$|P_1^1|,\dots,| P^L_{n_L}|$, then the genus \emph{further increases by $h$}.

\item $G_{P_1^1|\dots| P^{k-2}_{n_{k-2}}|nanq_1\dots
    q_{k-3}|P^k_1|\dots| P^L_{n_L}|}$\quad also has unchanged genus 
according to Proposition~\ref{Prop:genus}.iii) after connection of the
cycle of length $k\geq 3$ via $(\phi^2)_{pa}$. However, the final
cycle structure is more involved: Assume the vertex is
$\phi_{pn} \phi_{na} (\phi^{\otimes r})_{(b_i)} $ and to be
inserted into the cycle $J_{q_{k-1}q_k}J_{q_k q_1} J_{q_1 q_2}\cdots
J_{q_{k-2}q_{k-1}}$. We first contract $\phi_{pn}$ with
$J_{q_{k-2}q_{k-1}}$, giving a $\delta_{nq_{k-2}}\delta_{pq_{k-1}}$. 
According to Proposition~\ref{Prop:genus}.i), the
legs of the vertex insert into the cycle, giving 
\begin{align}
J_{p q_k} J_{q_k q_1}J_{q_1 q_2}\cdots J_{q_{k-3} n} \phi_{na} (\phi^{\otimes
  r})_{(b_i)} \;.
\tag{*}
\end{align}
In the second step $\phi_{na}$ is contracted with $ J_{pq_k}$,
giving a $\delta_{ap}\delta_{nq_k}$. Since $k\geq 3$ and $r\geq
1$, the two legs $\phi_{na}$ and $ J_{pq_{k}}$ are not neighbours,
so that according to Proposition~\ref{Prop:genus}.iii) the cycle (*) is
split into two cycles $J_{n q_1}J_{q_1 q_2}\cdots J_{q_{k-3} n}$ and
$(\phi^{\otimes  r})_{(b_i)}$. When $G_{P_1^1|\dots|
  P^{k-2}_{n_{k-2}}|nanq_1\dots q_{k-3}|P^k_1|\dots| P^L_{n_L}|}$ is
used in a Schwinger-Dyson equation, the $J$-derivatives encoded by
$(\phi^{\otimes  r})_{(b_i)}$ necessarily connect to $h\geq 1$ 
different $J$-cycles, and this increases the genus by $h$. 

\item $G_{|a|P_1^1|\dots|P^L_{n_L}|}G_{|a|Q_1^1|\dots|Q^{L'}_{m_{L'}}|}$
  arises by connecting two length-1 cycles of different subgraphs via
  $(\phi^2)_{pa}$. According to Proposition~\ref{Prop:genus}.i), the
  genus is unchanged. The remaining part $(\phi^{\otimes r})_{(b_i)}$
  of the total vertex $(\phi^2)_{pa}(\phi^{\otimes r})_{(b_i)}$ forms
  a cycle of its own. In Schwinger-Dyson equations, this cycle
  connects to $h\geq 1$ other cycles in
  $G_{|a|P_1^1|\dots|P^L_{n_L}|}G_{|a|Q_1^1|\dots|Q^{L'}_{m_{L'}}|}$, thus
  increasing the genus by $h$.

\item $J_{pn}\frac{\partial \mathcal{Z}[J]}{\partial J_{an}}
{-}J_{na}\frac{\partial\mathcal{Z}[J]}{\partial J_{np}}$.\quad
Here we have to expand $Z$ as polynomial in connected functions $G$
and apply the above discussion case by case.

\end{enumerate}

For the single-cycle two-point Schwinger-Dyson equation (\ref{G2-regular}) 
we thus have:
\begin{align}
G^{(g)}_{|ab|} &=\frac{\delta_{g0}}{E_a+E_b} -\frac{\lambda_4}{(E_a+E_b)} 
\Big( \frac{1}{V} \sum_{g'+g''+1=g} G^{(g')}_{|ab|} G^{(g'')}_{|a|a|}
+\frac{1}{V} \sum_{n\in I}\sum_{g'+g''=g} G^{(g')}_{|ab|} G^{(g'')}_{|an|}
\nonumber
\\*
&\qquad\qquad\qquad\qquad 
+ \frac{1}{V^2}\Big(G^{(g-2)}_{|a|a|ab|}+\sum_{n\in I} G^{(g-1)}_{|an|ab|}
+ G^{(g-1)}_{|aaab|}+G^{(g-1)}_{|baba|}\Big)\Big)
\nonumber
\\*
&
+\frac{\lambda_4}{(E_a+E_b)} \Big(
\frac{1}{V} \sum_{p \in I} \frac{G^{(g)}_{|ab|}- G^{(g)}_{|pb|}}{E_a-E_p} 
+\frac{1}{V} \frac{G^{(g-1)}_{|a|b|}- G^{(g-1)}_{|b|b|}}{E_a-E_b} 
\Big)\;.
\label{Gab:B=1}
\end{align}
The very last term $\frac{G^{(g-1)}_{|a|b|}-
  G^{(g-1)}_{|b|b|}}{E_a-E_b}$ has genus increased by $1$ because two
different cycles are connected by the external vertex.  Similarly, the
equation~(\ref{GN-regular}) for the single-cycle $(N\geq 4)$-point
function has the expansion

\begin{align}
&G^{(g)}_{|a b_1\dots b_{N-1}|}
\nonumber
\\*
&= \frac{(-\lambda_4)}{E_a+E_{b_1}}\Big\{
\frac{1}{V} \sum_{g'+g''=g} \sum_{n\in I} G_{|an|}^{(g')} 
G^{(g'')}_{|ab_1\dots b_{N-1}|}
+\frac{1}{V} \sum_{g'+g''+1=g} G^{(g')}_{|a|a|} G^{(g'')}_{|ab_1\dots b_{N-1}|}
\nonumber
\\
&
+ \frac{1}{V^2} \Big( G^{(g-2)}_{|a|a|ab_1\dots b_{N-1}|}
+ \sum_{n\in I} G^{(g-1)}_{|an|ab_1\dots b_{N-1}|}
+ G^{(g-1)}_{|aaab_1\dots b_{N-1}|}
+\sum_{k=1}^{N-1} 
G^{(g-1)}_{|b_kab_kb_{k+1}\dots b_{N-1}a b_1\dots b_{k-1}|}\Big)
\nonumber
\\
& - \frac{1}{V} \sum_{p\in I}
\frac{G^{(g)}_{|pb_1\dots b_{N-1}|}-G^{(g)}_{|ab_1\dots b_{N-1}|}}{E_p-E_a} 
- \frac{1}{V} \sum_{k=1}^{N-1}\frac{G^{(g-1)}_{|b_1\dots b_k|b_{k+1}\dots
    b_{N-1}b_k|}-
G^{(g-1)}_{|b_1\dots b_k|b_{k+1}\dots b_{N-1}a|}
}{E_{b_k}-E_a}
\nonumber
\\
&
-\sum_{l=1}^{\frac{N-2}{2}} \sum_{g'+g''=g} G^{(g')}_{|b_1\dots b_{2l}|}
\frac{G^{(g'')}_{|b_{2l+1}\dots  b_{N-1}b_{2l}|}
-G^{(g'')}_{|b_{2l+1}\dots b_{N-1}a|} 
}{E_{b_{2l}}-E_a}
\Big\}\;.
\label{SD-GN-regular-g}
\end{align}
The same expansion extends to (\ref{GN-algebraic-E}). 
In particular, we have proved 
\begin{Proposition}
  The (unrenormalised) planar connected $N$-point functions
  $G^{(0)}_{|ab_1\dots b_{N-1}|}$ at $B=1$ in a real $\phi^4$-matrix model with
  injective external matrix $E$ satisfy the recursive system of equations
\begin{align}
G^{(0)}_{|ab|} &=  \frac{1}{E_a+E_b} -\frac{\lambda_4}{(E_a+E_b)V} G^{(0)}_{|ab|}
\sum_{p \in I} G^{(0)}_{|ap|}
+ \frac{\lambda_4}{(E_a+E_b)V} \sum_{p \in I}
\frac{ G^{(0)}_{|pb|}-G^{(0)}_{|ab|}}{E_p-E_a}\;,
\raisetag{2ex}
\label{SD-G}
\\
G^{(0)}_{|b_0 b_1\dots b_{N-1}|}
&= (-\lambda_4)
\sum_{l=1}^{\frac{N-2}{2}} 
\frac{G^{(0)}_{|b_0 b_1 \dots b_{2l-1}|} G^{(0)}_{|b_{2l}b_{2l+1}\dots b_{N-1}|} 
- G^{(0)}_{|b_{2l} b_1 \dots b_{2l-1}|} G^{(0)}_{|b_0 b_{2l+1}\dots b_{N-1}|} 
}{(E_{b_0}-E_{b_{2l}})(E_{b_1}-E_{b_{N-1}})}\;.
\label{SD-GN}
\end{align}
In the scaling limit $(V\to
\infty, \frac{1}{V}\sum_{p\in I}\,\text{finite})$, the full function 
and the genus-0 function satisfy the same equation so that
$\lim_{\di{V\to \infty}{\frac{1}{V}\sum_{p\in I}\,\text{finite}}} 
G^{(0)}_{|a b_1\dots b_{N-1}|}=\lim_{\di{V\to \infty}{
\frac{1}{V}\sum_{p\in I}\,\text{finite}}} 
G_{|a b_1\dots b_{N-1}|}$. \hfill $\square$%
\label{prop:planar}
\end{Proposition}
The equation (\ref{SD-G}) is a
remarkable \emph{closed equation for the planar connected regular
  two-point function alone}.  This was already established in
\cite{Grosse:2009pa} by a different method, and versions of
(\ref{SD-G}) for self-dual noncommutative $\phi^4$-theory at $a=b=0$
have already been derived in \cite{Disertori:2006nq}. We have
generalised the ideas of \cite{Disertori:2006nq} to a large class of
matrix models and obtained a self-consistent equation for $G^{(0)}_{|ab|}$
for \emph{any} matrix indices $a,b$. Usual Schwinger-Dyson techniques
give an \emph{open} system of equations relating the 2-point
function to the 4-point function, the 4-point function to the
6-point function, and so on.  In our setting, the Ward identity
allowed us to break this system apart. There is now one closed 
non-linear equation for the 2-point function alone and then a
recursive formula for the exact solution of all $(N> 2)$-point functions. 

We stress that the two-point function is by definition symmetric,
$G^{(0)}_{|ab|}=G^{(0)}_{|ba|}$, although this is not manifest in
(\ref{SD-G})!

\medskip

We observe in Appendix~\ref{appendix:B=2} that planar functions with
$B=2$ cycles scale with $\frac{1}{V}$ for $V\to \infty$.  Inserted
into (\ref{Gab:B=1}) and (\ref{SD-GN-regular-g}) we see that functions
with $(g=1,B=1)$ then scale with $\frac{1}{V^2}$. This holds in
general: the whole non-planar sector with genus $g\geq 1$ is scaled
away for $V\to \infty$.
\begin{Proposition}
In the scaling limit $(V\to
\infty, \frac{1}{V}\sum_{p\in I}\,\text{finite})$, 
the ($N_1{+}\dots {+}N_B$)-point function of genus $g$ has a scaling 
\begin{equation}
G^{(g)}_{|b_1^1\dots b^1_{N_1}|\dots |b_1^B\dots b^B_{N_B}|}
= V^{1-B-2g} 
G^g_{b_1^1\dots b^1_{N_1}|\dots 
|b_1^B\dots b^B_{N_B}} \;,
\label{scaling-GN:B:g}
\end{equation}
where $G^g_{b_1^1\dots b^1_{N_1}|\dots |b_1^B\dots b^B_{N_B}}$ has a
finite limit for $V\to \infty$.
\label{prop:genus}
\end{Proposition}
\emph{Proof.} We proceed by induction in $g$ and $B$. The step from
$(B=1,g=0)$ to $(B=2,g=0)$ is verified in (\ref{Gabc-B=2-odd}) and
(\ref{Gabc-B=2-even}). Then (\ref{Gab:B=1}) and
(\ref{SD-GN-regular-g}) verify the step from $(B=1,g=0)$ and
$(B=2,g=0)$ to $(B=2,g=0)$. Knowing that these matrix models have a
universal scaling degree in the combination $B+2g$
\cite{Grosse:2004yu}) we confirm (\ref{scaling-GN:B:g}) in the general
case. \hspace*{\fill}$\square$%

\bigskip

Recall that the genus expansion reads $G_{|b_1^1\dots b^1_{N_1}|\dots
  |b_1^B\dots b^B_{N_B}|} =\sum_{g=0}^\infty G^{(g)}_{|b_1^1\dots
  b^1_{N_1}|\dots |b_1^B\dots b^B_{N_B}|}$. Expressing $G^{(g)}$ in
terms of the functions $G^{g}$ on the lhs of (\ref{scaling-GN:B:g})
which have a limit for $V\to \infty$ we see that a genus-$g$
contribution is suppressed by a factor $V^{-2g}$. There is even a
better argument why the whole non-planar sector with $g>0$ is scaled
away: As noticed in Propositions~\ref{prop:planar},
\ref{prop:planar-B=2-odd} and \ref{prop:planar-B=2-even}, the whole
function and its genus-0 part satisfy the same equation.

Note that (\ref{scaling-GN:B:g}) also suggests that $N$-point
functions with $B>1$ are suppressed by a factor $V^{-(B-1)}$
over the one-cycle functions. This is definitely true for matrix
model correlation functions. We give in section~\ref{sec:misc} an
outlook to ongoing work in which we study for the $\phi^4$-model on
Moyal space the passage from matrix representation to position
space. It turns out that all planar functions 
with $(B\geq 1,g=0)$ do contribute to position space correlation
functions, whereas the non-planar sector remains suppressed.

\subsection{Graphical solution of the recursion formula in the planar sector}

\label{sec:graphics}

In the limit $V\to \infty$ the second line of (\ref{GN-algebraic-E}) is
absent, so that we obtain an algebraic recursion formula in the $B=1$
sector alone. For $N\in \{4,6\}$ we thus obtain
\begin{align}
&G^{(0)}_{|abcd|} = (-\lambda_4)
\frac{G^{(0)}_{|ab|} G^{(0)}_{|cd|}-G^{(0)}_{|ad|}G^{(0)}_{|bc|}}{
(E_a-E_c)(E_b-E_d)}\;,
\label{G4-new}
\\
&G^{(0)}_{|b_0\dots b_5|} 
= (-\lambda_4)
\Big\{ \frac{G^{(0)}_{|b_0b_1|}G^{(0)}_{|b_2b_3b_4b_5|}
  {-}G^{(0)}_{|b_1b_2|}G^{(0)}_{|b_3b_4b_5b_0|}}{(E_{b_0}-E_{b_2})(E_{b_1}-E_{b_5})}
+  \frac{G^{(0)}_{|b_0b_1b_2b_3|}G^{(0)}_{|b_4b_5|}
  {-}G^{(0)}_{|b_1b_2b_3b_4|}G^{(0)}_{|b_5b_0|}}{
(E_{b_0}-E_{b_4})(E_{b_1}-E_{b_5})}\Big\}
\nonumber
\\
&= 
(-\lambda_4)^2\Big\{
\frac{G^{(0)}_{|b_0b_1|}G^{(0)}_{|b_2b_5|}G^{(0)}_{|b_4b_3|}}{
e_{b_0b_2} e_{b_2b_4} {\cdot} e_{b_1b_5} e_{b_5b_3}}
+ 
\frac{G^{(0)}_{|b_0b_3|}G^{(0)}_{|b_2b_1|}G^{(0)}_{|b_4b_5|}}{
e_{b_4 b_0} e_{b_0b_2}{\cdot} e_{b_5b_3}e_{b_3b_1}}
+ 
\frac{G^{(0)}_{|b_0b_5|}G^{(0)}_{|b_2b_3|}G^{(0)}_{|b_4b_1|}}{
e_{b_2b_4}e_{b_4b_0}{\cdot} e_{b_3b_1}e_{b_1b_5}}
\nonumber
\\
& + G^{(0)}_{|b_0b_1|}G^{(0)}_{|b_2b_3|}G^{(0)}_{|b_4b_5|}
\Big( \frac{1}{e_{b_0b_2}e_{b_2b_4} {\cdot} e_{b_1b_3}e_{b_3b_5}}
+ \frac{1}{e_{b_0b_4}e_{b_4b_2} {\cdot} e_{b_1b_3}e_{b_3b_5}}
+ \frac{1}{e_{b_0b_4} e_{b_4b_2} {\cdot} e_{b_1b_5}e_{b_5b_3}}\Big)
\nonumber
\\
& + 
G^{(0)}_{|b_0b_5|}G^{(0)}_{|b_2b_1|}G^{(0)}_{|b_4b_3|}\Big(
\frac{1}{e_{b_0b_4} e_{b_4b_2} {\cdot} e_{b_5b_3} e_{b_3b_1}}
+ \frac{1}{e_{b_0b_2}e_{b_2b_4} {\cdot} e_{b_5b_3}e_{b_3b_1}}
+ \frac{1}{e_{b_0b_2}e_{b_2b_4} {\cdot} e_{b_5b_1}e_{b_1b_3}}
\Big)\Big\}\;,
\label{N=6}
\end{align}
where $e_{b_ib_j}:=E_{b_i}-E_{b_j}$. The denominators arising in
  (\ref{N=6}) are not unique; we rearrange them using the identity
\begin{align}
\frac{1}{e_{b_ib_j}e_{b_jb_k}}+\frac{1}{e_{b_jb_k}e_{b_kb_i}}+
\frac{1}{e_{b_kb_i}e_{b_ib_j}}=0\;.
\label{ident:bijk}
\end{align}

We introduce a graphical representation for a contribution to
$G^{(0)}_{|b_0\dots b_{N-1}|}$. We put the
indices $b_0,\dots,b_{N-1}$ in cyclic order on a circle.
A factor $G^{(0)}_{|b_ib_j|}$ is drawn as a \emph{thick} chord
between $b_i$ and $b_j$, and a factor $\frac{1}{E_{b_m}-E_{b_n}}$ as a
\emph{thin} arrow from $b_m$ to $b_n$. Reversing the arrow thus
changes the sign.  Later on we connect the arrows to trees where it
will be necessary to specify the root as a thick dot. The
corresponding graphical description of (\ref{G4-new}) reads
\begin{align}
G^{(0)}_{|b_0b_1b_2b_3|}
=  (-\lambda_4) \left\{
\ifpdf\includegraphics[viewport=0 25 60 60]{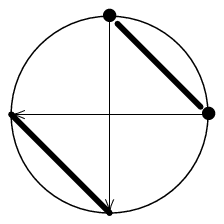}\else
\input fig-4a.xy
% \mbox{\xy <1cm,0cm>:
% (0,0)*\cir<1cm>{},
% (1,0)="B0",
% (0,1)="B1",
% (-1,0)="B2",
% (0,-1)="B3",
% \ar @{*{\bullet}{-}*{\bullet}}@*{[|(4)]} "B0";"B1",
% \ar @{-}@*{[|(4)]} "B2";"B3",
% \ar @*{[|(0.5)]} "B0";"B2",
% \ar @*{[|(0.5)]} "B1";"B3",
% \endxy}
\fi
+ 
\ifpdf\includegraphics[viewport=0 25 60 60]{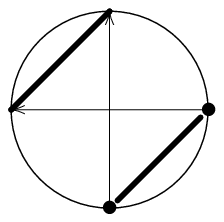}\else
\input fig-4b.xy
% \mbox{\xy <1cm,0cm>:
% (0,0)*\cir<1cm>{},
% (1,0)="B0",
% (0,1)="B1",
% (-1,0)="B2",
% (0,-1)="B3",
% \ar  @{*{\bullet}{-}*{\bullet}}@*{[|(4)]}@*{[|(4)]} "B0";"B3",
% \ar @{-}@*{[|(4)]} "B2";"B1",
% \ar @*{[|(0.5)]} "B0";"B2",
% \ar @*{[|(0.5)]} "B3";"B1",
% \endxy}
\fi
\right\}\;.
\end{align}
The two graphs are turned into each other by a rotation of
$\frac{2\pi}{4}$ and possibly reversion of both arrows. 

Our rules give the following graphical representation of the
six-point function (\ref{N=6}):
\begin{align}
\def\halfrootthree=0.8660254{}
G^{(0)}_{|b_0\dots b_5|} & = (-\lambda_4)^2
\left\{
\ifpdf\includegraphics[viewport=0 25 60 60]{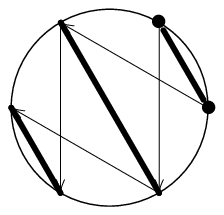}\else
\input fig-6a.xy
% \mbox{\xy <1cm,0cm>:
% (0,0)*\cir<1cm>{},
% (1,0)="B0",
% (0.5,\halfrootthree)="B1",
% (-0.5,\halfrootthree)="B2",
% (-1,0)="B3",
% (-0.5,-\halfrootthree)="B4",
% (0.5,-\halfrootthree)="B5",
% \ar @{*{\bullet}{-}*{\bullet}}@*{[|(4)]} "B0";"B1",
% \ar @{-}@*{[|(4)]} "B2";"B5",
% \ar @{-}@*{[|(4)]} "B4";"B3",
% \ar @*{[|(0.5)]} "B0";"B2",
% \ar @*{[|(0.5)]} "B2";"B4",
% \ar @*{[|(0.5)]} "B1";"B5",
% \ar @*{[|(0.5)]} "B5";"B3",
% \endxy}
\fi
{}~~+ \mbox{\begin{tabular}{c}
two rotations 
\\ by $\frac{2\pi k}{6}$ \\
$k\in \{1,2\}$
\end{tabular}}
\right\}
\nonumber
\\
& + (-\lambda_4)^2  \left\{ 
\left( 
\ifpdf\includegraphics[viewport=0 25 60 60]{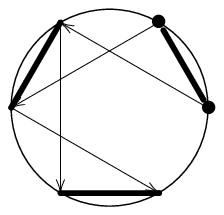}\else
\input fig-6b.xy
% \mbox{\xy <1cm,0cm>:
% (0,0)*\cir<1cm>{},
% (1,0)="B0",
% (0.5,\halfrootthree)="B1",
% (-0.5,\halfrootthree)="B2",
% (-1,0)="B3",
% (-0.5,-\halfrootthree)="B4",
% (0.5,-\halfrootthree)="B5",
% \ar @{*{\bullet}{-}*{\bullet}}@*{[|(4)]} "B0";"B1",
% \ar @{-}@*{[|(4)]} "B2";"B3",
% \ar @{-}@*{[|(4)]} "B4";"B5",
% \ar @*{[|(0.5)]} "B0";"B2",
% \ar @*{[|(0.5)]} "B2";"B4",
% \ar @*{[|(0.5)]} "B1";"B3",
% \ar @*{[|(0.5)]} "B3";"B5",
% \endxy}
\fi
+ 
\ifpdf\includegraphics[viewport=0 25 60 60]{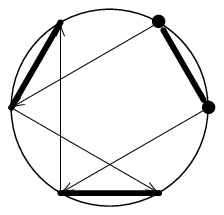}\else
\input fig-6c.xy
% \mbox{\xy <1cm,0cm>:
% (0,0)*\cir<1cm>{},
% (1,0)="B0",
% (0.5,\halfrootthree)="B1",
% (-0.5,\halfrootthree)="B2",
% (-1,0)="B3",
% (-0.5,-\halfrootthree)="B4",
% (0.5,-\halfrootthree)="B5",
% \ar @{*{\bullet}{-}*{\bullet}}@*{[|(4)]} "B0";"B1",
% \ar @{-}@*{[|(4)]} "B2";"B3",
% \ar @{-}@*{[|(4)]} "B4";"B5",
% \ar @*{[|(0.5)]} "B0";"B4",
% \ar @*{[|(0.5)]} "B4";"B2",
% \ar @*{[|(0.5)]} "B1";"B3",
% \ar @*{[|(0.5)]} "B3";"B5",
% \endxy}
\fi
+ 
\ifpdf\includegraphics[viewport=0 25 60 60]{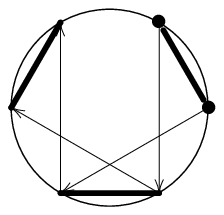}\else
\input fig-6d.xy
% \mbox{\xy <1cm,0cm>:
% (0,0)*\cir<1cm>{},
% (1,0)="B0",
% (0.5,\halfrootthree)="B1",
% (-0.5,\halfrootthree)="B2",
% (-1,0)="B3",
% (-0.5,-\halfrootthree)="B4",
% (0.5,-\halfrootthree)="B5",
% \ar @{*{\bullet}{-}*{\bullet}}@*{[|(4)]} "B0";"B1",
% \ar @{-}@*{[|(4)]} "B2";"B3",
% \ar @{-}@*{[|(4)]} "B4";"B5",
% \ar @*{[|(0.5)]} "B0";"B4",
% \ar @*{[|(0.5)]} "B4";"B2",
% \ar @*{[|(0.5)]} "B1";"B5",
% \ar @*{[|(0.5)]} "B5";"B3",
% \endxy}
\fi
\right) + 
\mbox{\begin{tabular}{@{}c@{}}
one \\ rotation 
\\ by $\frac{2\pi}{6}$
\end{tabular}}
\right\}. 
\end{align}
We first notice that only the 5 \emph{non-crossing pairings} of 
the cyclic indices $b_0b_1b_2b_3b_4b_5$ appear. The remaining pairing
$G_{b_0b_3}G_{b_1b_4}G_{b_2b_5}$ has a self-crossing. The 
non-crossing pairings of $2n$ cyclic indices are counted by the
Catalan number $C_n= \frac{(2n)!}{n!(n+1)!}$. 

Note that the 4 possible products of even/odd paths starting with root
$b_0$/$b_1$ are distributed over the pairings in such a manner that
the paths intersect the pairings only in the vertices 
and every product of paths
appears exactly once. Further rotation of the first line by
$\frac{2\pi\cdot 3}{6}$ and of the second line by $\frac{2\pi\cdot
  2}{6}$ give the same graphs after common reversal of the arrows and
use of the identity (\ref{ident:bijk}).

These observations generalise to any planar $N$-point function at $B=1$:
\begin{Proposition}
\label{Prop:graph-GN}
The $N$-point function $G^{(0)}_{|b_0\dots b_{N-1}|}$ is a sum with prefactors
$\pm (-\lambda_4)^{\frac{N}{2}-1}$ of
graphs with vertices $b_0,\dots,b_{N-1}$ put in cyclic order on a
circle and (fat) edges between $b_k,b_l$ representing a factor
$G^{(0)}_{|b_kb_l|}$ and (thin) arrows from $b_k$ to $b_l$ representing a
factor $\frac{1}{E_{b_k}-E_{b_l}}$. We call $b_{2i}$ an even vertex and
$b_{2i+1}$ an odd vertex.  In every such graph we have (or can achieve
via (\ref{ident:bijk})):
\begin{enumerate}\itemsep 0pt
\item Every even vertex is paired (by a fat chord) with exactly
  one odd vertex, and these chords do not cross.

\item Taking any even vertex as a root, there is a an oriented rooted
  tree between all even vertices. This even tree has no
  self-intersection and intersects the fat chord only at the vertices.

\item Taking any odd vertex as a root, there is a an oriented rooted
  tree between all odd vertices. This odd tree has no self-intersection
  and intersects the fat chords
  only at the vertices. 
\end{enumerate}
\end{Proposition}
\emph{Proof.} By (\ref{SD-GN}), the $N$-point function
$G^{(0)}_{|b_0\dots b_{N-1}|}$ arises by connections of $2l$-point functions
with ($N-2l$)-point functions via an even arrow and an odd arrow.  We
proceed by induction and assume validity of the Proposition for the $2l$-
and ($N-2l$)-point functions, starting with $G^{(0)}_{|b_0b_1|}$ which is
represented by one chord without any arrows.  We can symbolise
(\ref{SD-GN}) (without prefactor $(-\lambda_4)$) as follows:
\[
G^{(0)}_{|b_0\dots b_{N-1}|}= \sum_{l=1}^{\frac{N-2}{2}}
\Bigg(
\parbox{54mm}{\begin{picture}(45,48)
\put(10,25){\ifpdf\includegraphics[viewport=1 50 92 92]{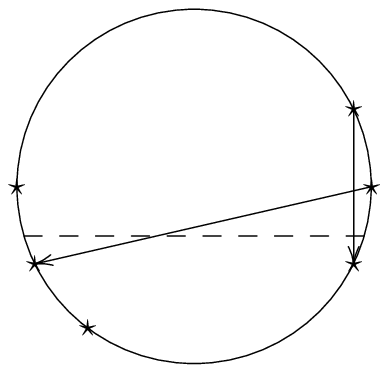}\else
\input fig-na.xy
% \mbox{\xy <1.8cm,0cm>:
% (0,0)*\cir<1.8cm>{},
% (1,0)="B0",
% (0.96,-0.28)="B0N",
% (0.9,0.43589)="B1",
% (0.9,-0.43589)="BNminus1",
% (-0.96,-0.28)="Bell-ellminus1",
% (-0.9,-0.43589)="Bell",
% (-1,0)="Bellminus1",
% (-0.8,-0.6)="Bell-ellplus1",
% (-0.6,-0.8)="Bellplus1",
% "B0"*{\star},
% "B1"*{\star},
% "BNminus1"*{\star},
% "Bell"*{\star},
% "Bellminus1"*{\star},
% "Bellplus1"*{\star},
% \ar @{--} "B0N";"Bell-ellminus1",
% \ar @{->} "B0";"Bell",
% \ar @{->} "B1";"BNminus1",
% \endxy}
\fi
}
\put(47,32){\mbox{\small$b_1$}}
\put(49,24){\mbox{\small$b_0$}}
\put(47,15){\mbox{\small$b_{N-1}$}}
\put(2,25){\mbox{\small$b_{2l-1}$}}
\put(7,18){\mbox{\small$b_{2l}$}}
\put(8,10){\mbox{\small$b_{2l+1}$}}
\put(17,30){\mbox{\scriptsize$G^{(0)}_{|b_0\dots b_{2l-1}|}$}}
\put(19,14){\mbox{\scriptsize$G^{(0)}_{|b_{2l}\dots b_{N-1}|}$}}
\end{picture}}
-
\parbox{54mm}{\begin{picture}(45,48)
\put(10,25){\ifpdf\includegraphics[viewport=1 50 90 90]{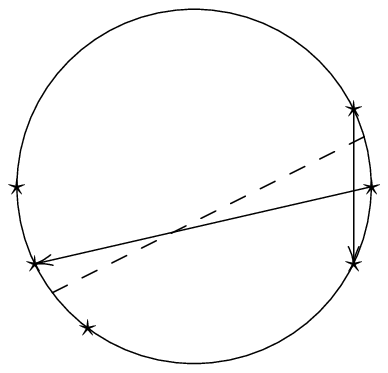}\else
\input fig-nb.xy
% \mbox{\xy <1.8cm,0cm>:
% (0,0)*\cir<1.8cm>{},
% (1,0)="B0",
% (0.96,0.28)="B01",
% (0.9,0.43589)="B1",
% (0.9,-0.43589)="BNminus1",
% (-0.96,-0.28)="Bell-ellminus1",
% (-0.9,-0.43589)="Bell",
% (-1,0)="Bellminus1",
% (-0.8,-0.6)="Bell-ellplus1",
% (-0.6,-0.8)="Bellplus1",
% "B0"*{\star},
% "B1"*{\star},
% "BNminus1"*{\star},
% "Bell"*{\star},
% "Bellminus1"*{\star},
% "Bellplus1"*{\star},
% \ar @{--} "B01";"Bell-ellplus1",
% \ar @{->} "B0";"Bell",
% \ar @{->} "B1";"BNminus1",
% \endxy}
\fi
}
\put(47,33){\mbox{\small$b_1$}}
\put(49,24){\mbox{\small$b_0$}}
\put(47,15){\mbox{\small$b_{N-1}$}}
\put(2,25){\mbox{\small$b_{2l-1}$}}
\put(7,18){\mbox{\small$b_{2l}$}}
\put(8,10){\mbox{\small$b_{2l+1}$}}
\put(17,30){\mbox{\scriptsize$G^{(0)}_{|b_1\dots b_{2l}|}$}}
\put(20,15){\mbox{\scriptsize$G^{(0)}_{|b_{2l+1}\dots b_{N{-}1}b_0|}$}}
\end{picture}} \Bigg)
\]
Here, the $\star$ locates the vertices and the dashed line separates
the $2l$-point function from the $(N-2l)$-point function of which we
do not show their inner chords and trees. 

\begin{enumerate} \itemsep 0pt
\item[1)] By induction in one of
these smaller functions every even vertex is paired with exactly one
odd vertex. These pairings do not cross in the smaller function and
since they do not cross the dashed cut, there cannot be a crossing
between chords of the two smaller functions. 

\item[2)] All even vertices in the $2l$-point function and in the
  $(N-2l)$-point functions are connected by a tree which by induction
  has no self-intersection and intersects the chords only in the
  vertices.  The additional arrow from $b_0$ to $b_l$ connects both
  trees of $l$ and $\frac{N}{2}-l$ vertices to a larger tree of
  $\frac{N}{2}$ vertices. It is geometrically clear that the
  additional arrow intersects both the two smaller even trees and the
  chords precisely in the vertices $b_0$ and $b_l$.

\item[3)] All odd vertices in the $2l$-point function and in the
  $(N-2l)$-point functions are connected by a tree which by induction
  has no self-intersection and intersects the chords only in the
  vertices.  The additional arrow from $b_1$ to $b_{N-1}$ connects
  both trees of $l$ and $\frac{N}{2}-l$ vertices to a larger tree of
  $\frac{N}{2}$ vertices. This larger tree does not have
  self-intersections, but in general intersects the chord emanating
  from $b_0$. Since a cyclic permutation of the indices interchanges
  even and odd vertices and hence even and odd trees, by 2) it must be
  possible to use the identity (\ref{ident:bijk}) to rearrange the odd
  tree in such a way that the possible intersection with the chord
  emanating from $b_0$ can be avoided. \hfill $\square$%
\end{enumerate}

Proposition~\ref{Prop:graph-GN} determines the planar connected
$N$-point function up to the problem to identify the product of even
and odd trees associated with a given non-crossing chord diagram. Due to
(\ref{ident:bijk}) this product cannot be unique. We leave this
question of canonical trees for future investigation. We finish this
section with the eight-point function which shows that not all
possible trees do actually arise:
\begin{align}
&G^{(0)}_{|b_0\dots b_7|} 
\nonumber
\\*[-1mm]
&= (-\lambda)\big\{ 
\frac{G^{(0)}_{|b_0b_1|}G^{(0)}_{|b_2b_3b_4b_5b_6b_7|}
  {-}G^{(0)}_{|b_1b_2|}G^{(0)}_{|b_3b_4b_5b_6b_7b_0|}}{
(E_{b_0}-E_{b_2})(E_{b_1}-E_{b_5})}
+  \frac{G^{(0)}_{|b_0b_1b_2b_3|}G^{(0)}_{|b_4b_5b_6b_7|}
  {-}G^{(0)}_{|b_1b_2b_3b_4|}G^{(0)}_{|b_5b_6b_7b_0|}}{
(E_{b_0}-E_{b_4})(E_{b_1}-E_{b_5})}
\nonumber
\\*
&\qquad\qquad +  \frac{G^{(0)}_{|b_0b_1b_2b_3b_4b_5|}G^{(0)}_{|b_6b_7|}
  {-}G^{(0)}_{|b_1b_2b_3b_4b_5b_6|}G^{(0)}_{|b_7b_0|}}{
(E_{b_0}-E_{b_6})(E_{b_1}-E_{b_5})}\Big\}
\nonumber
\\
& =(-\lambda)^3
\Big\{ G^{(0)}_{|b_0b_1|}G^{(0)}_{|b_2b_7|} G^{(0)}_{|b_4b_5|} G^{(0)}_{|b_6b_3|} 
\big( \tfrac{1}{e_{b_0b_2}e_{b_2b_6}e_{b_6b_4} \cdot 
e_{b_1b_7}e_{b_7b_3}e_{b_3b_5}}\big)
\nonumber
\\
&+  G^{(0)}_{|b_0b_3|}G^{(0)}_{|b_2b_1|} G^{(0)}_{|b_4b_7|} G^{(0)}_{|b_6b_5|}
\big( \tfrac{1}{e_{b_2b_0}e_{b_0b_4}e_{b_4b_6} \cdot 
e_{b_1b_3}e_{b_3b_7}e_{b_7b_5}}\big)
\nonumber
\\
&
+ 
 G^{(0)}_{|b_0b_5|}G^{(0)}_{|b_2b_3|} G^{(0)}_{|b_4b_1|}  G^{(0)}_{|b_6b_7|}
\big(\tfrac{1}{e_{b_2b_4}e_{b_4b_0}e_{b_0b_6} \cdot
e_{b_3b_1}e_{b_1b_5}e_{b_5b_7}} \big)
\nonumber
\\
&+ 
 G^{(0)}_{|b_0b_7|}G^{(0)}_{|b_2b_5|}G^{(0)}_{|b_4b_3|}  G^{(0)}_{|b_6b_1|} 
\big( \tfrac{1}{e_{b_4b_2}e_{b_2b_6}e_{b_6b_0}
\cdot e_{b_3b_5}e_{b_5b_1}e_{b_1b_7}} \big)
\nonumber
\\
& +  G^{(0)}_{|b_0b_1|}G^{(0)}_{|b_2b_7|} G^{(0)}_{|b_4b_3|} G^{(0)}_{|b_6b_5|}
\big(\tfrac{1}{e_{b_0b_2}e_{b_2b_4}e_{b_4b_6}  \cdot 
e_{b_1b_7}e_{b_7b_3}e_{b_3b_5}} 
+\tfrac{1}{e_{b_0b_2}e_{b_2b_4}e_{b_4b_6} \cdot 
e_{b_1b_7}e_{b_7b_5}e_{b_5b_3}} 
\nonumber
\\
& \hspace*{5cm}
+\tfrac{1}{e_{b_0b_2}e_{b_2b_6}e_{b_6b_4} \cdot 
e_{b_1b_7}e_{b_7b_5}e_{b_5b_3}} \big)
\nonumber
\\
&+  G^{(0)}_{|b_0b_3|}G^{(0)}_{|b_2b_1|} G^{(0)}_{|b_4b_5|} G^{(0)}_{|b_6b_7|} 
\big( \tfrac{1}{e_{b_2b_0}e_{b_0b_4}e_{b_4b_6} \cdot 
e_{b_1b_3}e_{b_3b_5}e_{b_5b_7}} 
+\tfrac{1}{e_{b_2b_0}e_{b_0b_6}e_{b_6b_4} \cdot
e_{b_1b_3}e_{b_3b_5}e_{b_5b_7}} 
\nonumber
\\
& \hspace*{5cm}
+\tfrac{1}{e_{b_2b_0}e_{b_0b_6}e_{b_6b_4}
\cdot e_{b_1b_3}e_{b_3b_7}e_{b_7b_5}} \big)
\nonumber
\\
& +  G^{(0)}_{|b_0b_7|} G^{(0)}_{|b_2b_3|} G^{(0)}_{|b_4b_1|} G^{(0)}_{|b_6b_5|}
\big(\tfrac{1}{e_{b_2b_4}e_{b_4b_6}e_{b_6b_0}
\cdot e_{b_3b_1}e_{b_1b_5}e_{b_5b_7}}
+ \tfrac{1}{e_{b_2b_4}e_{b_4b_6}e_{b_6b_0} \cdot 
e_{b_3b_1}e_{b_1b_7}e_{b_7b_5}}
\nonumber
\\
& \hspace*{5cm}
+\tfrac{1}{e_{b_2b_4}e_{b_4b_0}e_{b_0b_6} \cdot 
e_{b_3b_1}e_{b_1b_7}e_{b_7b_5}}\big)
\nonumber
\\
& + G^{(0)}_{|b_0b_1|}G^{(0)}_{|b_2b_5|} G^{(0)}_{|b_4b_3|} G^{(0)}_{|b_6b_7|}
\big( \tfrac{1}{e_{b_4b_2}e_{b_2b_6}e_{b_6b_0} \cdot 
e_{b_3b_5}e_{b_5b_7}e_{b_7b_1}}
+\tfrac{1}{e_{b_4b_2}e_{b_2b_0}e_{b_0b_6} \cdot 
e_{b_3b_5}e_{b_5b_7}e_{b_7b_1}}
\nonumber
\\
& \hspace*{5cm}
+\tfrac{1}{e_{b_4b_2}e_{b_2b_0}e_{b_0b_6} \cdot 
e_{b_3b_5}e_{b_5b_1}e_{b_1b_7}}\big)
\nonumber
\\
& + G^{(0)}_{|b_0b_7|}G^{(0)}_{|b_2b_1|} G^{(0)}_{|b_4b_5|} G^{(0)}_{|b_6b_3|} 
\big( \tfrac{1}{e_{b_4b_6}e_{b_6b_0}e_{b_0b_2} \cdot
e_{b_5b_3}e_{b_3b_7}e_{b_7b_1}}
+\tfrac{1}{e_{b_4b_6}e_{b_6b_0}e_{b_0b_2} \cdot
e_{b_5b_3}e_{b_3b_1}e_{b_1b_7}}
\nonumber
\\
& \hspace*{5cm}
+\tfrac{1}{e_{b_4b_6}e_{b_6b_2}e_{b_2b_0}\cdot 
e_{b_5b_3}e_{b_3b_1}e_{b_1b_7}}\big)
\nonumber
\\
& +  G^{(0)}_{|b_0b_1|}G^{(0)}_{|b_2b_3|} G^{(0)}_{|b_4b_7|} G^{(0)}_{|b_6b_5|}
\big( \tfrac{1}{e_{b_6b_4}e_{b_4b_0}e_{b_0b_2} \cdot 
e_{b_5b_7}e_{b_7b_1}e_{b_1b_3}}
+ \tfrac{1}{e_{b_6b_4}e_{b_4b_2}e_{b_2b_0} \cdot 
e_{b_5b_7}e_{b_7b_1}e_{b_1b_3}}
\nonumber
\\
& \hspace*{5cm}
+\tfrac{1}{e_{b_6b_4}e_{b_4b_2}e_{b_2b_0} \cdot e_{b_5b_7}e_{b_7b_3}e_{b_3b_1}} \big)
\nonumber
\\
& + G^{(0)}_{|b_0b_5|}G^{(0)}_{|b_2b_1|} G^{(0)}_{|b_4b_3|} G^{(0)}_{|b_6b_7|} 
\big( \tfrac{1}{e_{b_6b_0}e_{b_0b_2}e_{b_2b_4}\cdot 
e_{b_7b_5}e_{b_5b_1}e_{b_1b_3}}
+\tfrac{1}{e_{b_6b_0}e_{b_0b_2}e_{b_2b_4}\cdot
e_{b_7b_5}e_{b_5b_3}e_{b_3b_1}}
\nonumber
\\
& \hspace*{5cm}
+\tfrac{1}{e_{b_6b_0}e_{b_0b_4}e_{b_4b_2}\cdot 
e_{b_7b_5}e_{b_5b_3}e_{b_3b_1}}\big)
\nonumber
\\
& + G^{(0)}_{|b_0b_7|}G^{(0)}_{|b_2b_3|} G^{(0)}_{|b_4b_5|} G^{(0)}_{|b_6b_1|} 
\big( \tfrac{1}{e_{b_0b_6}e_{b_6b_2}e_{b_2b_4} \cdot
e_{b_7b_1}e_{b_1b_3}e_{b_3b_5}}
+\tfrac{1}{e_{b_0b_6}e_{b_6b_4}e_{b_4b_2}\cdot 
e_{b_7b_1}e_{b_1b_3}e_{b_3b_5}}
\nonumber
\\
& \hspace*{5cm}
+\tfrac{1}{e_{b_0b_6}e_{b_6b_4}e_{b_4b_2}\cdot 
e_{b_7b_1}e_{b_1b_5}e_{b_5b_3}}\big)
\nonumber
\\
&+  G^{(0)}_{|b_0b_1|}G^{(0)}_{|b_2b_3|} G^{(0)}_{|b_4b_5|} G^{(0)}_{|b_6b_7|} 
\big(\tfrac{1}{e_{b_0b_6}e_{b_6b_4}e_{b_4b_2} \cdot
e_{b_1b_7}e_{b_7b_5}e_{b_5b_3}}  
+\tfrac{1}{e_{b_0b_6}e_{b_6b_4}e_{b_4b_2} \cdot 
e_{b_1b_7}e_{b_7b_3}e_{b_3b_5}}  
\nonumber
\\
&\qquad\quad
+\tfrac{1}{e_{b_0b_6}e_{b_6b_4}e_{b_4b_2} \cdot 
e_{b_1b_3}e_{b_3b_7}e_{b_7b_5}}  
+ \tfrac{1}{e_{b_0b_2}e_{b_2b_4}e_{b_4b_6}\cdot 
e_{b_1b_3}e_{b_3b_5}e_{b_5b_7}}
+ \tfrac{1}{e_{b_0b_2}e_{b_2b_6}e_{b_6b_4}\cdot 
e_{b_1b_3}e_{b_3b_5}e_{b_5b_7}}
\nonumber
\\
&\qquad\quad
+ \tfrac{1}{e_{b_0b_6}e_{b_6b_2}e_{b_2b_4}\cdot 
e_{b_1b_3}e_{b_3b_5}e_{b_5b_7}}
+ \tfrac{1}{e_{b_0b_2}e_{b_2b_6}e_{b_6b_4}\cdot
e_{b_1b_3}e_{b_3b_7}e_{b_7b_5}}
+ \tfrac{1}{e_{b_0b_6}e_{b_6b_2}e_{b_2b_4}\cdot
e_{b_1b_7}e_{b_7b_3}e_{b_3b_5}}
\nonumber
\\
&\qquad\quad
+ \tfrac{1}{e_{b_0b_6}e_{b_6b_2}e_{b_2b_4}\cdot
e_{b_1b_3}e_{b_3b_7}e_{b_7b_5}}
+ \tfrac{1}{e_{b_0b_6}e_{b_0b_4}e_{b_4b_2}\cdot
e_{b_1b_3}e_{b_1b_5}e_{b_5b_7}}
\big)
\nonumber
\\
& +  G^{(0)}_{|b_0b_7|}G^{(0)}_{|b_2b_1|} G^{(0)}_{|b_4b_3|} G^{(0)}_{|b_6b_5|}
\big(\tfrac{1}{e_{b_0b_6}e_{b_6b_4}e_{b_4b_2}\cdot 
e_{b_7b_5}e_{b_5b_3}e_{b_3b_1}} 
+\tfrac{1}{e_{b_0b_6}e_{b_6b_2}e_{b_2b_4}\cdot
e_{b_7b_5}e_{b_5b_3}e_{b_3b_1}} 
\nonumber
\\
&\qquad\quad 
+\tfrac{1}{e_{b_0b_2}e_{b_2b_6}e_{b_6b_4} \cdot
e_{b_7b_5}e_{b_5b_3}e_{b_3b_1}} 
+ \tfrac{1}{e_{b_0b_2}e_{b_2b_4}e_{b_4b_6}\cdot 
e_{b_7b_1}e_{b_1b_3}e_{b_3b_5}}
+ \tfrac{1}{e_{b_0b_2}e_{b_2b_4}e_{b_4b_6}\cdot 
e_{b_7b_1}e_{b_1b_5}e_{b_5b_3}}
\nonumber
\\
&\qquad\quad 
+ \tfrac{1}{e_{b_0b_2}e_{b_2b_4}e_{b_4b_6}\cdot 
e_{b_7b_5}e_{b_5b_1}e_{b_1b_3}}
+ \tfrac{1}{e_{b_0b_2}e_{b_2b_6}e_{b_6b_4}\cdot
e_{b_7b_1}e_{b_1b_5}e_{b_5b_3}}
+ \tfrac{1}{e_{b_0b_6}e_{b_6b_2}e_{b_2b_4}\cdot
e_{b_7b_5}e_{b_5b_1}e_{b_1b_3}}
\nonumber
\\
&\qquad\quad 
+ \tfrac{1}{e_{b_0b_2}e_{b_2b_6}e_{b_6b_4}\cdot
e_{b_7b_5}e_{b_5b_1}e_{b_1b_3}}
+ \tfrac{1}{e_{b_0b_2}e_{b_0b_4}e_{b_4b_6}\cdot
e_{b_7b_5}e_{b_7b_3}e_{b_3b_1}}
\big)\Big\}\;.
\label{G8}
\end{align}
We have used (\ref{ident:bijk}) to bring the
denominators for the last two products of $G_{b_ib_j}$ into a 
form which (in the sum) is symmetric in the even and odd indices and
has a common sign. A novel feature compared with the six-point function 
is the appearance of trees with branches  
$\tfrac{1}{e_{b_0b_6}e_{b_0b_4}e_{b_4b_2}}\cdot\tfrac{1}{e_{b_1b_3}e_{b_1b_5}e_{b_5b_7}}$ and 
$\tfrac{1}{e_{b_0b_2}e_{b_0b_4}e_{b_4b_6}}\cdot\tfrac{1}{e_{b_7b_5}e_{b_7b_3}e_{b_3b_1}}$.
Trees
become unavoidable in the 12-point function where some pairings do not 
permit a non-crossing rooted path. It seems natural to require that
paired roots carry the same number of branches. If we furthermore require
that in 8-point functions only the root has more than one branch, then
the representation given in (\ref{G8}) seems distinguished. Its
graphical representation reads
\begin{align}
G^{(0)}_{|b_0\dots b_7|}
&= (-\lambda)^3
\left\{
\ifpdf\includegraphics[viewport=0 25 60 60]{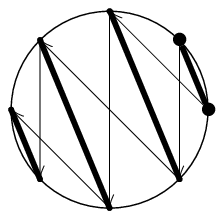}\else
\input fig-8a.xy
% \mbox{\xy <1cm,0cm>:
% (0,0)*\cir<1cm>{},
% (1,0)="B0",
% (\halfroottwo,\halfroottwo)="B1",
% (0,1)="B2",
% (-\halfroottwo,\halfroottwo)="B3",
% (-1,0)="B4",
% (-\halfroottwo,-\halfroottwo)="B5",
% (0,-1)="B6",
% (\halfroottwo,-\halfroottwo)="B7",
% \ar @{*{\bullet}{-}*{\bullet}}@*{[|(4)]} "B0";"B1",
% \ar @{-}@*{[|(4)]} "B2";"B7",
% \ar @{-}@*{[|(4)]} "B4";"B5",
% \ar @{-}@*{[|(4)]} "B3";"B6",
% \ar @*{[|(0.5)]} "B0";"B2",
% \ar @*{[|(0.5)]} "B2";"B6",
% \ar @*{[|(0.5)]} "B6";"B4",
% \ar @*{[|(0.5)]} "B1";"B7",
% \ar @*{[|(0.5)]} "B7";"B3",
% \ar @*{[|(0.5)]} "B3";"B5",
% \endxy}
\fi
+ \mbox{\begin{tabular}{c}
three rotations 
\\
by $\frac{2\pi k}{8}$,
\\
$k\in \{1,2,3\}$
\end{tabular}}\right\}
\nonumber
\\
& + (-\lambda)^3 \left\{\left(
\ifpdf\includegraphics[viewport=0 25 60 60]{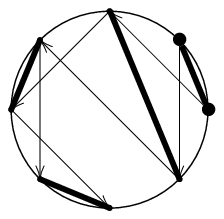}\else
\input fig-8b.xy
% \mbox{\xy <1cm,0cm>:
% (0,0)*\cir<1cm>{},
% (1,0)="B0",
% (\halfroottwo,\halfroottwo)="B1",
% (0,1)="B2",
% (-\halfroottwo,\halfroottwo)="B3",
% (-1,0)="B4",
% (-\halfroottwo,-\halfroottwo)="B5",
% (0,-1)="B6",
% (\halfroottwo,-\halfroottwo)="B7",
% \ar @{*{\bullet}{-}*{\bullet}}@*{[|(4)]} "B0";"B1",
% \ar @{-}@*{[|(4)]} "B2";"B7",
% \ar @{-}@*{[|(4)]} "B4";"B3",
% \ar @{-}@*{[|(4)]} "B5";"B6",
% \ar @*{[|(0.5)]} "B0";"B2",
% \ar @*{[|(0.5)]} "B2";"B4",
% \ar @*{[|(0.5)]} "B4";"B6",
% \ar @*{[|(0.5)]} "B1";"B7",
% \ar @*{[|(0.5)]} "B7";"B3",
% \ar @*{[|(0.5)]} "B3";"B5",
% \endxy}
\fi
+ 
\ifpdf\includegraphics[viewport=0 25 60 60]{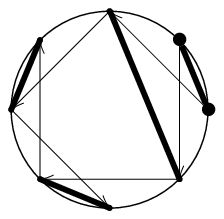}\else
\input fig-8c.xy
% \mbox{\xy <1cm,0cm>:
% (0,0)*\cir<1cm>{},
% (1,0)="B0",
% (\halfroottwo,\halfroottwo)="B1",
% (0,1)="B2",
% (-\halfroottwo,\halfroottwo)="B3",
% (-1,0)="B4",
% (-\halfroottwo,-\halfroottwo)="B5",
% (0,-1)="B6",
% (\halfroottwo,-\halfroottwo)="B7",
% \ar @{*{\bullet}{-}*{\bullet}}@*{[|(4)]} "B0";"B1",
% \ar @{-}@*{[|(4)]} "B2";"B7",
% \ar @{-}@*{[|(4)]} "B4";"B3",
% \ar @{-}@*{[|(4)]} "B5";"B6",
% \ar @*{[|(0.5)]} "B0";"B2",
% \ar @*{[|(0.5)]} "B2";"B4",
% \ar @*{[|(0.5)]} "B4";"B6",
% \ar @*{[|(0.5)]} "B1";"B7",
% \ar @*{[|(0.5)]} "B7";"B5",
% \ar @*{[|(0.5)]} "B5";"B3",
% \endxy}
\fi
+ 
\ifpdf\includegraphics[viewport=0 25 60 60]{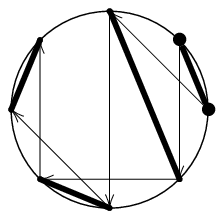}\else
\input fig-8d.xy
% \mbox{\xy <1cm,0cm>:
% (0,0)*\cir<1cm>{},
% (1,0)="B0",
% (\halfroottwo,\halfroottwo)="B1",
% (0,1)="B2",
% (-\halfroottwo,\halfroottwo)="B3",
% (-1,0)="B4",
% (-\halfroottwo,-\halfroottwo)="B5",
% (0,-1)="B6",
% (\halfroottwo,-\halfroottwo)="B7",
% \ar @{*{\bullet}{-}*{\bullet}}@*{[|(4)]} "B0";"B1",
% \ar @{-}@*{[|(4)]} "B2";"B7",
% \ar @{-}@*{[|(4)]} "B4";"B3",
% \ar @{-}@*{[|(4)]} "B5";"B6",
% \ar @*{[|(0.5)]} "B0";"B2",
% \ar @*{[|(0.5)]} "B2";"B6",
% \ar @*{[|(0.5)]} "B6";"B4",
% \ar @*{[|(0.5)]} "B1";"B7",
% \ar @*{[|(0.5)]} "B7";"B5",
% \ar @*{[|(0.5)]} "B5";"B3",
% \endxy}
\fi
\right)
+  \mbox{\begin{tabular}{@{}c@{}}
seven rotations 
\\
by $\frac{2\pi k}{8}$,
\\
$k{\in} \{1,2,3,4,5,6,7\}$
\end{tabular}}\right\}
\nonumber
\\
&  +(-\lambda)^3 \left\{\left(
\ifpdf\includegraphics[viewport=0 25 60 60]{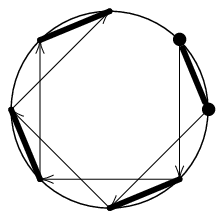}\else
\input fig-8e.xy
% \mbox{\xy <1cm,0cm>:
% (0,0)*\cir<1cm>{},
% (1,0)="B0",
% (\halfroottwo,\halfroottwo)="B1",
% (0,1)="B2",
% (-\halfroottwo,\halfroottwo)="B3",
% (-1,0)="B4",
% (-\halfroottwo,-\halfroottwo)="B5",
% (0,-1)="B6",
% (\halfroottwo,-\halfroottwo)="B7",
% \ar @{*{\bullet}{-}*{\bullet}}@*{[|(4)]} "B0";"B1",
% \ar @{-}@*{[|(4)]} "B2";"B3",
% \ar @{-}@*{[|(4)]} "B4";"B5",
% \ar @{-}@*{[|(4)]} "B6";"B7",
% \ar @*{[|(0.5)]} "B0";"B6",
% \ar @*{[|(0.5)]} "B6";"B4",
% \ar @*{[|(0.5)]} "B4";"B2",
% \ar @*{[|(0.5)]} "B1";"B7",
% \ar @*{[|(0.5)]} "B7";"B5",
% \ar @*{[|(0.5)]} "B5";"B3",
% \endxy}
\fi
+ 
\ifpdf\includegraphics[viewport=0 25 60 60]{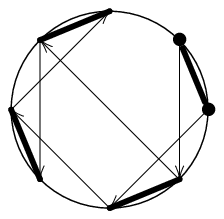}\else
\input fig-8f.xy
% \mbox{\xy <1cm,0cm>:
% (0,0)*\cir<1cm>{},
% (1,0)="B0",
% (\halfroottwo,\halfroottwo)="B1",
% (0,1)="B2",
% (-\halfroottwo,\halfroottwo)="B3",
% (-1,0)="B4",
% (-\halfroottwo,-\halfroottwo)="B5",
% (0,-1)="B6",
% (\halfroottwo,-\halfroottwo)="B7",
% \ar @{*{\bullet}{-}*{\bullet}}@*{[|(4)]} "B0";"B1",
% \ar @{-}@*{[|(4)]} "B2";"B3",
% \ar @{-}@*{[|(4)]} "B4";"B5",
% \ar @{-}@*{[|(4)]} "B6";"B7",
% \ar @*{[|(0.5)]} "B0";"B6",
% \ar @*{[|(0.5)]} "B6";"B4",
% \ar @*{[|(0.5)]} "B4";"B2",
% \ar @*{[|(0.5)]} "B1";"B7",
% \ar @*{[|(0.5)]} "B7";"B3",
% \ar @*{[|(0.5)]} "B3";"B5",
% \endxy}
\fi
+ 
\ifpdf\includegraphics[viewport=0 25 60 60]{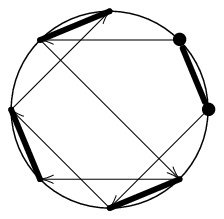}\else
\input fig-8g.xy
% \mbox{\xy <1cm,0cm>:
% (0,0)*\cir<1cm>{},
% (1,0)="B0",
% (\halfroottwo,\halfroottwo)="B1",
% (0,1)="B2",
% (-\halfroottwo,\halfroottwo)="B3",
% (-1,0)="B4",
% (-\halfroottwo,-\halfroottwo)="B5",
% (0,-1)="B6",
% (\halfroottwo,-\halfroottwo)="B7",
% \ar @{*{\bullet}{-}*{\bullet}}@*{[|(4)]} "B0";"B1",
% \ar @{-}@*{[|(4)]} "B2";"B3",
% \ar @{-}@*{[|(4)]} "B4";"B5",
% \ar @{-}@*{[|(4)]} "B6";"B7",
% \ar @*{[|(0.5)]} "B0";"B6",
% \ar @*{[|(0.5)]} "B6";"B4",
% \ar @*{[|(0.5)]} "B4";"B2",
% \ar @*{[|(0.5)]} "B1";"B3",
% \ar @*{[|(0.5)]} "B3";"B7",
% \ar @*{[|(0.5)]} "B7";"B5",
% \endxy} 
\fi
\right.\right.
\nonumber
\\
&
\qquad\qquad+ 
\ifpdf\includegraphics[viewport=0 25 60 60]{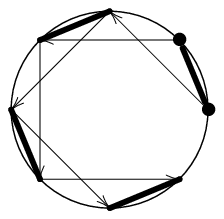}\else
\input fig-8h.xy
% \mbox{\xy <1cm,0cm>:
% (0,0)*\cir<1cm>{},
% (1,0)="B0",
% (\halfroottwo,\halfroottwo)="B1",
% (0,1)="B2",
% (-\halfroottwo,\halfroottwo)="B3",
% (-1,0)="B4",
% (-\halfroottwo,-\halfroottwo)="B5",
% (0,-1)="B6",
% (\halfroottwo,-\halfroottwo)="B7",
% \ar @{*{\bullet}{-}*{\bullet}}@*{[|(4)]} "B0";"B1",
% \ar @{-}@*{[|(4)]} "B2";"B3",
% \ar @{-}@*{[|(4)]} "B4";"B5",
% \ar @{-}@*{[|(4)]} "B6";"B7",
% \ar @*{[|(0.5)]} "B0";"B2",
% \ar @*{[|(0.5)]} "B2";"B4",
% \ar @*{[|(0.5)]} "B4";"B6",
% \ar @*{[|(0.5)]} "B1";"B3",
% \ar @*{[|(0.5)]} "B3";"B5",
% \ar @*{[|(0.5)]} "B5";"B7",
% \endxy}
\fi
+ \ifpdf\includegraphics[viewport=0 25 60 60]{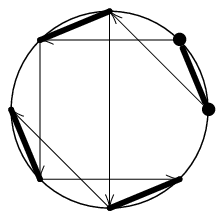}\else
\input fig-8i.xy
% \mbox{\xy <1cm,0cm>:
% (0,0)*\cir<1cm>{},
% (1,0)="B0",
% (\halfroottwo,\halfroottwo)="B1",
% (0,1)="B2",
% (-\halfroottwo,\halfroottwo)="B3",
% (-1,0)="B4",
% (-\halfroottwo,-\halfroottwo)="B5",
% (0,-1)="B6",
% (\halfroottwo,-\halfroottwo)="B7",
% \ar @{*{\bullet}{-}*{\bullet}}@*{[|(4)]} "B0";"B1",
% \ar @{-}@*{[|(4)]} "B2";"B3",
% \ar @{-}@*{[|(4)]} "B4";"B5",
% \ar @{-}@*{[|(4)]} "B6";"B7",
% \ar @*{[|(0.5)]} "B0";"B2",
% \ar @*{[|(0.5)]} "B2";"B6",
% \ar @*{[|(0.5)]} "B6";"B4",
% \ar @*{[|(0.5)]} "B1";"B3",
% \ar @*{[|(0.5)]} "B3";"B5",
% \ar @*{[|(0.5)]} "B5";"B7",
% \endxy}  
\fi
+ 
\ifpdf\includegraphics[viewport=0 25 60 60]{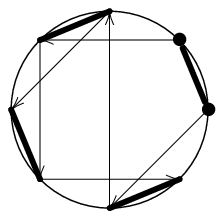}\else
\input fig-8j.xy
% \mbox{\xy <1cm,0cm>:
% (0,0)*\cir<1cm>{},
% (1,0)="B0",
% (\halfroottwo,\halfroottwo)="B1",
% (0,1)="B2",
% (-\halfroottwo,\halfroottwo)="B3",
% (-1,0)="B4",
% (-\halfroottwo,-\halfroottwo)="B5",
% (0,-1)="B6",
% (\halfroottwo,-\halfroottwo)="B7",
% \ar @{*{\bullet}{-}*{\bullet}}@*{[|(4)]} "B0";"B1",
% \ar @{-}@*{[|(4)]} "B2";"B3",
% \ar @{-}@*{[|(4)]} "B4";"B5",
% \ar @{-}@*{[|(4)]} "B6";"B7",
% \ar @*{[|(0.5)]} "B0";"B6",
% \ar @*{[|(0.5)]} "B6";"B2",
% \ar @*{[|(0.5)]} "B2";"B4",
% \ar @*{[|(0.5)]} "B1";"B3",
% \ar @*{[|(0.5)]} "B3";"B5",
% \ar @*{[|(0.5)]} "B5";"B7",
% \endxy}
\fi
\nonumber
\\
& \qquad\qquad\ifpdf\rule{0mm}{2cm}\fi \left.\left.
+
\ifpdf\includegraphics[viewport=0 25 60 60]{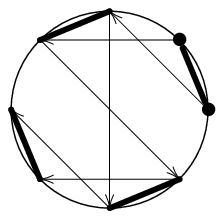}\else
\input fig-8k.xy
% \mbox{\xy <1cm,0cm>:
% (0,0)*\cir<1cm>{},
% (1,0)="B0",
% (\halfroottwo,\halfroottwo)="B1",
% (0,1)="B2",
% (-\halfroottwo,\halfroottwo)="B3",
% (-1,0)="B4",
% (-\halfroottwo,-\halfroottwo)="B5",
% (0,-1)="B6",
% (\halfroottwo,-\halfroottwo)="B7",
% \ar @{*{\bullet}{-}*{\bullet}}@*{[|(4)]} "B0";"B1",
% \ar @{-}@*{[|(4)]} "B2";"B3",
% \ar @{-}@*{[|(4)]} "B4";"B5",
% \ar @{-}@*{[|(4)]} "B6";"B7",
% \ar @*{[|(0.5)]} "B0";"B2",
% \ar @*{[|(0.5)]} "B2";"B6",
% \ar @*{[|(0.5)]} "B6";"B4",
% \ar @*{[|(0.5)]} "B1";"B3",
% \ar @*{[|(0.5)]} "B3";"B7",
% \ar @*{[|(0.5)]} "B7";"B5",
% \endxy}
\fi
+
\ifpdf\includegraphics[viewport=0 25 60 60]{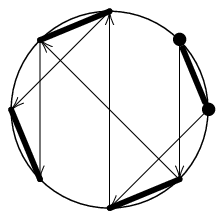}\else
\input fig-8l.xy
% \mbox{\xy <1cm,0cm>:
% (0,0)*\cir<1cm>{},
% (1,0)="B0",
% (\halfroottwo,\halfroottwo)="B1",
% (0,1)="B2",
% (-\halfroottwo,\halfroottwo)="B3",
% (-1,0)="B4",
% (-\halfroottwo,-\halfroottwo)="B5",
% (0,-1)="B6",
% (\halfroottwo,-\halfroottwo)="B7",
% \ar @{*{\bullet}{-}*{\bullet}}@*{[|(4)]} "B0";"B1",
% \ar @{-}@*{[|(4)]} "B2";"B3",
% \ar @{-}@*{[|(4)]} "B4";"B5",
% \ar @{-}@*{[|(4)]} "B6";"B7",
% \ar @*{[|(0.5)]} "B0";"B6",
% \ar @*{[|(0.5)]} "B6";"B2",
% \ar @*{[|(0.5)]} "B2";"B4",
% \ar @*{[|(0.5)]} "B1";"B7",
% \ar @*{[|(0.5)]} "B7";"B3",
% \ar @*{[|(0.5)]} "B3";"B5",
% \endxy}
\fi
+
\ifpdf\includegraphics[viewport=0 25 60 60]{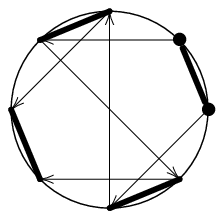}\else
\input fig-8m.xy
% \mbox{\xy <1cm,0cm>:
% (0,0)*\cir<1cm>{},
% (1,0)="B0",
% (\halfroottwo,\halfroottwo)="B1",
% (0,1)="B2",
% (-\halfroottwo,\halfroottwo)="B3",
% (-1,0)="B4",
% (-\halfroottwo,-\halfroottwo)="B5",
% (0,-1)="B6",
% (\halfroottwo,-\halfroottwo)="B7",
% \ar @{*{\bullet}{-}*{\bullet}}@*{[|(4)]} "B0";"B1",
% \ar @{-}@*{[|(4)]} "B2";"B3",
% \ar @{-}@*{[|(4)]} "B4";"B5",
% \ar @{-}@*{[|(4)]} "B6";"B7",
% \ar @*{[|(0.5)]} "B0";"B6",
% \ar @*{[|(0.5)]} "B6";"B2",
% \ar @*{[|(0.5)]} "B2";"B4",
% \ar @*{[|(0.5)]} "B1";"B3",
% \ar @*{[|(0.5)]} "B3";"B7",
% \ar @*{[|(0.5)]} "B7";"B5",
% \endxy}
\fi
+
\ifpdf\includegraphics[viewport=0 25 60 60]{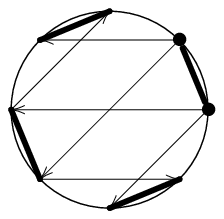}\else
\input fig-8n.xy
% \mbox{\xy <1cm,0cm>:
% (0,0)*\cir<1cm>{},
% (1,0)="B0",
% (\halfroottwo,\halfroottwo)="B1",
% (0,1)="B2",
% (-\halfroottwo,\halfroottwo)="B3",
% (-1,0)="B4",
% (-\halfroottwo,-\halfroottwo)="B5",
% (0,-1)="B6",
% (\halfroottwo,-\halfroottwo)="B7",
% \ar @{*{\bullet}{-}*{\bullet}}@*{[|(4)]} "B0";"B1",
% \ar @{-}@*{[|(4)]} "B2";"B3",
% \ar @{-}@*{[|(4)]} "B4";"B5",
% \ar @{-}@*{[|(4)]} "B6";"B7",
% \ar @*{[|(0.5)]} "B0";"B6",
% \ar @*{[|(0.5)]} "B0";"B4",
% \ar @*{[|(0.5)]} "B4";"B2",
% \ar @*{[|(0.5)]} "B1";"B3",
% \ar @*{[|(0.5)]} "B1";"B5",
% \ar @*{[|(0.5)]} "B5";"B7",
% \endxy}
\fi
\right)
+ \!\!\! 
  \mbox{\begin{tabular}{c}
one \\ rotation 
\\
by $\frac{2\pi}{8}$
\end{tabular}} \!\!\!\right\}.
\label{G8-graph}
\end{align}
Again we obtain all $C_4=14$ non-crossing pairings of the cyclic
external indices $b_0,\dots,b_7$ which give the products of
$G^{(0)}_{|b_ib_j|}$.

\section{Self-dual noncommutative  $\phi^4_4$-theory}

\subsection{Definition of the matrix model}

\label{sec:GW}

In order to improve the problems of four-dimensional quantum field
theory it was suggested to include ``gravity effects'' through
deforming space-time. A simple example for such a deformation is the
Moyal space, which arises by deformation of the Fr\'echet algebra of 
Schwarz class functions by the action of the translation 
group \cite{Rieffel:1993},
\begin{align}
(f\star g)(x)=\int_{\mathbb{R}^d\times \mathbb{R}^d}
\frac{dy\,dk}{(2\pi)^d} 
f\big(x+\tfrac{1}{2} \Theta k\big) \,g\big(x+y\big)\, e^{\mathrm{i}
  \langle k,y\rangle}\;,\qquad f,g\in \mathcal{S}(\mathbb{R}^d)\;.
\end{align}
Here, $\Theta=-\Theta^t\in M_d(\mathbb{R})$ is a skew-symmetric
matrix. Although not required by the general theory, for our purpose
we need $\Theta$ to be of full rank, which implies that $d$ is even.
The algebra $\mathcal{A}_\Theta=(\mathcal{S}(\mathbb{R}^d),\star)$ is a
pre-$C^*$-algebra \cite[Prop.~2.14]{Gayral:2003dm}. The
standard $\mathbb{R}^d$-Lebesgue integral defines a linear
functional which is positive $\int_{\mathbb{R}^d} (f\star
\bar{f})(x)\geq 0$ and tracial 
$\int_{\mathbb{R}^d} (f\star g)(x)= 
\int_{\mathbb{R}^d} (g\star f)(x)$.
Therefore, GNS-construction gives rise to a Hilbert space
$H=H_1\otimes H_2$ on which there are commuting representations
$\pi\otimes \mathrm{id}$ of $\mathcal{A}_\Theta$ and
$\mathrm{id}\otimes \pi^{op}$ of the opposite algebra
$\mathcal{A}_\Theta^{op}$. Restricted to $H_1$, any element of
$\mathcal{A}_\Theta$ is a compact operator on $H_1$, in fact even a trace-class
operator.

This means that we can define field-theoretical matrix models on
$\mathcal{A}_\Theta\ni \phi$ by specifying the polynomial $P[\phi]$
and the external matrix $E$. We take $d=4$ and 
$P[\phi]=\frac{\lambda_4}{4} \phi^4$. The unbounded
operator $E$ should play the r\^ole of the Laplacian. However, the
Laplacian $\Delta$ on $\mathbb{R}^d$ does not have compact resolvent. This
non-compactness of $\Delta$ is the reason why field theories based on
action functionals such as 
\[
\int_{\mathbb{R}^d} dx \Big( \frac{1}{2} \phi (-\Delta+\mu^2)\phi 
+\frac{\lambda_4}{4} \phi\star\phi\star\phi\star\phi\Big)(x)
\]
have bad properties (UV/IR-mixing \cite{Minwalla:1999px}).

In our previous work \cite{Grosse:2004yu} we found a way to handle
this problem. We realised that extending the Laplacian to the harmonic
oscillator Hamiltonian (which has compact resolvent!), the field theory
defined by the action
\begin{align}
S=64\pi^2
\int d^4x\Big(
\frac{1}{2} \phi (-\Delta+ \Omega^2 \|2\Theta^{-1}x\|^2 +\mu^2) \phi
+ \frac{\lambda}{4} \phi\star \phi\star \phi \star \phi \Big)(x)
\label{GW}
\end{align}
is renormalisable to all orders of perturbation theory. 
The perturbative renormalisability was proved by renormalisation group 
techniques \cite{Grosse:2004yu} and multi-scale analysis, both in  
matrix representation \cite{Rivasseau:2005bh} and position space
\cite{Gurau:2005gd}. The model is covariant under the Langmann-Szabo 
duality transformation \cite{Langmann:2002cc}. It becomes self-dual at 
$\Omega=1$, a point at which the $\beta$-function vanishes to all 
orders in perturbation theory \cite{Disertori:2006nq}. Certain variants 
of this model have also been treated, see \cite{Rivasseau:2007ab} for 
a review.

Writing the action functional (\ref{GW}) in the matrix base of the
Moyal space, see \cite{Grosse:2004yu,Grosse:2003aj}, one obtains a
\emph{non-local} matrix model in the sense that the $\phi$-bilinear
term is a tensor product $\mathrm{tr}(\hat{E}\cdot (\phi \otimes
\phi))$. We have studied the power-counting behaviour of such matrix
models in \cite{Grosse:2003aj}. By a sophisticated analysis of the
tensor-product external matrix $\hat{E}$ we proved in
\cite{Grosse:2004yu} the perturbative renormalisability of the model
(\ref{GW}).  The action simplifies enormously at the self-duality
point $\Omega=1$ where it becomes an ordinary matrix model in the
sense of section~\ref{sec:matrixmodel} with diagonal external matrix
$E$. In terms of the \emph{bare} quantities, which involves the bare
mass $\mu_{bare}$ and the wave function renormalisation $\phi \mapsto
Z^{\frac{1}{2}}\phi$, the action functional becomes 
\cite[eqs.\ (2.5)+(2.6)]{Grosse:2004yu}
\begin{subequations}
\begin{align}
S[\phi]&=  V\Big( \sum_{\under{m},\under{n} \in \mathbb{N}^2_{\mathcal{N}} }
E_{\under{m}} \phi_{\under{m}\under{n}} \phi_{\under{n}\under{m}} 
+ \frac{S_{int}[\phi]}{V}\Big)\;,\qquad V:=\Big(\frac{\theta}{4}\Big)^2
\label{theta-volume}
\\
E_{\under{m}}
&= Z\Big( \frac{4}{\theta} |\under{m}|
+\frac{\mu_{bare}^2}{2} \Big) \;, 
\qquad
\frac{S_{int}[\phi]}{V}= \frac{Z^2\lambda}{4}
\sum_{\under{m},\under{n},\under{k},\under{l} 
\in \mathbb{N}^2_{\mathcal{N}} } 
\phi_{\under{m}\under{n}} \phi_{\under{n}\under{k}} 
\phi_{\under{k}\under{l}} \phi_{\under{l}\under{m}}\;.
\label{Vphi}
\end{align}
\end{subequations}
We have made use of the fact that, after an appropriate coordinate
transformation in $\mathbb{R}^4$, the only non-vanishing components of
$\Theta$ are
$\Theta_{12}=-\Theta_{21}=\Theta_{34}=-\Theta_{43}=:\theta$.  We have
also absorbed the ground energy of the harmonic oscillator into a
redefinition of $\mu_{bare}$. The prefactor $64\pi^2$ in (\ref{GW}) is
introduced in order to identify the volume $V=(\frac{\theta}{4})^2$
instead of the usual factor $(2\pi\theta)^2$ arising from the integral
of the matrix basis.  All summation indices
$\under{m},\under{n},\dots$ belong to $\mathbb{N}^2$, with
$|\under{m}|:=m_1+m_2$. The symbol $\mathbb{N}^2_{\mathcal{N}}$ refers
to a cut-off in the matrix size which later on will be made more
precise. The scalar field is real, $\phi_{\under{m}\under{n}}
=\overline{\phi_{\under{n}\under{m}}}$. We then know from
Theorem~\ref{Thm:beta=0} that there is no need of a renormalisation
$\lambda =Z_4\lambda_{ren}$ of the coupling constant.

\subsection{Renormalisation of the two-point function}

We now focus on the Schwinger-Dyson equation (\ref{SD-G}) for the
planar regular connected two-point function
$G^{(0)}_{|\under{a}\under{b}|}$.  From (\ref{Vphi}) we read off
$\lambda_4= Z^2 \lambda$.  The summation over
$\under{p}\in I:=\mathbb{N}^2_{\mathcal{N}}$ in (\ref{SD-G}) diverges if we
remove the cut-off $\mathbb{N}^2_{\mathcal{N}} \to \mathbb{N}^2$ and,
therefore, requires renormalisation. Renormalisation is defined by
\emph{normalisation conditions} for the first Taylor coefficients of
the \emph{one-particle irreducible (1PI)} function
$\Gamma_{\under{a}\under{b}}$ related to $G^{(0)}_{|\under{a}\under{b}|}$
according to
\begin{align}
G^{(0)}_{|\under{a}\under{b}|}= 
(H_{\under{a}\under{b}}-\Gamma_{\under{a}\under{b}})^{-1}\;,
\qquad H_{\under{a}\under{b}}:=E_{\under{a}}+E_{\under{b}} =
Z\Big(  \frac{4}{\theta}(|\under{a}|+|\under{b}|)+\mu_{bare}^2\Big) \;.
\label{Eab}
\end{align}
It is then an easy exercise to rewrite (\ref{SD-G}) into the following
equation for $\Gamma_{\under{a}\under{b}}$:
\begin{align}
\Gamma_{\under{a}\under{b}}=  
-\frac{\lambda Z^2}{V}   \sum_{\under{p} \in
\mathbb{N}^2_{\mathcal{N}} }  \Big(\frac{1}{H_{\under{a}\under{p}}
-\Gamma_{\under{a}\under{p}}}
+\frac{1}{H_{\under{p}\under{b}}-\Gamma_{\under{p}\under{b}}}\Big)
+ \frac{\lambda Z}{V}  \sum_{\under{p} \in \mathbb{N}^2_{\mathcal{N}} }
\frac{1}{(H_{\under{p}\under{b}}-\Gamma_{\under{p}\under{b}})} 
\frac{ \Gamma_{\under{p}\under{b}} -\Gamma_{\under{a}\under{b}}}{
\frac{4}{\theta}(|\under{p}|{-}|\under{a}|)}\;.
\label{SD-Gamma}
\end{align}
We require 
\begin{align}
\Gamma_{\under{a}\under{b}}= Z \mu_{bare}^2 -\mu^2 
+ (Z-1) \frac{4}{\theta} (|\under{a}|+|\under{b}|) 
+\Gamma^{ren}_{\under{a}\under{b}}\;, 
\nonumber
\\
\Gamma^{ren}_{\under{0}\under{0}}=0 \qquad 
(\partial \Gamma^{ren})_{\under{0}\under{0}}=0\:,
\label{normalisation}
\end{align}
where $\partial \Gamma^{ren}$ is any of the (discrete) derivatives 
with respect to 
$a_1,a_2,b_1,b_2$. This implies
\begin{align}
\frac{1}{G^{(0)}_{|\under{a}\under{b}|}}
= H_{\under{a}\under{b}}-\Gamma_{\under{a}\under{b}}
= \frac{4}{\theta}(|\under{a}|+|\under{b}|) 
+\mu^2-\Gamma_{\under{a}\under{b}}^{ren}\;.
\label{G-inverse}  
\end{align}
Hence, $\mu$ is the renormalised mass, and both
$G^{(0)}_{|\under{a}\under{b}|}$ and $\Gamma^{ren}_{\under{a}\under{b}}$ should be
regular if the cut-off in the matrix indices is removed.  Inserted
into (\ref{SD-Gamma}) we find
\begin{align}
  &Z \mu_{bare}^2 -\mu^2 +
  (Z-1) \tfrac{4}{\theta}(|\under{a}|+|\under{b}|) 
+\Gamma^{ren}_{\under{a}\under{b}}
  \nonumber
  \\
  &= -\frac{\lambda}{V}  \sum_{\under{p} \in
\mathbb{N}^2_{\mathcal{N}} } 
\frac{Z^2}{\frac{4}{\theta}(|\under{a}|+|\under{p}|)+\mu^2
-\Gamma^{ren}_{\under{a}\under{p}}}
  \nonumber
  \\
  &
-\frac{\lambda}{V} \sum_{\under{p} \in
\mathbb{N}^2_{\mathcal{N}} }\Big( 
\frac{Z}{\frac{4}{\theta}(|\under{b}|+|\under{p}|) 
+\mu^2 -\Gamma^{ren}_{\under{p}\under{b}}} 
- \frac{Z}{\frac{4}{\theta}(|\under{b}|+|\under{p}|)+\mu^2 
-\Gamma^{ren}_{\under{p}\under{b}}} \:
  \frac{\Gamma^{ren}_{\under{p}\under{b}}-\Gamma^{ren}_{\under{a}\under{b}}}{
\frac{4}{\theta}(|\under{p}|-|\under{a}|)}\Big)\;.
\label{Gamma-total}
\end{align}
Notice the different exponents of $Z$ in the two tadpoles
$\sum_{\under{p}} \frac{1}{\frac{4}{\theta}(|\under{\bullet}|+|\under{p}|)+\mu^2 
-\Gamma^{ren}_{\under{p}\under{\bullet}}}$! 
Separating the first Taylor term (by putting $|\under{a}|=|\under{b}|=0$)  
we obtain
\begin{align*}
&Z \mu_{bare}^2 -\mu^2 
\\*
 &= -\frac{\lambda}{V} \sum_{\under{p} \in
\mathbb{N}^2_{\mathcal{N}} } 
\frac{Z^2}{\frac{4}{\theta}|\under{p}|+\mu^2 -\Gamma^{ren}_{\under{0}\under{p}}} 
-\frac{\lambda}{V} \sum_{\under{p} \in
\mathbb{N}^2_{\mathcal{N}} } 
\Big( \frac{Z}{\frac{4}{\theta}|\under{p}|+\mu^2 
-\Gamma^{ren}_{\under{p}\under{0}}} 
- \frac{Z}{\frac{4}{\theta}|\under{p}|+\mu^2 -\Gamma^{ren}_{\under{p}\under{0}}}
\:  \frac{\Gamma^{ren}_{\under{p}\under{0}}}{\frac{4}{\theta}|\under{p}|}\Big)\;,
\end{align*}
which inserted into (\ref{Gamma-total}) gives
\begin{align}
(Z-1) \tfrac{4}{\theta} &(|\under{a}|+|\under{b}|) 
+\Gamma^{ren}_{\under{a}\under{b}}
\nonumber
\\
  &= -\frac{\lambda}{V} \sum_{\under{p} \in
\mathbb{N}^2_{\mathcal{N}} } \Big( 
\frac{Z^2}{\frac{4}{\theta}(|\under{a}|+|\under{p}|)+\mu^2
-\Gamma^{ren}_{\under{a}\under{p}}}
-\frac{Z^2}{\frac{4}{\theta}|\under{p}|+\mu^2
-\Gamma^{ren}_{\under{0}\under{p}}}\Big)
  \nonumber
  \\[-0.5ex]
  &
-\frac{\lambda}{V} \sum_{\under{p} \in
\mathbb{N}^2_{\mathcal{N}} } \Big( 
\frac{Z}{\frac{4}{\theta}(|\under{b}|+|\under{p}|)+\mu^2 
-\Gamma^{ren}_{\under{p}\under{b}}} 
-\frac{Z}{\frac{4}{\theta}|\under{p}|+\mu^2-\Gamma^{ren}_{\under{p}\under{0}}}
\nonumber
\\[-0.5ex]
& \qquad\quad - \frac{Z}{\frac{4}{\theta}(|\under{b}|+|\under{p}|)
+\mu^2 -\Gamma^{ren}_{\under{p}\under{b}}}
 \: \frac{\Gamma^{ren}_{\under{p}\under{b}}-\Gamma^{ren}_{\under{a}\under{b}}}{
\frac{4}{\theta}(|\under{p}|-|\under{a}|)}
+ \frac{Z}{\frac{4}{\theta}|\under{p}|+\mu^2 -\Gamma^{ren}_{\under{p}\under{0}}}
\:  \frac{\Gamma^{ren}_{\under{p}\under{0}}}{\frac{4}{\theta}|\under{p}|}\Big)\;.
\label{ZG}
\end{align}
The next step consists in differentiating (\ref{ZG}) at $0$ with
respect to $a_i$ and $b_i$ in order to get a self-consistent system of
equations for $Z,\Gamma_{\under{a}\under{b}}^{ren}$.  It is now
crucial that (\ref{ZG}) depends only on the sums
$|\under{a}|=a_1{+}a_2$, $|\under{b}|=b_1{+}b_2$ and $|\under{p}|=p_1{+}p_2$
of indices. This degeneracy is first passed on to the bare two-point
function $\Gamma_{\under{a}\under{b}}$. But then the derivatives
$\frac{\partial}{\partial a_i}|_{|\under{a}|,|\under{b}|=0}=
\frac{\partial}{\partial |\under{a}|}|_{|\under{a}|,|\under{b}|=0}$
and $\frac{\partial}{\partial b_i}|_{|\under{a}|,|\under{b}|=0}=
\frac{\partial}{\partial |\under{b}|}|_{|\under{a}|,|\under{b}|=0}$
respect the degeneracy, so that also the renormalised two-point
function $\Gamma^{red}_{\under{a}\under{b}}$ depends only on the 
1-norms $|\under{a}|=a_1{+}a_2$ and $|\under{b}|=b_1{+}b_2$. Consequently,
we can replace the index sums by $\sum_{\under{p} \in
  \mathbb{N}^2_{\mathcal{N}} } f(|\under{p}|) =
\sum_{|\under{p}|=0}^{\mathcal{N}}  (|\under{p}|{+}1)f(|\under{p}|)$, because
there are $|\under{p}|+1$ possibilities\footnote{In $d$ dimensions the
  measure would be
  $\binom{|\under{p}|+\frac{d}{2}-1}{\frac{d}{2}-1}$.}  to write
$|\under{p}|$ as a sum $|\under{p}|=p_1+p_2$ with $p_1,p_2\in
\mathbb{N}$. The cut-off $|\under{p}|\leq \mathcal{N} \in \mathbb{N}$ then
specifies the previously unclear symbol $\under{p} \in 
\mathbb{N}^2_{\mathcal{N}} $.
In summary, (\ref{ZG}) takes with $V=(\frac{\theta}{4})^2$ from 
(\ref{theta-volume}) the form
\begin{align}
(Z-1) &\tfrac{4}{\theta} (|\under{a}|+|\under{b}|) 
+\Gamma^{ren}_{\under{a}\under{b}}
\nonumber
\\
  &= -\lambda \Big(\frac{4}{\theta}\Big)^2 \sum_{|\under{p}| =0}^{\mathcal{N}} 
(|\under{p}|+1) \Big( 
\frac{Z^2}{\frac{4}{\theta}(|\under{a}|+|\under{p}|)+\mu^2
-\Gamma^{ren}_{\under{a}\under{p}}}
-\frac{Z^2}{\frac{4}{\theta}|\under{p}|+\mu^2
-\Gamma^{ren}_{\under{0}\under{p}}}\Big)
  \nonumber
  \\[-0.5ex]
  &
-\lambda\Big(\frac{4}{\theta}\Big)^2 \sum_{|\under{p}|=0}^{\mathcal{N}} 
(|\under{p}|+1) \Big( 
\frac{Z}{\frac{4}{\theta}(|\under{b}|+|\under{p}|)+\mu^2 
-\Gamma^{ren}_{\under{p}\under{b}}} 
-\frac{Z}{\frac{4}{\theta}|\under{p}|+\mu^2-\Gamma^{ren}_{\under{p}\under{0}}}
\nonumber
\\[-0.5ex]
& \qquad\quad - \frac{Z}{\frac{4}{\theta}(|\under{b}|+|\under{p}|)
+\mu^2 -\Gamma^{ren}_{\under{p}\under{b}}}\:
  \frac{\Gamma^{ren}_{\under{p}\under{b}}-\Gamma^{ren}_{\under{a}\under{b}}}{
\frac{4}{\theta}(|\under{p}|-|\under{a}|)}
+ \frac{Z}{\frac{4}{\theta}|\under{p}|+\mu^2 -\Gamma^{ren}_{\under{p}\under{0}}}
\:  \frac{\Gamma^{ren}_{\under{p}\under{0}}}{\frac{4}{\theta}|\under{p}|}\Big)\;.
\label{ZG-norm}
\end{align}

\subsection{Integral representation}

\label{sec:Integral}

We study a particular limit in which the self-dual noncommutative
$\phi^4_4$-model converges to a large-$\mathcal{N}$ limit of a certain matrix
model. This limit is defined by sending $\theta,\mathcal{N}\to \infty$ 
such that 
\begin{equation}
\frac{4}{\theta} \mathcal{N}  = \mu^2  (1+\mathcal{Y})\Lambda^2
= \mathrm{const}\;.
\label{limit}
\end{equation}
Here $\mathcal{Y}>-1$ is a real number which we identify later in
order to simplify our equation. The number $(1+\mathcal{Y})$ can be
seen as a finite wavefunction renormalisation.
This large-$\mathcal{N}$ limit (\ref{limit}) plays the r\^ole of a
\emph{thermodynamic} (infinite volume\footnote{See
  \cite{Gayral:2011vu,Grosse:2007jy} where it is shown that the
  spectral action behind the model under consideration yields for
  $\Omega=1$ an 8-dimensional finite volume proportional to $\theta^4$.})
\emph{limit} because it turns the discrete norms
$|\under{a}|,|\under{b}|,|\under{p}|$ of matrix indices into real
numbers $a,b,p \in [0,\Lambda^2]$ defined by
\begin{align}
&\tfrac{4}{\theta} |\under{a}|=:\mu^2 (1+\mathcal{Y})a\;,\quad
\tfrac{4}{\theta} |\under{b}|=:\mu^2 (1+\mathcal{Y})b\;,\quad
\tfrac{4}{\theta} |\under{p}|=:\mu^2 (1+\mathcal{Y})p\;,\quad
\Gamma^{ren}_{\under{a}\under{b}} =: \mu^2 \Gamma_{ab} \;.
\label{under-a-to-a}
\end{align}
In the very end we also have remove the cut-off $\Lambda^2$ 
in a \emph{continuum limit $\Lambda\to \infty$} (thus relaxing 
$\Lambda=\text{const}$ in (\ref{limit})). 
It is important that $\Gamma^{ren}_{\under{a}\under{b}}$ only depends
on $|\under{a}|$ and $|\under{b}|$ so that it converges to a function
$[0,\Lambda^2]^2 \ni (a,b) \mapsto \Gamma_{ab} \in \mathbb{R}$.
In the limit (\ref{limit}), a sum
over $|\under{p}|$ converges to 
\[
\tfrac{4}{\theta}\sum_{|\under{p}|=0}^{\mathcal{N}} f(\tfrac{4}{\theta}
|\under{p}|) \stackrel{\eqref{limit}}{\longrightarrow} 
\mu^2(1+\mathcal{Y}) \int_0^{\Lambda^2} dp \;f(\mu^2(1+\mathcal{Y})
p)\;.
\]
Therefore, the equation (\ref{ZG-norm}) converges to the following
\emph{integral equation} where the mass $\mu^2$ drops out (we work
with densities):
\begin{align}               
&(Z-1)(1+\mathcal{Y})(a+b) +\Gamma_{ab}
\nonumber
\\
&= -\lambda(1+\mathcal{Y})^2
\int_0^{\Lambda^2} p\,dp \;\Big( 
\frac{Z^2}{(a+p)(1+\mathcal{Y})+1 -\Gamma_{ap}} 
-\frac{Z^2}{p(1+\mathcal{Y})+1 -\Gamma_{0p}} 
\Big)
\nonumber
\\*
& -\lambda(1+\mathcal{Y})^2
\int_0^{\Lambda^2} p\,dp \; \Big( 
\frac{Z}{(b+p)(1+\mathcal{Y})+1 -\Gamma_{pb}} 
-\frac{Z}{p(1+\mathcal{Y})+1 -\Gamma_{p0}} 
  \nonumber
  \\*
  &
\qquad\qquad 
- \frac{Z}{(b+p)(1+\mathcal{Y})+1 -\Gamma_{pb}} 
  \:\frac{\Gamma_{pb}-\Gamma_{ab}}{
(1+\mathcal{Y})(p-a)}
+ \frac{Z}{p(1+\mathcal{Y})+1 -\Gamma_{p0}} 
\:  \frac{\Gamma_{p0}}{p(1+\mathcal{Y})}
\Big)\;.
\label{ZG-inteq}
\end{align}
The powerful analytical tools available for integral equations are a
huge advantage over the discrete equations. On the other hand, in
regions of the space of continuous parameters where $\Gamma$ varies
slowly we can expect the integral equation to be a good approximation
for the discrete equation~(\ref{ZG-norm}) also for finite $\theta$.

It turns out that the terms with $Z^2$-coefficient in (\ref{ZG-inteq}) 
need a subtle discussion. It is 
highly convenient to eliminate them via the equation resulting from  
(\ref{ZG-inteq}) at $b=0$:
\begin{align}               
&(Z-1)(1+\mathcal{Y})a +\Gamma_{a0}
\nonumber
\\
&= -\lambda(1+\mathcal{Y})^2
\int_0^{\Lambda^2} p\,dp \;\Big( 
\frac{Z^2}{(a+p)(1+\mathcal{Y})+1 -\Gamma_{ap}} 
-\frac{Z^2}{p(1+\mathcal{Y})+1 -\Gamma_{0p}} 
\Big)
\nonumber
\\
& +\lambda(1+\mathcal{Y})
\int_0^{\Lambda^2} dp \; 
\frac{Z}{p(1+\mathcal{Y})+1 -\Gamma_{p0}} \:
  \frac{a\Gamma_{p0}-p\Gamma_{a0}}{(p-a)}\;.
\label{ZG-inteq:b=0}
\end{align}
The difference between (\ref{ZG-inteq}) and
(\ref{ZG-inteq:b=0}), expressed in terms of the dimensionless function 
\begin{equation}
G_{ab}:=\mu^2 G^{(0)}_{|ab|} = \frac{1}{(a+b)(1+\mathcal{Y})+1-\Gamma_{ab}}\;,
\label{def:Gab}
\end{equation}
reads 
\begin{align}               
Z(1+\mathcal{Y})b - \frac{1}{G_{ab}}+\frac{1}{G_{a0}}
&= \lambda(1+\mathcal{Y})
\int_0^{\Lambda^2} p\,dp \; 
Z \frac{\frac{G_{pb}}{G_{ab}}-\frac{G_{p0}}{G_{a0}}}{p-a}\;.
\label{ZG-inteq:diffeq}
\end{align}
{}From (\ref{ZG-inteq:diffeq}) we obtain the desired equation for $Z^{-1}$ 
by putting $a=0$, dividing by $Z(1+\mathcal{Y})b$ and going to 
the limit $b\to 0$, where  
$\lim_{b\to 0} \frac{1}{b}(\frac{1}{G_{0b}}-1)=1+\mathcal{Y}$ and
$\lim_{b\to 0} G_{0b}=1$ are used:
\begin{align*}
Z^{-1}
&= 1-\lambda(1+\mathcal{Y})
\int_0^{\Lambda^2} dp \;G_{p0} 
 -\lambda
\lim_{b\to 0} 
\int_0^{\Lambda^2} dp \; 
\frac{G_{pb}-G_{p 0}}{b}\;.
\end{align*}
It will be convenient to choose $\mathcal{Y}$ as 
\begin{align}               
\mathcal{Y}:=
 -\lambda 
\lim_{b\to 0} 
\int_0^{\Lambda^2} dp \; 
\frac{G_{pb}-G_{p 0}}{b} \quad \Rightarrow \quad 
Z^{-1}
= (1+\mathcal{Y})\Big( 1-\lambda
\int_0^{\Lambda^2} dp \;G_{p0} \Big)\;.
\label{ZZ2}
\end{align}
Of course it remains to verify that $-1 <\mathcal{Y}<\infty$.
Multiplication of (\ref{ZG-inteq:diffeq}) by 
$\frac{G_{ab}}{b(1+\mathcal{Y})Z}$ and elimination of 
$Z^{-1}$ via (\ref{ZZ2}) gives 
\begin{align}
(G_{ab}-G_{a0})\Big(1 + \frac{1}{bG_{a0}} \Big)+ G_{a0}
&= \frac{\lambda}{b}
\int_0^{\Lambda^2} \!\! dp\; \Big( 
\frac{p(G_{pb}-G_{p0} )
-a(G_{ab}-G_{a0}) \frac{G_{p0} }{G_{a0}}}{p-a}
\Big)\;.
\label{G-ab-0}
\end{align}

\subsection{The Carleman equation}

We make now the crucial assumption that $G_{ab}$ 
is \emph{H\"older-continuous}, i.e.\ 
\begin{align}
|G_{pb}-G_{ab}|\leq C_b |p-a|^{\eta_b}
\qquad \forall \,0\leq a \neq  p \leq \Lambda^2\;,
\label{assump:Hoelder}
\end{align}
for some constants $C_b,\eta_b>0$. Under this
assumption we may replace the singular integrals by their Cauchy
principal values:
\begin{align}
\int_0^{\Lambda^2} dp \;\frac{G_{pb}-G_{ab}}{p-a}
&= \lim_{\epsilon\to 0} \Big(\int_0^{a-\epsilon}+
\int_{a+\epsilon}^{\Lambda^2} \Big) dp \;\frac{G_{pb}-G_{ab}}{p-a}\;.
\end{align}
Introducing the \emph{finite Hilbert transform}
\begin{align}
\mathcal{H}_a^{\!\Lambda} [f(\bullet)]:=\frac{1}{\pi} 
\lim_{\epsilon\to 0} \Big(\int_0^{a-\epsilon}+
\int_{a+\epsilon}^{\Lambda^2}\Big) \frac{f(p)}{p-a}\;,
\label{Hilbert}
\end{align}
we can rewrite the equation (\ref{G-ab-0}) as 
\begin{subequations}
\begin{align}
\Big(\frac{b}{a} + \frac{1+\lambda\pi a
\mathcal{H}_a^{\!\Lambda}[G_{\bullet 0}]}{a G_{a0}} 
\Big) D_{ab} 
-\lambda\pi \mathcal{H}_a^{\!\Lambda}[D_{\bullet b}]
&= -G_{a0}
\;,
\label{D-Hilbert}
\\*
\label{GGDab}
D_{ab} &:= a \frac{G_{ab}-G_{a0}}{b}\;.
\end{align}
\end{subequations}
The equation (\ref{D-Hilbert}) is now a standard \emph{singular
  integral equation of Carleman type} \cite{Carleman, Tricomi}. 
We cite from Tricomi's book \cite{Tricomi}:
\begin{Proposition}[{\cite[\S 4.4]{Tricomi}}]
\label{Prop:Carleman}
Let $h\in \mathcal{C}({]{-}1,1[})$ and $f\in L^p({]{-}1,1[})$ for some $p>1$.
Then the singular integral equation 
\begin{align}
h(x) \varphi(x)-\lambda\pi \mathcal{H}_a[\varphi(\bullet)]=f(x)\;,
\qquad x\in {]{-}1,1[}\;,
\label{Carleman}
\end{align}
has the solution
\begin{align}
\varphi(x)&= \frac{\sin (\vartheta(x))}{\lambda \pi} \Big(
f(x)  \cos(\vartheta(x))
+ e^{\mathcal{H}_x[\vartheta]}
\mathcal{H}_x\big[e^{-\mathcal{H}_\bullet[\vartheta]}
f(\bullet) \sin(\vartheta(\bullet)) \big]
+\frac{Ce^{\mathcal{H}_x[\vartheta]}}{1-x}\Big)\;,
\label{Solution:Carleman}
\\
\vartheta(x)&=\di{\raisebox{-1.2ex}{\mbox{\normalsize$\arctan$}}}{
\mbox{\scriptsize$[0,\pi]$}}
\Big(\frac{\lambda\pi}{h(x)}\Big)\;,~~
\sin
(\vartheta(x))=\frac{|\lambda\pi|}{\sqrt{(h(x))^2
+(\lambda\pi)^2}}\;,~~
\cos (\vartheta(x))=\frac{\sin(\vartheta(x))}{
\tan (\vartheta(x))}\;,
\nonumber
\end{align}
where the Hilbert transform integrates over ${]-1,1[}$, and $C$ is an
arbitrary constant.

The angle $\vartheta(x)$ obeys the identities \textup{\cite[\S
  4.4(28)]{Tricomi}} and \textup{\cite[\S 4.4(18)]{Tricomi}},
\begin{subequations}
\begin{align}
e^{-\mathcal{H}_x[\vartheta]}\cos(\vartheta(x))
+ \mathcal{H}_x\big[
e^{-\mathcal{H}_\bullet[\vartheta]}\sin(\vartheta(\bullet)\big]
&=1\;,  \label{Tricomi-28}
\\
e^{\mathcal{H}_x[\vartheta]}\cos (\vartheta(x)) 
-\mathcal{H}_x\big[e^{\mathcal{H}_\bullet[\vartheta]} \sin (\vartheta(\bullet))\big]
&= 1\;.
\label{Tricomi-18}
\end{align}
\end{subequations}
\end{Proposition}
The solution (\ref{Solution:Carleman}) and the identities 
(\ref{Tricomi-28}) and (\ref{Tricomi-18}) are 
unchanged if transformed via $p=\frac{\Lambda^2}{2}(1+x)$ to $p\in
[0,\Lambda^2]$ instead of $x\in[-1,1]$. We make the following decisive 
\begin{Assumption}
$C=0\;.$
\label{Assump-C=0}
\end{Assumption}
This assumption (or rather choice) will be discussed at the end of
Sec.~\ref{sec:Master} and in Sec.~\ref{sec:Conclusion}.  
Under this assumption the solution of (\ref{D-Hilbert}) is
\begin{subequations}
\begin{align}
D_{ab} &= - \frac{\sin (\vartheta_b(a))}{\lambda\pi}  \Big(
G_{a0}  \cos(\vartheta_b(a))
+ e^{\mathcal{H}_a^{\!\Lambda}[\vartheta_b]}
\mathcal{H}_a^{\!\Lambda}\big[
e^{-\mathcal{H}_\bullet^{\!\Lambda}[\vartheta_b]}
G_{\bullet 0} \sin(\vartheta_b(\bullet)) \big]
\Big)\;,
\label{solution-Dab}
\\
\vartheta_b(a) &=\di{\raisebox{-1.2ex}{\mbox{\normalsize$\arctan$}}}{
\mbox{\scriptsize$[0,\pi]$}}
\Big(\frac{ \lambda\pi a G_{a0} }{
1 + b G_{a0}+\lambda \pi a \mathcal{H}_a^{\!\Lambda}
\big[G_{\bullet 0} \big] }\Big)\;.
\label{arctan-0}
\end{align}
\end{subequations}%
The form (\ref{solution-Dab}) of the solution of (\ref{D-Hilbert}) is
not very useful. We can simplify it enormously noting that
(\ref{arctan-0}) is, for $b=0$, also a Carleman-type singular integral
equation
\begin{align}
\lambda\pi \cot \vartheta_{0}(a) G_{a0}
- \lambda\pi \mathcal{H}_a^{\!\Lambda}[G_{a0}]
=\frac{1}{a}\;.
\label{Galpha0-Carleman}
\end{align}
This equation has the following solution:
\begin{Lemma}
\label{Lemma:G-alpha0-vartheta}
$\displaystyle \qquad 
G_{a0} = 
e^{\mathcal{H}_a^{\!\Lambda}[\vartheta_0]-\mathcal{H}_0^{\!\Lambda}[\vartheta_0]}
\frac{\sin (\vartheta_0(a))}{|\lambda| \pi a} \;.
$
\end{Lemma}
\emph{Proof.} By (\ref{Solution:Carleman}) the solution of
(\ref{Galpha0-Carleman}) is for $C=0$
\begin{align*}
G_{a0} = 
\frac{\sin (\vartheta_0(a))}{\lambda \pi} \Big(
\frac{\cos(\vartheta_0(a))}{a}
+ e^{\mathcal{H}_a^{\!\Lambda}[\vartheta_0]}\mathcal{H}_a^{\!\Lambda}\Big[
\frac{e^{-\mathcal{H}_\bullet^{\!\Lambda}[\vartheta_0]}
\sin(\vartheta_0(\bullet))}{\bullet}
\Big]\Big)\;.
\end{align*}
Rational fraction expansion $\displaystyle
\mathcal{H}_a^{\!\Lambda}
\Big[\frac{f(\bullet)}{\bullet}\Big] = \frac{1}{a}\Big(
\mathcal{H}_a^{\!\Lambda} \big[f(\bullet)\big] 
- \mathcal{H}_0^{\!\Lambda}\big[f(\bullet)\big]\Big)$ and the 
identity (\ref{Tricomi-28}) yield
the assertion, where $\vartheta_0(0)=0$ for $\lambda>0$ and
$\vartheta_0(0)=\pi$ for $\lambda<0$, hence
$\cos \vartheta_0(0)=\mathrm{sign}(\lambda)$,  are used.  \hfill $\square$%

\bigskip

\noindent
For the Hilbert transform occurring in (\ref{solution-Dab}) and for
the investigation of the four-point function later on we need the
following addition theorem:
\begin{Lemma}
\label{Lemma:add-thm}
For all $0 \leq a,b,d \leq \Lambda^2$ one has 
\begin{align}
\lambda \pi a \sin\big( \vartheta_d(a)-\vartheta_b(a)\big)
=  (b-d) \sin \vartheta_b(a)\sin \vartheta_d(a)\;.
\label{add-thm}
\end{align}
\end{Lemma}
\emph{Proof.} This follows from insertion of (\ref{arctan-0}) into 
$\cot \vartheta_b(a)-\cot \vartheta_d(a)$.
\hfill $\square$%

\bigskip 

\noindent
This Lemma has several important consequences:
\begin{Corollary}
\label{Cor:varthetamonotone}
\begin{enumerate}

\item For $\lambda>0$, the function $b\mapsto
  \vartheta_b(a)$ is monotonously \underline{decreasing} for any fixed
  $a>0$. In particular, $\vartheta_b(a)<\vartheta_0(a)$
  for any $0<a,b \leq \Lambda^2$.

\item For $\lambda<0$,  the function $b\mapsto
  \vartheta_b(a)$ is monotonously \underline{increasing} for any fixed
  $a>0$. In particular, $\vartheta_b(a)>\vartheta_0(a)$
  for any $0<a,b \leq \Lambda^2$.
\end{enumerate}
\end{Corollary}
\emph{Proof.} 
By continuity we can choose $|b-d|$ small enough
so that $\tau_{abd}
:=\vartheta_d(a)-\vartheta_b(a) \in
{]-\frac{\pi}{2},\frac{\pi}{2}[}$. Then 
$\tau_{abd}$ and $\sin \tau_{abd}$ have the same sign.
Let $b>d$. Then from (\ref{add-thm}) we conclude 
$\vartheta_d > \vartheta_b$ for $\lambda>0$ and
$\vartheta_d < \vartheta_b$ for $\lambda<0$. \hfill $\square$%

\bigskip

The next step is the analogue of Lemma~\ref{Lemma:G-alpha0-vartheta} 
for the Hilbert transform occurring in (\ref{solution-Dab}):%
\begin{Lemma}
\label{Lemma:Gbullet0-Hilbert}
\begin{align*}
\mathcal{H}_a^{\!\Lambda}\big[
e^{-\mathcal{H}_\bullet^{\!\Lambda}[\vartheta_b]}
G_{\bullet 0} \sin(\vartheta_b(\bullet)) \big]
&= 
\frac{\mathrm{sign}(\lambda) }{b} 
e^{-\mathcal{H}_0^{\!\Lambda}[\vartheta_0]}
\Big(
e^{\mathcal{H}_a^{\!\Lambda}[\vartheta_0-\vartheta_b]}
\cos \big(\vartheta_0(a){-}\vartheta_b(a)\big)
- 1
\Big).
\end{align*}
\end{Lemma}
\emph{Proof.}
We insert $G_{\bullet 0}$ from Lemma~\ref{Lemma:G-alpha0-vartheta} into 
$\mathcal{H}_a^{\!\Lambda}\big[e^{-\mathcal{H}_\bullet^{\!\Lambda}[\vartheta_b]}
G_{\bullet 0} \sin(\vartheta_b(\bullet)) \big]$ and obtain with 
(\ref{add-thm}):
\begin{align*}
& \mathcal{H}_a^{\!\Lambda}
\big[e^{-\mathcal{H}_a^{\!\Lambda}[\vartheta_b]}
G_{\bullet 0} \sin(\vartheta_b(\bullet)) \big]
=  \frac{\mathrm{sign}(\lambda)}{b}
e^{-\mathcal{H}_0^{\!\Lambda} [\vartheta_0]} 
\mathcal{H}^{\!\Lambda}_a\Big[
e^{\mathcal{H}^{\!\Lambda}_\bullet[\vartheta_0-\vartheta_b]}
\sin\big( \vartheta_0(\bullet){-}\vartheta_b(\bullet)\big)
\Big]\;.
\end{align*}
Now the identity (\ref{Tricomi-18}) yields the assertion.
\hfill $\square$%
\\[\bigskipamount]
This allows us to prove the following remarkable formula for 
$G_{ab}$:
\begin{Theorem}
\label{Thm:Gab}
 The renormalised planar connected two-point function
  $G_{ab}$ of self-dual noncommutative $\phi^4_4$-theory
  (with continuous indices arising as limit $\theta \to \infty$) 
satisfies (under the assumption~\ref{Assump-C=0}) the equation
\begin{align}
G_{ab}
&= 
\frac{\sin(\vartheta_b(a))}{|\lambda| \pi a}
e^{\mathcal{H}_a^{\!\Lambda}[\vartheta_b]-\mathcal{H}_0^{\!\Lambda}[\vartheta_0]}
\;.
\label{Gab-vartheta}
\end{align}
In particular, $G_{ab}\geq 0$ for all $a,b
\in [0,\Lambda^2]$. 
\end{Theorem}
\emph{Proof.}
Insertion of (\ref{GGDab}) into (\ref{solution-Dab}) gives 
\begin{align*}
G_{ab}
&= G_{a0} 
- b \frac{\sin (\vartheta_b(a))}{\lambda \pi a} \Big(
G_{a0}  \cos(\vartheta_b(a))
+ e^{\mathcal{H}_a^{\!\Lambda}[\vartheta_b]}
\mathcal{H}_a^{\!\Lambda}\Big[e^{-\mathcal{H}_\bullet^{\!\Lambda}[\vartheta_b]}
G_{\bullet 0} \sin(\vartheta_b(\bullet)) \Big]\Big)\;.
\end{align*}
We insert Lemma \ref{Lemma:G-alpha0-vartheta} and Lemma
\ref{Lemma:Gbullet0-Hilbert}:
\begin{align*}
G_{ab}
&= e^{-\mathcal{H}_0^{\!\Lambda}[\vartheta_0]}
\frac{\sin (\vartheta_0(a))}{|\lambda| \pi a} 
\bigg\{
e^{\mathcal{H}_a^{\!\Lambda}[\vartheta_0]}
- b 
\frac{\sin (\vartheta_b(a))}{\lambda \pi a} 
\Big\{
e^{\mathcal{H}_a^{\!\Lambda}[\vartheta_0]}
\cos(\vartheta_b(a))
\nonumber
\\*
& + e^{\mathcal{H}_a^{\!\Lambda}[\vartheta_b]}
\frac{\lambda \pi a}{b \sin (\vartheta_0(a))} 
\Big(
e^{\mathcal{H}_a^{\!\Lambda}[\vartheta_0-\vartheta_b]}
\cos \big(\vartheta_0(a){-}\vartheta_b(a)\big) -1 \Big)
\Big\}\bigg\}\;.
\end{align*}
We express $\cos \big(\vartheta_0(a){-}\vartheta_b(a)\big)$ by
its addition theorem and combine the $\sin$-$\sin$ part with 
$e^{\mathcal{H}_a^{\!\Lambda}[\vartheta_0]}$ after the first ``$\{$'':
\begin{align*}
G_{ab}
&= 
 e^{-\mathcal{H}_0^{\!\Lambda}[\vartheta_0]}
\frac{\sin (\vartheta_0(a))}{|\lambda| \pi a} 
\bigg\{
e^{\mathcal{H}_a^{\!\Lambda}[\vartheta_0]} \cos^2 (\vartheta_b(a))
+ e^{\mathcal{H}_a^{\!\Lambda}[\vartheta_b]}
\frac{\sin (\vartheta_b(a))}{
\sin (\vartheta_0(a))} 
\nonumber
\\
& 
-
e^{\mathcal{H}_a^{\!\Lambda}[\vartheta_0]}
\sin (\vartheta_b(a))\cos(\vartheta_b(a))
\Big(
\frac{b}{\lambda \pi a}  +\cot \vartheta_0(a) \Big)\bigg\}\;.
\end{align*}
Now (\ref{arctan-0}) implies 
$\frac{b}{\lambda \pi a}  +\cot \vartheta_0(a) 
=\cot \vartheta_b(a)$, and then the above
equation collapses to (\ref{Gab-vartheta}).
\hfill $\square$%

\begin{Lemma}
One has 
\begin{equation}
\mathcal{Y}
= \lambda \int_0^{\Lambda^2} dp\; \frac{(G_{p0})^2}{
\big(\lambda \pi p G_{p0}\big)^2 + \big(1
+ \lambda \pi p 
  \mathcal{H}_p^{\!\Lambda}[G_{\bullet 0}]\big)^2}\;.
\label{Y-int}
\end{equation}
\label{Lemma:Y-int}
\end{Lemma}
\emph{Proof.}
To compute 
$\mathcal{Y}= \displaystyle -\lambda
\int_0^{\Lambda^2}\frac{dp}{p} \;D_{p0}=-\lambda\pi
\mathcal{H}_0^{\!\Lambda}[D_{\bullet 0}]$ in (\ref{ZZ2})
we use the \emph{convolution
  theorem} for the finite Hilbert transform \cite[\S 4.3(4)]{Tricomi}
\begin{align}
\mathcal{H}_x\Big[ 
\phi_1(\bullet) \mathcal{H}_\bullet[\phi_2]
+\phi_2(\bullet) \mathcal{H}_\bullet[\phi_1]\Big]
= \mathcal{H}_x[\phi_1(\bullet)] \mathcal{H}_x[\phi_2(\bullet)]
-\phi_2(x) \phi_1(x)\;.
\end{align}
With $D_{p0}$ given by (\ref{solution-Dab}), we  
have to identify $x=0$, $\phi_1(a)= 
e^{\mathcal{H}_a^{\!\Lambda}[\vartheta_0]} \sin  \vartheta_0(a)$
and 
$\phi_2(a)=e^{-\mathcal{H}_a^{\!\Lambda}[\vartheta_0]}
G_{a0} \sin(\vartheta_0(a))$. With 
$\sin \vartheta_0(0)=0$ we obtain
\begin{align*}
\mathcal{Y}
&= 
\mathcal{H}_0^{\!\Lambda}\Big[ 
\sin (\vartheta_0(\bullet))
G_{\bullet 0}  \cos(\vartheta_0(\bullet))\Big]
- \mathcal{H}_0^{\!\Lambda}\Big[ 
\mathcal{H}_\bullet^{\!\Lambda}\Big[\sin (\vartheta_0)
e^{\mathcal{H}_\bullet^{\!\Lambda}[\vartheta_0]}\Big]
e^{-\mathcal{H}_\bullet^{\!\Lambda}[\vartheta_0]} G_{\bullet 0} 
\sin(\vartheta_0(\bullet)) \Big]
\nonumber
\\*
& + \mathcal{H}_0^{\!\Lambda}\Big[\sin (\vartheta_0(\bullet))
e^{\mathcal{H}_\bullet^{\!\Lambda}[\vartheta_0]}\Big]
\mathcal{H}_0^{\!\Lambda}\Big[e^{-\mathcal{H}_\bullet^{\!\Lambda}[\vartheta_0]} 
G_{\bullet 0} \sin(\vartheta_0(\bullet)) \Big]\;.
\end{align*}
Using (\ref{Tricomi-18}) this equation collapses with $\cos \vartheta_b(0)
=\mathrm{sign}(\lambda)$ to 
\begin{align*}
\mathcal{Y}
&= 
\mathrm{sign}(\lambda)\,  \mathcal{H}_0^{\!\Lambda}\Big[
e^{-\mathcal{H}_\bullet^{\!\Lambda}[\vartheta_0]+
\mathcal{H}_0[\vartheta_0]} G_{\bullet 0} 
\sin(\vartheta_0(\bullet)) \Big]
\\
&= \frac{1}{\lambda\pi^2} 
\int_0^{\Lambda^2} dp\; \frac{\sin^2(\vartheta_0(p))}{p^2}
= \lambda \int_0^{\Lambda^2} \frac{d p}{
(\lambda \pi p)^2 \big(1+\cot^2 (\vartheta_0(\rho)\big)}\;,
\end{align*}
where the second line follows from  Lemma \ref{Lemma:G-alpha0-vartheta}.
Now the assertion follows with (\ref{arctan-0}).
\hspace*{\fill}$\square$%

\bigskip

It follows that $\mathcal{Y}>0$ for $\lambda>0$ and $-1<\mathcal{Y}<
0$ for $\lambda<0$ with $|\lambda|$ small enough. Now observe that
(\ref{ZZ2}), Lemma~\ref{Lemma:G-alpha0-vartheta} and (\ref{Tricomi-18})
give
\begin{align}
Z^{-1}
&=(1+\mathcal{Y})\Big( 1- \mathrm{sign}(\lambda)
\frac{e^{-\mathcal{H}_0^{\!\Lambda}[\vartheta_0]}}{\pi}\int_0^{\Lambda^2} \frac{dp}{p}
\;\sin \vartheta_0(p)e^{\mathcal{H}_p^{\!\Lambda}[\vartheta_0]}\Big)
\nonumber
\\
&=(1+\mathcal{Y})\Big( 1- \mathrm{sign}(\lambda)
e^{-\mathcal{H}_0^{\!\Lambda}[\vartheta_0]} \mathcal{H}_0^{\!\Lambda}[
\sin \vartheta_0(\bullet)e^{\mathcal{H}_\bullet^{\!\Lambda}[\vartheta_0]}]\Big)
= (1+\mathcal{Y}) \mathrm{sign}(\lambda)
e^{-\mathcal{H}_0^{\!\Lambda}[\vartheta_0]}\;.
\label{eq:Z}
\end{align}
It therefore follows that
$\mathrm{sign}(Z^{-1})=\mathrm{sign}(\lambda)$ so that the case
$\lambda<0$ leads to an unstable action functional.
\emph{We therefore discard the case $\lambda<0$ in the sequel}.

\subsection{The master equation}

\label{sec:Master}

The identity (\ref{Gab-vartheta}) allows us to compute $G_{ab}$ once
$G_{a0}$ is known, but (\ref{Gab-vartheta}) \emph{alone is not
  sufficient} to determine $G_{a0}$. This is because we ignored so far
the equation (\ref{ZG-inteq:b=0}) which in terms of (\ref{def:Gab})
and after insertion of (\ref{ZZ2}) reads
\begin{align}               
a - \frac{1}{G_{a0}} + 1
&= -\lambda (1+\mathcal{Y})Z 
\int_0^{\Lambda^2} \! p\,dp \;( G_{ap}-G_{0 p})
-\lambda \int_0^{\Lambda^2} dp \; 
\frac{a-a\frac{G_{p0}}{G_{a0}}}{(p-a)}\;.
\end{align}
We introduce the Hilbert transform and 
insert  (\ref{Gab-vartheta}) and
(\ref{eq:Z}):
\begin{align}               
1+ a -\lambda\pi a \cot \vartheta_0(a) 
&={-} 
\int_0^{\Lambda^2} \!\!\! p\,dp \,\Big( 
\frac{\sin \vartheta_p(a)}{\pi a} e^{\mathcal{H}_a^{\!\Lambda}[\vartheta_p]}
-\lim_{a\to 0} 
\frac{\sin \vartheta_p(a)}{\pi a} e^{\mathcal{H}_0^{\!\Lambda}[\vartheta_p]}
\Big)
- \lambda\pi a \mathcal{H}_a^{\!\Lambda}[1]\;.
\end{align}
We write $\sin \vartheta_p(a)=(1+\cot^2 \vartheta_p(a))^{-\frac{1}{2}}
= \big(1+(\frac{p}{\lambda \pi a}+\cot
\vartheta_0(a))^2\big)^{-\frac{1}{2}}$ and use the following limit for $\lambda>0$:
\begin{align}
 \lim_{a\to 0}
\Big(\frac{\sin(\vartheta_p(a))}{ \pi a}\Big) 
= \cos \vartheta_{p}(0)
\lim_{a\to 0} \frac{\vartheta_p(a)}{a}
=\frac{\lambda}{(1+p)} \;.
\label{lim-sinvartheta}
\end{align}
This gives:
\begin{Proposition}
The function 
$\mathcal{T}_a:= 
\lambda \pi a \cot \vartheta_0(a)$, with $\mathcal{T}_0=1$,
is determined by
\begin{align}               
\mathcal{T}_a
=1+a+ \lambda
\lim_{\epsilon\to 0} \Big(\int_0^{a-\epsilon}\!\! +
\int_{a+\epsilon}^{\Lambda^2} \Big)  dp \; \bigg( 
\frac{a}{p-a}
&+\frac{p\, \exp\Big(\mathcal{H}_a^{\!\Lambda}\Big[\arctan
  \frac{\lambda \pi \bullet}{p+\mathcal{T}_\bullet} \Big]\Big)}{
\sqrt{(\lambda\pi a)^2+ (p +\mathcal{T}_a)^2}}
\nonumber
\\
&
-\frac{p\, \exp\Big(\mathcal{H}_0^{\!\Lambda}\Big[\arctan
  \frac{\lambda \pi \bullet}{p+\mathcal{T}_\bullet} \Big]\Big)}{1+p}
\bigg)\;.
\label{master-Ta}
\\*[-2ex]
\tag*{\mbox{$\square$}}
\end{align}
\end{Proposition}
This is a non-linear self-consistency equation whose solution
yields the input data $\vartheta_0(a)$ and $\vartheta_b(a)$ to evaluate the
two-point function $G_{ab}$ by Theorem~\ref{Thm:Gab}.

Unfortunately, (\ref{master-Ta}) is too complicated for further
analysis. We therefore derive an alternative equation from the
symmetry $G_{b0}=\lim_{a\to 0} G_{ab}$.  This requirement together
with Theorem~\ref{Thm:Gab} gives:
\begin{Proposition}
\label{Prop:G-beta0}
The renormalised planar regular two-point function $G_{b 0}$ is
determined by the self-consistency equation
\begin{align}
G_{b0} 
= \frac{1}{1+b}
\exp\Bigg(- \lambda
\int_0^b \!\! dt \int_0^{\Lambda^2} \!\!\! dp\;
\frac{(G_{p0})^2}{\big(\lambda \pi p G_{p0}\big)^2 
+\big( 1+ tG_{p0} +\lambda\pi p
 \mathcal{H}^{\!\Lambda}_p[G_{\bullet 0}]\big)^2} 
\Bigg) \;,
\label{G-beta0}
\end{align}
which is well-defined in the class of continuously differentiable
functions on $[0,\Lambda^2]$.
\end{Proposition}
\emph{Proof.} The limit $a\to 0$ in (\ref{Gab-vartheta}) 
yields with (\ref{lim-sinvartheta})
\begin{align}
G_{b0}\equiv \lim_{a\to 0} G_{ab} 
&= \frac{1}{1+b}
e^{-\mathcal{H}_0^{\!\Lambda}[\vartheta_0-\vartheta_b]}
=\frac{1}{1+b} 
\exp\Big(-\frac{1}{\pi} \int_0^{\Lambda^2} \frac{dp}{p} 
\big(\vartheta_0(p)-\vartheta_b(p)\big)\Big)\;.
\label{G-beta0-vartheta}
\end{align}
Together with Corollary \ref{Cor:varthetamonotone} this formula already
establishes that for $\lambda>0$ the function 
$b \mapsto G_{b0}$ is monotonously decreasing and maps $[0,\Lambda^2]$ into
$[0,1]$. 
The explicit formula (\ref{G-beta0}) follows from
\begin{align*}
-\frac{1}{\pi} \int_0^{\Lambda^2} \frac{dp}{p} 
\big(\vartheta_0(p)-\vartheta_b(p)\big)
= \frac{1}{\pi} \int_0^{\Lambda^2} \frac{dp}{p} 
\int_0^b dt \; \frac{d \vartheta_t(p)}{dt} 
\end{align*}
and (\ref{arctan-0}) after exchange of the $p,t$-integrals.

The equation (\ref{G-beta0}) is meaningful on the class of
continuously differentiable functions. Namely, if $G_{\bullet 0}\in
\mathcal{C}^1[0,\Lambda^2]$, then  
\begin{align*}
\lambda\pi p \mathcal{H}_p^{\!\Lambda}[G_{\bullet 0}]
&=\lambda p
\lim_{\epsilon\to 0}\Big(\int_0^{p-\epsilon}\!\!\! +\int_{p+\epsilon}^{\Lambda^2}\Big)
dq \; G_{q0} \frac{d}{dq} \log |q-p|
\nonumber
\\*
&=\lambda p \Big(
G_{\Lambda^2 0} \log(\Lambda^2-p) -\log p 
- \int_0^{\Lambda^2} dq \; G'_{q0} \log |q-p|\Big)\;.
\end{align*}
This shows that $p\mapsto \lambda\pi p
\mathcal{H}_p^{\!\Lambda}[G_{\bullet 0}]$ is a continuous function on
${[0,\Lambda^2[}$. Since $\lim_{p\to \Lambda^2} \lambda \pi p
\mathcal{H}_p^{\!\Lambda}[G_{\bullet 0}] =-\infty$ and $\lambda\pi p
\mathcal{H}_p^{\!\Lambda}[G_{\bullet 0}]$ appears in the denominator
together with bounded functions, the integrand in (\ref{G-beta0})
extends to a continuous function on $[0,b]\times[0,\Lambda^2]\ni
(t,p)$.  By the fundamental theorem of calculus we have
\begin{align}
\frac{d G_{b0}}{db}
=-G_{b0}\bigg(\frac{1}{1+b}
+ \lambda
\int_0^{\Lambda^2} \!\!\! dp\;
\frac{(G_{p0})^2}{\big(\lambda\pi p G_{p0}\big)^2 
  +\big( 1+ b G_{p0} + \lambda \pi p
 \mathcal{H}_p^{\!\Lambda}[G_{\bullet 0}]\big)^2} \bigg)\;.
\label{G-beta0-deriv}
\end{align}
Since the rhs is a smooth function of $b \in {[0,\Lambda^2]}$,
continuous derivatives of $G_{b0}$ exists inductively to any order.
\hfill $\square$%
\\[\bigskipamount]
Observe from (\ref{G-beta0-deriv}) and (\ref{Y-int}) that 
$\frac{d G_{b0}}{db}\big|_{b=0}=-(1+\mathcal{Y})$. 

The next task is to prove that the master equation
(\ref{G-beta0}) has a solution for any $\lambda>0$ and $\Lambda^2$
sufficently large, including the continuum limit $\Lambda \to
\infty$. This will be achieved by the Schauder fixed point theorem:
\begin{Theorem}[Schauder]
\label{Thm:Schauder}
Let $X$ be a Banach space, $K\subset X$ a convex subset  and $T$ a
continuous mapping of $K$ into itself. If $T(K)$ is contained in a
compact subset of $K$, then $T$ has a fixed point.
\end{Theorem}
The main road to establish compactness for $X$ being the space of
continuous functions on a compact Hausdorff space is the
Arzel\`a-Ascoli theorem.  There exist generalisations to
differentiable functions and to locally compact Hausdorff spaces. The
setup relevant for our situation is contained in
\cite{Cianciaruso:2011??} that we specify to our case.  Let
\begin{align*}
\mathcal{C}_0(\mathbb{R}_+)&:=\{
f:\mathbb{R}_+\to \mathbb{R}  \text{ continuous and
  bounded}\;,~
\lim_{t\to \infty}f(t)=0\}
\\
\mathcal{C}_0^1(\mathbb{R}_+)&:=\{
f:\mathbb{R}_+\to \mathbb{R}\;:~ f,f' \text{ continuous and
  bounded}\;,~
\lim_{t\to \infty}f(t)=\lim_{t\to \infty}f'(t)=0\}
\end{align*}
be the spaces of continuous and continuously differentiable functions on 
$\mathbb{R}_+$, respectively, that vanish 
(together with their derivative) at
$\infty$. These are Banach spaces with norms 
$\|f\|_{\infty}=\sup_{t\in \mathbb{R}_+} |f(t)|$ and  
$\|f\|_{\infty}'=\sup_{t\in \mathbb{R}_+} |f(t)|+\sup_{t\in \mathbb{R}_+}
|f'(t)|$, respectively. One has
\begin{Proposition}[{\cite[Lemme 1]{Avramescu:1969??}}]
\label{Prop:Avramescu}
A subset $K\subset \mathcal{C}_0(\mathbb{R}_+)$ is relatively compact
if and only if it is uniformly bounded, equicontinuous on
$\mathbb{R}_+$ and uniformly convergent at $\infty$. 
\end{Proposition}
We recall that a subset $K\subset X$ is relatively compact in the
topology induced from $X$ if its closure $\bar{K}$ is compact. A
subset $K$ of continuous functions on a metric space $(M,d)$ is
equicontinuous if for every $\epsilon>0$ there is a uniform $\delta>0$
such that $|f(t)-f(s)|<\epsilon$ for all $f\in K$ and all $s,t\in M$
with $d(s,t)<\delta$. For subsets of $\mathcal{C}_0^1(\mathbb{R}_+)$
one has:
\begin{Proposition}[{\cite[Thm.~3.1]{Cianciaruso:2011??}}]
\label{Prop:Cianciaruso} 
A subset $K\subset \mathcal{C}^1_0(\mathbb{R}_+)$ is relatively compact
if and only if 
\begin{enumerate}
\item $K':=\{ f'\;:~f\in K\}$ is relatively compact in 
$\mathcal{C}_0(\mathbb{R}_+)$,

\item For every $\epsilon>0$ there is a $L>0$ with 
$\big\|f\big|_{[L,\infty[}\big\|_\infty < \epsilon$ for all $f\in K$.
\end{enumerate}
\end{Proposition}

The following operator $T_\lambda$ is well-defined on the subset of
positive functions in 
$\mathcal{C}^1_0(\mathbb{R}_+) \cap
L^q(\mathbb{R}_+)$, for any $1\leq q<\infty$:
\begin{align}
(T_\lambda f)(b) := 
\frac{1}{1+b}
\exp\Bigg(- \lambda
\int_0^b \!\! dt \int_0^\infty \!\!\! 
\frac{dp}{\big(\lambda\pi p\big)^2 
+\big( t + \frac{1+ \lambda\pi p
 {\mathcal{H}_p}^{\!\!\!\!\infty}[ f(\bullet)]}{f(p)}\big)^2} 
\Bigg) \;.
\label{Tf}
\end{align}
The Hilbert transform maps differentiable functions to continuous
functions on $]0,\infty[$, with logarithmic divergence at $p=0$. To
see this we write for $p>0$ 
\begin{align}
{\mathcal{H}_p}^{\!\!\!\!\infty}[f(\bullet)]
= \frac{1}{\pi}\int_0^{2p} dq \;\frac{f(q)-f(p)}{q-p}
+ \frac{1}{\pi}\int_{2p}^\infty dq \;\frac{f(q)}{q-p}\;,
\label{Hilbert-infty}
\end{align}
using the fact that $\mathcal{H}_p^{\sqrt{2p}}[1]=0$. The second
integral is bounded by H\"older's inequality, and its derivative with
respect to $p$ exists. In the first integral we have
\begin{align*}
& \int_0^{2(p+\delta)} dq \;\frac{f(q)-f(p+\delta)}{q-(p+\delta)}
- \int_0^{2p} dq \;\frac{f(q)-f(p)}{q-p}
\\*
&=  \int_{2p}^{2(p+\delta)} dq \;\frac{f(q)-f(p+\delta)}{q-(p+\delta)}
+ \int_0^{2p} dq
\;\Big(\frac{\delta}{q-(p+\delta)} \frac{f(q)-f(p)}{q-p}
+\frac{f(p+\delta)-f(p)}{q-(p+\delta) }\Big)\;.
\end{align*}
Using the mean value theorem $f(p)-f(q)=(p-q) f'(\xi)$ for
some $\xi$ between $p$ and $q$, this expression tends to zero for
$\delta\to 0$. Since $\lim_{p\to 0} p{\mathcal{H}_p}^{\!\!\!\!\infty}[
f(\bullet)]=0$, the integrand in (\ref{Tf}) is a rational function of
positive continuous functions on $\mathbb{R}_+$ with denominator
separated from zero. Therefore the $p$-integral exists and is
convergent for large $p$. The fundamental theorem of calculus
guarantees that $(T_\lambda f)$ is differentiable. 

The following even stronger result holds:
\begin{Lemma}
\label{Lemma:Tf-deriv-boundend}
Let $\lambda>0$ and assume that 
$f\in \mathcal{C}^1_0(\mathbb{R}_+)$ satisfies
\begin{itemize}
\item[\textup{(A1)}]  $f(0) =1$,

\item[\textup{(A2)}]  $0 < f(b) \leq\frac{1}{1+b}$, 

\item[\textup{(A3)}] 
$ 0\leq -\big(\frac{1}{b+1}+\frac{f'(b)}{f(b)}\big)\leq C_\lambda$, 

\end{itemize}
for some $C_\lambda\geq 0$. Let
$P_\lambda >0$ be the unique solution of 
$2\lambda P_\lambda^2 (1+C_\lambda) e^{C_\lambda P_\lambda}=1$.  
Then 
\begin{enumerate}
\item[\textup{(C1)}] $(T_\lambda f)(0)=1$,

\item[\textup{(C2)}] $0 < (T_\lambda f)(b)\leq \frac{1}{1+b}$,
\item[\textup{(C3)}]
$ 0\leq -\big(\frac{1}{b+1}+\frac{(T_\lambda f)'(b)}{(T_\lambda
  f)(b)}\big) \leq \frac{1}{2} + \frac{1}{\lambda\pi^2 P_\lambda}$, 

\item[\textup{(C4)}] $\big|\frac{(T_\lambda f)''(b)}{(T_\lambda f)(b)}\big| \leq 
\frac{23}{4}+\frac{2}{\pi}+\frac{7+8\pi}{2}
\frac{1}{(\lambda\pi^2 P_\lambda)^2}$.
\end{enumerate}
\end{Lemma}
\emph{Proof.}
(C1) and (C2) are obvious from (\ref{Tf}). 
Integrating (A3) with initial condition (A1) one has
$\frac{e^{-C_\lambda b} }{1+b} \leq f(b) \leq\frac{1}{1+b}$ which is
even stronger than (A2). In analogy to (\ref{G-beta0-deriv}) we have
\begin{align}
 -\Big(\frac{(T_\lambda f)'(b)}{(Tf)(b)} + \frac{1}{1+b}\Big)
= \lambda
\int_0^\infty 
\frac{dp}{\big(\lambda\pi p \big)^2 
  +\big( b+ \frac{1+ \lambda\pi p
 {\mathcal{H}_p}^{\!\!\!\!\infty}[f(\bullet)]}{f(p)}\big)^2} \;.
\label{Tf-deriv}
\end{align}
The rhs of (\ref{Tf-deriv}) is positive, and it remains to show
the bound (C3). Since $f>0$ and $f'\leq 0$ we have the 
following inequalities for the integrals in (\ref{Hilbert-infty}):
\begin{align*}
0 &\leq \int_{2p}^\infty dq \;\frac{f(q)}{q-p} \leq \frac{1}{p+1} 
\log \frac{1+2p}{p}\;,\qquad 
\\
0 &\geq \int_0^{2p} dq \;\frac{f(q)-f(p)}{q-p} \geq 2p 
\inf_{q\in [0,2p]} f'(q)  \geq -2p(C_\lambda+1)\;.
\end{align*}
Collecting these results and using $\frac{e^{-C_\lambda b} }{1+b} \leq
f(b) \leq\frac{1}{1+b}$ we conclude
\begin{subequations}
\begin{align}
\frac{1+ \lambda\pi p
 {\mathcal{H}_p}^{\!\!\!\!\infty}[f(\bullet)]}{f(p)}
\geq (1+p)
\Big(1- 2\lambda p^2 (1+C_\lambda) e^{C_\lambda  p}\Big)\;.
\end{align}
For fixed $C_\lambda \geq 0$ the continuous function $\mathbb{R}_+\ni
p \mapsto 2\lambda p^2(1+C_\lambda) e^{C_\lambda p}$
is monotonously increasing and maps $\mathbb{R}_+$ to itself.  Let
$P_\lambda$ be the unique solution of $2\lambda P_\lambda^2
(1+C_\lambda) e^{C_\lambda P_\lambda}=1$. By the implicit
function theorem, this gives rise to a globally defined differentiable
function $\mathbb{R}_+\ni C_\lambda \mapsto P_\lambda$ which is
monotonously decreasing. These monotonicity properties lead to the
inequality
\begin{align}
0<p<P_\lambda\quad\Rightarrow \quad 
\frac{1+ \lambda\pi p
 {\mathcal{H}_p}^{\!\!\!\!\infty}[f(\bullet)]}{f(p)}
\geq 1-\frac{p^2}{P_\lambda^2} \geq 0\;.
\label{estim-Plambda}
\end{align}
\end{subequations}

We now split the integral (\ref{Tf-deriv}) at $P_\lambda$ where for 
$0\leq p\leq P_\lambda$ we use (\ref{estim-Plambda}) whereas for 
$ p \geq P_{\lambda}$ we simply use
$\big( b+ \frac{1+ \lambda\pi p
 {\mathcal{H}_p}^{\!\!\!\!\infty}[f(\bullet)]}{f(p)}\big)^2\geq
0$:
\begin{align}
0 \leq  -\Big(\frac{(T_\lambda f)'(b)}{(Tf)(b)} &+ \frac{1}{1+b}\Big)
\leq 
\lambda
\int_0^{P_\lambda} 
\frac{dp}{\big(\lambda\pi p \big)^2 
  +
\big(1-\frac{p^2}{P_\lambda^2}\big)^2}
+ \lambda
\int_0^{P_\lambda} 
\frac{dp}{\big(\lambda\pi p \big)^2 }
\nonumber
\\
& =  
\lambda P_\lambda
\int_0^1 \frac{dx}{(\lambda\pi P_\lambda)^2 
x^2 +(1-x^2)^2} + \frac{1}{\lambda\pi^2 P_\lambda} 
\nonumber
\\
&\leq 
\lambda P_\lambda
\int_0^1 \frac{dx\;(1+x^2)}{(\lambda\pi P_\lambda)^2 
x^2 +(1-x^2)^2} + \frac{1}{\lambda\pi^2 P_\lambda} 
= \frac{1}{2}+ \frac{1}{\lambda\pi^2 P_\lambda}\;.
\end{align}
We have used the integral $\displaystyle \int_0^1 dx
\;\frac{1+x^2}{a^2 x^2+(1-x^2)^2}= \frac{1}{2} \int_0^\infty dt
\;\frac{\cosh \frac{t}{2}}{(\frac{a^2}{2}-1)+\cosh t}= \frac{\pi}{2a}$
which can be looked up in \cite[\S 3.517.1,\S
8.702]{Gradsteyn:1994??}. This finishes the proof of (C3).

\medskip

Another derivative of (\ref{Tf-deriv}) gives
\begin{align}
(T_\lambda f)''(b)
=(Tf)(b) \bigg\{&
\bigg(\frac{1}{1+b}
+ \lambda
\int_0^\infty \!\!\! 
\frac{dp}{\big(\lambda\pi p\big)^2 
  +\big( b+ \frac{1+  \lambda\pi p
{\mathcal{H}_p}^{\!\!\!\!\infty}[f(\bullet)]}{f(p)}\big)^2} \bigg)^2 
\nonumber
\\*
& + \frac{1}{1+b}
+ \lambda
\int_0^\infty \!\!\! dp\;
\frac{2 \big( b+ \frac{1+  \lambda\pi p
{\mathcal{H}_p}^{\!\!\!\!\infty}[f(\bullet)]}{f(p)}\big)}{
\Big(\big(\lambda\pi p\big)^2 
  +\big( b+ \frac{1+  \lambda\pi p
{\mathcal{H}_p}^{\!\!\!\!\infty}[f(\bullet)]}{f(p)}\big)^2\Big)^2} 
\bigg\}
\;.
\label{Tf-deriv-2}
\end{align}
We thus get with (C3) the estimate 
\begin{align*}
\frac{\big|(T_\lambda f)''(b)\big|}{(T_\lambda f)(b)}
& \leq \Big(\frac{3}{2}+\frac{1}{\lambda\pi^2 P_\lambda}\Big)^2+1
+ 2 \lambda
\int_0^\infty \!\!\! 
\frac{dp}{
\Big(\big(\lambda\pi p\big)^2 
  +\big( b+ \frac{1+  \lambda\pi p
{\mathcal{H}_p}^{\!\!\!\!\infty}[f(\bullet)]}{f(p)}\big)^2\Big)^{\frac{3}{2}}} 
\\
& \leq 
\Big(\frac{3}{2}+\frac{1}{\lambda \pi^2 P_\lambda}\Big)^2+1
+4\pi \Big(\frac{1}{\lambda\pi^2 P_\lambda}\Big)^2+
2\lambda P_\lambda
\int_0^1 \!\!\! 
\frac{dx}{
\big((\lambda\pi P_\lambda)^2 
x^2  +(1-x^2)^2 \big)^{\frac{3}{2}}} \;.
\end{align*}
We have with \cite[\S 3.517.1, \S 8.702, \S 9.131.1, 
\S 9.122.1]{Gradsteyn:1994??}
\begin{align*}
\int_0^1 \!\!\! 
\frac{(1+x^4)\;dx}{
\big(a^2 x^2  +(1-x^2)^2 \big)^{\frac{3}{2}}}
&= \frac{1}{\sqrt{8}}
\int_0^\infty dt\; \frac{\cosh t}{((\frac{a}{2}-1)+\cosh
  t)^{\frac{3}{2}}}
\\
&=
\frac{3\pi}{4a^2}\;{}_2F_1\Big(\di{-\frac{1}{2},
\frac{3}{2}}{2}\Big|1-\frac{a^2}{4}\Big)
=
\frac{3\pi}{8a}\;{}_2F_1\Big(\di{-\frac{1}{2},
\frac{1}{2}}{2}\Big|1-\frac{4}{a^2}\Big)
\leq \frac{1+a}{a^2}\;.
\end{align*}
Together with $2\alpha \leq 1+\alpha^2$ 
we confirm (C4). \hfill $\square$%

\bigskip

It is of fundamental importance that we can always choose $C_\lambda$
such that $\frac{1}{2} + \frac{1}{\lambda\pi^2 P_\lambda} <
C_\lambda$. Namely,
\begin{subequations}
\label{select-Plambda}
\begin{align}
\frac{1}{2} + \frac{1}{\lambda\pi^2 P_\lambda} < C_\lambda
\qquad& \Leftrightarrow\qquad 
2\lambda P_\lambda^2
\Big(\frac{3}{2} + \frac{1}{\lambda\pi^2 P_\lambda}\Big) 
e^{(\frac{1}{2} + \frac{1}{\lambda\pi^2 P_\lambda})P_\lambda}
< 2\lambda P_\lambda^2
(1+C_\lambda)e^{C_\lambda P_\lambda}
:= 1
\nonumber
\\
\qquad &\Leftrightarrow\qquad 
P_\lambda e^{\frac{1}{\lambda\pi^2}} \cdot 
\Big(3\lambda P_\lambda + \frac{2}{\pi^2}\Big) 
e^{\frac{P_\lambda}{2}} < 1 \;.
\end{align}
The condition is satisfied for
\begin{align}
P_\lambda = \frac{e^{-\frac{1}{\lambda\pi^2}}}{\sqrt{1+4 \lambda}}
\qquad \Rightarrow\qquad 
P_\lambda e^{\frac{1}{\lambda\pi^2}} \cdot 
\Big(3\lambda P_\lambda + \frac{2}{\pi^2}\Big) 
e^{\frac{P_\lambda}{2}} < 0.8303  <1\;.
\end{align}
\end{subequations}
Note that the inverse solution for $C_\lambda$ 
is huge for $0< \lambda \ll 1$.

We thus consider the following subset of $\mathcal{C}^1_0(\mathbb{R}_+)$:
\begin{align}
\mathcal{K}_\lambda:= 
\Big\{ f\in \mathcal{C}^1_0(\mathbb{R}_+)\;:~
& f(0)=1\;,\quad 
0 <  f(b)\leq \frac{1}{1+b}\;,\quad
0\leq -f'(b) \leq \big( \tfrac{1}{1+b}+C_\lambda\big) f(b) \Big\}
\end{align}
with $C_\lambda$ the solution of $2\lambda P_\lambda^2
(1+C_\lambda)e^{C_\lambda P_\lambda}
= 1$ at $P_\lambda = \frac{\exp(-\frac{1}{\lambda\pi^2})}{\sqrt{1+4\lambda}}$.
We have established the following facts:
\begin{itemize}
\item[(K1)] $T_\lambda$ maps $\mathcal{K}_\lambda$ into
  $\mathcal{K}_\lambda$ by
  Lemma~\ref{Lemma:Tf-deriv-boundend}.(C1)+(C2)+(C3) and
  (\ref{select-Plambda}).

\item[(K2)] $\big(T_\lambda(\mathcal{K}_\lambda)\big)' \subset 
\mathcal{C}_0(\mathbb{R}_+)$ is relatively compact. 
Namely, $\big(T_\lambda(\mathcal{K}_\lambda)\big)' \subset 
\mathcal{C}_0(\mathbb{R}_+)$ is equicontinuous by
Lemma~\ref{Lemma:Tf-deriv-boundend}.(C4), uniformly bounded as subset of 
$\mathcal{K}_\lambda$ and uniformly convergent by
Lemma~\ref{Lemma:Tf-deriv-boundend}.(C3)+(C2), thus 
relatively compact by Proposition~\ref{Prop:Avramescu}. 

\item[(K3)] $\big\|g\big|_{[\frac{1}{\epsilon},\infty[}\big\|_\infty <
  \epsilon$ for all $g\in T_\lambda(\mathcal{K}_\lambda)$ 
by Lemma~\ref{Lemma:Tf-deriv-boundend}.(C2). 

\item[(K4)] For the closure we have
  $\overline{T_\lambda(\mathcal{K}_\lambda)} \subset
  \mathcal{K}_\lambda$: Functions $g=\lim_{k\to \infty} g_k$ with
  $g_k\in T_\lambda(\mathcal{K}_\lambda)$ on the closure can at most
  exceed $\mathcal{K}_\lambda$ in $g(b)=0$ for some $b$. But $0\leq
  -g'(b)\leq (\frac{1}{1+b}+C_\lambda) g(b)$ and $g(0)=1$ imply
  $\frac{e^{-C_\lambda b}}{1+b} \leq g(b)\leq \frac{1}{1+b}$ so that
  also $g$ is strictly positive.

\item[(K5)] $\mathcal{K}_\lambda$ is convex: We have $\mu f_1+(1-\mu)f_2\in
\mathcal{K}_\lambda$ for any $f_1,f_2\in \mathcal{K}_\lambda$ and $0\leq
\mu \leq 1$ by definition of $\mathcal{K}_\lambda$. 

\item[(K6)] $T_\lambda:\mathcal{K}_\lambda\to \mathcal{K}_\lambda$ is
  continuous, i.e.\ $\|T_\lambda f-T_\lambda \tilde{f}\|'_\infty
  <\epsilon$ for $\|f-\tilde{f}\|'_\infty <\delta$: By the uniform
  convergence Lemma~\ref{Lemma:Tf-deriv-boundend}.(C2)+(C3) with
  $\frac{1}{2}+\frac{1}{\lambda\pi^2 P_\lambda} \leq C_\lambda$ it
  suffices to restrict the sup-norms to the compact interval
  $[0,\frac{(6+3C_\lambda)}{\epsilon}]$. Since continuous functions on
  compact intervals are uniformly continuous, it suffices that for
  fixed $t$ the map \mbox{$\displaystyle \mathcal{K}_\lambda \ni f \mapsto
  \int_0^\infty \frac{dp}{\big(\lambda\pi p \big)^2 +\big(
    t+ \frac{1+ \lambda\pi p
      {\mathcal{H}_p}^{\!\!\!\!\infty}[ f(\bullet)]}{f(p)}\big)^2} \in
  \mathbb{R}_+$} is continuous.  Since the integrand is bounded by
  $(\frac{1}{\lambda\pi p})^2$, we can for $\lambda>0$ restrict the
  integral to a compact interval $[0,\Lambda^2(\epsilon)]$. Since the
  Hilbert transform is continuous according to the discussion in
  (\ref{Hilbert-infty}) and rational functions of continuous functions
  with denominator separated from zero are continuous, the whole
  function $\mathcal{K}_\lambda \ni f \mapsto \frac{1}{(\lambda\pi p)^2 
  +\big( t+ \frac{1+\lambda\pi p
      \mathcal{H}_p^{\!\infty}[ f(\bullet)]}{f(p)}\big)^2} \in
  \mathbb{R}_+$ is continuous for every $p$ and hence uniformly
  continuous on $[0, \Lambda^2(\epsilon)]$. This implies $\|T_\lambda
  f-T_\lambda \tilde{f}\|'_\infty <\epsilon$ for $f,\tilde{f}$
  sufficiently close in $\mathcal{K}_\lambda$.

\end{itemize}
(K2) and (K3) are the conditions in Proposition~\ref{Prop:Cianciaruso}
which imply that $T_\lambda(\mathcal{K}_\lambda) \subset
\mathcal{C}^1_0(\mathbb{R}_+) $ is relatively compact.  Hence
$\overline{T_\lambda(\mathcal{K}_\lambda)} \subset
\mathcal{C}^1_0(\mathbb{R}_+) $ is compact, with
$\overline{T_\lambda(\mathcal{K}_\lambda)} \subset
\mathcal{K}_\lambda$ by (K1) and (K4). This means that $T_\lambda$
maps $\mathcal{K}_\lambda$ into a compact subset of
$\mathcal{K}_\lambda$. The properties (K5) and (K6) verify the other
conditions in the Schauder fixed point theorem
(Theorem~\ref{Thm:Schauder}), which thus guarantees that the map
$T_\lambda:\mathcal{K}_\lambda\to \mathcal{K}_\lambda$ has a fixed
point $G_{b0}$:
\begin{Theorem}
\label{Thm:existence}
For any $\lambda>0$, the equation 
\begin{align}
G_{b0} 
&= \frac{1}{1+b}
\exp\Bigg(- \lambda
\int_0^b \!\! dt \int_0^\infty \!\!\! dp\;
\frac{(G_{p0})^2}{\big(\lambda\pi p G_{p0}\big)^2 
+\big( 1+ tG_{p0} +\lambda\pi p
 {\mathcal{H}_p}^{\!\!\!\!\infty}[G_{\bullet 0}]\big)^2} 
\Bigg) 
\label{G-beta0-infty}
\end{align}
has a solution $G_{b0}\in \mathcal{C}^1_0(\mathbb{R}_+)$.
This solution is automatically smooth, monotonously decreasing 
with uniformly bounded
derivative $-\big(\frac{1}{1+b}+C_\lambda\big) G_{b0} 
\leq \frac{dG_{b0}}{db} \leq 0$ 
and pointwise bounded by $0 < G_{b0}\leq \frac{1}{1+b}$.
At $b=0$ one has 
$G_{00}=1$ and $\frac{dG_{b0}}{db}\big|_{b=0}=-(1+\mathcal{Y})$.
\hfill $\square$%
\end{Theorem}
In retrospect the Theorem justifies the assumption 
(\ref{assump:Hoelder}) on H\"older
continuity of $G_{ab}$.

It is clear that by restricting to $[0,\Lambda^2]$ and ignoring the
behaviour at $\infty$, also (\ref{G-beta0}) has a solution
$G^\Lambda_{\bullet 0}\in \mathcal{C}^1[0,\Lambda^2]$. Note however that this
proof works for fixed $\Lambda^2$ (including $\Lambda^2=\infty$) but
without control on the limit. This means that our existence proof does not
imply the (highly plausible) existence of the limit 
$\lim_{\Lambda\to \infty} G^\Lambda_{b0}$ and the equality with $G_{b0}$.

\smallskip

The solution $G_{b0}$ of (\ref{G-beta0-infty}) provides $\vartheta_b(a)$ via
(\ref{arctan-0}) and then gives the complete two-point function
$G_{ab}$ according to (\ref{Gab-vartheta}) and all higher 
correlation function via Theorem~\ref{Thm:GN-algebraic}.

If we also knew uniqueness of the solution, then the resulting unique
$\vartheta_0$ would be the only candidate for a solution of
(\ref{master-Ta}). But before addressing the uniqueness question one
has to understand the possible non-trivial solutions of the
homogeneous Carleman equation which we ignored by
Assumption~\ref{Assump-C=0}. This is a project of its own.

\subsection{Higher correlation functions and effective coupling
  constant}

\label{sec:effective}

Higher correlation functions of the noncommutative $\phi^4_4$-model in
matrix representation are obtained from the algebraic recursion
formula (\ref{GN-algebraic-E}) after specification to the 
parameters and index sets (\ref{Vphi}). We are interested in the 
limit $V\to \infty$ subject to (\ref{limit}). According to
(\ref{under-a-to-a}) we have $\lim_{V\to \infty} 
E_{\under{a}}-E_{\under{b}}
=Z\mu^2(1+\mathcal{Y}) (a-b)$. This suggests to absorb the mass
dimension as follows:
\begin{equation}
G_{ab_1\dots b_{N-1}}:=
\lim_{V\to \infty} \mu^{3N-4} G^{(0)}_{|\under{a}\under{b}_1\dots \under{b}_{N-1}|}\;,
\label{GN-under-a-to-a}
\end{equation}
which for $N=2$ is compatible with (\ref{G-inverse}) and
(\ref{def:Gab}). With $\lambda_4=Z^2\lambda$ we obtain from 
(\ref{GN-algebraic-E}) in the limit $V\to \infty$ the recursion formula
\begin{align}
G_{b_0b_1\dots b_{N-1}}
&= \frac{(-\lambda)}{(1+\mathcal{Y})^2}
\sum_{l=1}^{\frac{N-2}{2}} 
\frac{G_{b_0 b_1 \dots b_{2l-1}} G_{b_{2l}b_{2l+1}\dots b_{N-1}} 
- G_{b_{2l} b_1 \dots b_{2l-1}} G_{b_0 b_{2l+1}\dots b_{N-1}} 
}{(b_0-b_{2l})(b_1-b_{N-1})}\;.
\label{GN-algebraic}
\end{align}
This gives 
\begin{align}
G_{abcd}
&= \frac{(-\lambda)}{(1+\mathcal{Y})^2}
\frac{G_{ab} G_{cd}-G_{ad}G_{bc}}{(a-c)(b-d)}
\label{G4}
\end{align}
and so on.

\medskip

Of particular interest is the \emph{effective coupling constant}
$\lambda_{\mathit{eff}} =- G_{0000}$.  Indeed, we have
$\lambda_{\mathit{eff}}=\Gamma_{0000}$ where the 1PI function
$\Gamma_{abcd}$ is obtained by amputation of the connected two-point
functions from $G_{abcd}$ and a change of sign, i.e.\ $\Gamma_{abcd}=-
\frac{G_{abcd}}{G_{ab}G_{bc}G_{cd}G_{da}}$. The direct computation
from (\ref{G4}) involves up to second derivatives of $G_{ab}$ at
$a=b=0$ which are difficult to control. We therefore use a different
method based on the $g=0$ sector of the original equation
(\ref{SD-GN-regular-g}).  Specifying to the parameters and index sets
(\ref{Vphi}) of self-dual noncommutative $\phi^4$-theory we obtain in
terms of (\ref{under-a-to-a}) and (\ref{GN-under-a-to-a}) the formula
\begin{align}
&\frac{Z^{-1}}{G_{ab_1}} G_{ab_1\dots b_{N-1}}
-  \lambda \Big(\frac{4}{\theta\mu^2}\Big)^2  
\sum_{|\under{p}|=0}^{\mathcal{N}} (|\under{p}|+1) 
\frac{G_{pb_1\dots b_{N-1}}- \frac{G_{pb_1}}{G_{ab_1}} 
G_{ab_1\dots b_{N-1}}}{\frac{4}{\theta \mu^2}(|\under{p}|-|\under{a}|)}
\nonumber
\\*[-2ex]
&= \lambda
\sum_{l=1}^{\frac{N-2}{2}} 
G_{b_1\dots b_{2l}}
\frac{G_{b_{2l+1}\dots b_{N-1}b_{2l}}-G_{b_{2l+1}\dots b_{N-1}a}}{
\frac{4}{\theta\mu^2}(|\under{b}_{2l}|-|\under{a}|)}\;,
\label{SD-GN-Moyal-a}
\end{align}
where $p,a,b_k$ are viewed as functions of $|\under{p}|,|\under{a}|,
|\under{b}_k|$ according to (\ref{under-a-to-a}). 
In the limit (\ref{limit}) this equation converges to an integral
equation which under the assumption that 
$G_{ab_1\dots b_{N-1}}$ is H\"older-continuous can be rearranged as 
\begin{align}
&\frac{G_{ab_1\dots b_{N-1}}}{G_{ab_1}} 
\Big( \frac{Z^{-1}}{1+\mathcal{Y}}
+ \lambda\pi \mathcal{H}_a^{\!\Lambda}
\big[\bullet G_{\bullet b_1}\big]
\Big)
-  \lambda\pi 
\mathcal{H}_a^{\!\Lambda}
\big[\bullet G_{\bullet b_1\dots b_{N-1}}\big]
\nonumber
\\
&= \frac{\lambda}{(1+\mathcal{Y})^2} 
\sum_{l=1}^{\frac{N-2}{2}} G_{b_1\dots b_{2l}}
\frac{G_{b_{2l+1}\dots b_{N-1}b_{2l}}-G_{b_{2l+1}\dots b_{N-1}a}}{b_{2l}-a}\;.
\label{inteq-GN}
\end{align}
Using the previous definitions and identities (\ref{GGDab}),
(\ref{D-Hilbert}) and (\ref{ZZ2}), the prefactor involving $Z^{-1}$ is
treated as follows:
\begin{align}
&\frac{Z^{-1}}{1+\mathcal{Y}}
+ \lambda\pi \mathcal{H}_a^{\!\Lambda}
\big[\bullet G_{\bullet b_1}\big]
\nonumber
\\
&= \frac{Z^{-1}}{1+\mathcal{Y}}
+ \lambda\pi b_1 \mathcal{H}_a^{\!\Lambda}\big[D_{\bullet b_1}\big]
+ \lambda\pi \mathcal{H}_a^{\!\Lambda}\big[\bullet G_{\bullet 0}\big]
\nonumber
\\*
&= 
b_1 D_{ab_1}\Big(\frac{b_1}{a}+
\frac{1+\lambda\pi a
\mathcal{H}_a^{\!\Lambda}[G_{\bullet 0}]}{a G_{a0}}\Big)  + b_1 G_{a0}
+ \frac{Z^{-1}}{1+\mathcal{Y}} + \lambda\pi \mathcal{H}_a^{\!\Lambda}
\big[\bullet G_{\bullet 0}\big]
\nonumber
\\*
&= 
a G_{ab_1}\Big(\frac{b_1}{a}+
\frac{1+\lambda\pi a
\mathcal{H}_a^{\!\Lambda}[G_{\bullet 0}]}{a G_{a0}}\Big)  
+ \underbrace{\frac{Z^{-1}}{1+\mathcal{Y}}
- 1
+ \lambda\pi \mathcal{H}_a^{\!\Lambda}
\big[(\bullet-a) G_{\bullet 0}\big]}_{=0}\;.
\label{ident-Z}
\end{align}
Inserted back into (\ref{inteq-GN}) we obtain for $(a G_{ab_1\dots
  b_{N-1}})$ again a Carleman equation
\begin{align}
&\Big(\frac{b_1}{a}+
\frac{1+\lambda\pi a
\mathcal{H}_a^{\!\Lambda}[G_{\bullet 0}]}{a G_{a0}}\Big) 
\big(a G_{ab_1\dots b_{N-1}}\big)
-  \lambda\pi \mathcal{H}_a^{\!\Lambda}
\big[\bullet G_{\bullet b_1\dots b_{N-1}}\big]
\nonumber
\\*
&= \frac{\lambda}{(1+\mathcal{Y})^2}
\sum_{l=1}^{\frac{N-2}{2}} G_{b_1\dots b_{2l}}
\frac{G_{b_{2l+1}\dots b_{N-1}b_{2l}}-G_{b_{2l+1}\dots b_{N-1}a}}{b_{2l}-a}\;.
\label{GN-Carleman}
\end{align}
Remarkably, the Carleman equation (\ref{GN-Carleman}) has the same
homogeneous part as the equation (\ref{D-Hilbert}) for the two-point
function $D_{ab_1}$, only
the inhomogeneity is different. Its solution is given by
Proposition~\ref{Prop:Carleman}, again under the assumption $C=0$, as: 
\begin{align}
G_{ab_1\dots b_{N-1}}
&= \frac{\lambda}{(1+\mathcal{Y})^2} \sum_{l=1}^{\frac{N-2}{2}} 
G_{b_1\dots b_{2l}} \frac{\sin \vartheta_{b_1}(a)}{\lambda \pi a}
\Big(
\frac{G_{b_{2l+1}\dots b_{N-1}b_{2l}}-G_{b_{2l+1}\dots b_{N-1}a}}{b_{2l}-a}
\cos (\vartheta_{b_1}(a))
\nonumber
\\
&\qquad\qquad\quad
+ e^{\mathcal{H}_a^{\!\Lambda}[\vartheta_{b_1}]}
\mathcal{H}_a^{\!\Lambda}\Big[
e^{-\mathcal{H}_\bullet^{\!\Lambda}[\vartheta_{b_1}]} \sin (\vartheta_{b_1}(\bullet))
\frac{G_{b_{2l+1}\dots b_{N-1}b_{2l}}-G_{b_{2l+1}\dots b_{N-1}\bullet}}{b_{2l}-\bullet}
\Big]\Big)\;.
\label{GN-Carleman-solution}
\end{align}
Consistency of our method implies that the solutions of 
(\ref{GN-Carleman-solution}) agree with the algebraic solutions 
(\ref{GN-algebraic}), a fact that we have explicitly
verified\footnote{The proof is given in appendix A of the
  preliminary versions v2,v3 of this paper on arXiv. We have
  suppressed this appendix in the final version.} for $N=4$.

We use (\ref{GN-Carleman-solution}) to compute 
$G_{0000}=\lim_{a\to 0} G_{a0000}$, taking 
$G_{00}=1$, $\lim_{\alpha \to 0} \frac{G_{a0}-1}{a}=-(1+\mathcal{Y})$ and
$\cos \vartheta_0(0)=\mathrm{sign}(\lambda)$ into account:
\begin{align*}
\lambda_{\mathit{eff}}
&= - \lim_{a\to 0} \frac{\sin \vartheta_0(a)}{(1+\mathcal{Y})^2 \pi a} 
\Big(
\frac{ G_{a0}-1}{a} \cos (\vartheta_0(a))
+ e^{\mathcal{H}_a^{\!\Lambda}[\vartheta_0]}
\mathcal{H}_a^{\!\Lambda}\Big[
e^{-\mathcal{H}_\bullet^{\!\Lambda}[\vartheta_0]} \sin (\vartheta_0(\bullet))
\frac{G_{\bullet 0}-1}{\bullet}
\Big]\Big)
\\*
&=\lim_{a\to 0} \frac{\sin \vartheta_0(a)}{(1+\mathcal{Y}) \pi a} 
\Big( \mathrm{sign}(\lambda)
+ 
\mathcal{H}_0^{\!\Lambda}\Big[
e^{\mathcal{H}_0^{\!\Lambda}[\vartheta_0]-\mathcal{H}_\bullet^{\!\Lambda}[\vartheta_0]} 
\sin (\vartheta_0(\bullet))
\frac{1-G_{\bullet 0}}{(1+\mathcal{Y}) \bullet}
\Big]\Big)\;.
\end{align*}
Recalling $\lim_{a\to 0} \frac{\sin \vartheta_0(a)}{\pi a}=
|\lambda|$ from  (\ref{lim-sinvartheta}) and 
Lemma~\ref{Lemma:G-alpha0-vartheta}, we obtain
\begin{align*}
\lambda_{\mathit{eff}}
&= \frac{\lambda}{1+\mathcal{Y}} + 
\frac{1}{(1+\mathcal{Y})^2}  \mathcal{H}_0^{\!\Lambda}\Big[
\frac{1-G_{\bullet 0}}{G_{\bullet 0}} \frac{\sin^2
  \vartheta_0(\bullet)}{\pi (\bullet)^2} \Big]\;.
\end{align*}
Spelling out the Hilbert transform, taking (\ref{Y-int}) into
account and going to the limit $\Lambda\to \infty$, we have proved:
\begin{Proposition}
The effective coupling constant 
$\lambda_\mathit{eff}=-G_{0000}$ of self-dual noncommutative
$\phi^4_4$-theory is given in terms of the bare coupling constant
$\lambda$ by the following equivalent formulae
\begin{align}
\lambda_\mathit{eff}
&= \frac{\lambda}{1+\mathcal{Y}} + \frac{\lambda^2}{(1+\mathcal{Y})^2}
\int_0^{\infty} dp \; 
\frac{1-G_{p0}}{p  G_{p0}}
\frac{\sin^2 \vartheta_0(p)}{(\lambda \pi p)^2}
\nonumber
\\*
&= \lambda \bigg\{ 1
+ \frac{\lambda}{(1+\mathcal{Y})}
\int_0^{\infty} \!\! dp \; 
\frac{
\Big(\dfrac{1-G_{p0}}{(1+\mathcal{Y}) p} - G_{p0}\Big)G_{p0}}{
\big(\lambda\pi pG_{p0}\big)^2 + 
\big(1+\lambda\pi p {\mathcal{H}_p}^{\!\!\!\!\infty} 
[G_{\bullet 0}]\big)^2}
\bigg\}\;.
\label{lambda-eff}
\end{align}
By Theorem~\ref{Thm:existence}, the change
$\lambda_{\mathit{eff}}\mapsto \lambda$ is only a finite
renormalisation of the bare coupling constant in response to an
infinite change of scales, which means that the QFT model has a
non-perturbatively vanishing $\beta$-function.  \hfill $\square$%
\end{Proposition}

We recall that vanishing of the $\beta$-function at any order of
perturbation theory was proved by Disertori, Gurau, Magnen and
Rivasseau in \cite{Disertori:2006nq}. A decisive step in their proof
was to neglect a term in \cite[eq.~(4.18)]{Disertori:2006nq} which in
our notation reads $\sum_{p\in \mathbb{N}^2_{\mathcal{N}}}
  \frac{1}{|\under{p}|} G^{(0)}_{\under{p}\under{m}\under{0}\under{m}}$. This is
perturbatively justified, but as far as we can see, this argument
cannot be used non-perturbatively; at least it is not obvious. Our
derivation of (\ref{lambda-eff}) shows that the omission of the
mentioned term is not necessary in the non-perturbative treatment. 
By Theorem~\ref{Thm:beta=0} much more is true: the $\beta$-function of
any renormalisable $\phi^4$-matrix model with action
$\mathrm{tr}(E\phi^2+\frac{\lambda}{4}\phi^4)$ vanishes identically.

\subsection{Miscellaneous remarks}

\label{sec:misc}

\begin{enumerate}
\item We have shown in (\ref{scaling-GN:B:g}) that $N$-point functions
  with $B>1$ seem to be suppressed by a factor $V^{-(B-1)}$ over the
  one-cycle functions.  We will show in a paper in preparation 
\cite{GW-xspace} that there
  is a reasonable (and interesting) limit in which the sector
  $(B>1,g=0)$ survives. In the matrix basis of the Moyal plane,
  individual matrix elements of correlation functions have Gau\ss{}ian
  decay $e^{-\|x\|^2/\theta}$ for $\|x\|\to \infty$. Such functions
  belong to Schwartz space where $V$ is the correct volume. At the end
  we are interested in a particle interpretation with plane wave
  asymptotics. These plane waves are in the unital algebra. According
  to \cite{Gayral:2011vu,Grosse:2007jy}, \emph{the spectral dimension
    of unital functions is doubled}, i.e.\ the volume is $V^2$ instead
  of $V$. The assembly of plane waves from Gau\ss{} packets will
  involve sums over matrix indices which increase the volume from $V$
  to $V^2$. We expect one such factor $V$ per cycle so that \emph{all}
  correlation functions arising from $\frac{1}{V^2} \log
  \mathcal{Z}[J]$ should have a finite limit for $V\to \infty$ as soon
  as we assemble plane waves. In this situation the contributions from
  $B>1$ will survive.

We recall
that the $\theta\to \infty$ limit of the model at $\Omega=0$ was
studied by Becchi, Giusto and Imbimbo
\cite{Becchi:2002kj,Becchi:2003dg}. They already established the
suppression of the non-planar sector (also noticed in
\cite{Minwalla:1999px}) and presence of functions $B>1$ (``swiss
cheese'') in the limit $\theta\to\infty$.

\item 
Functional derivatives of the original Ward identity (\ref{eq:Ward})
express the index integral of an $N$-point function in terms of other
functions. Putting $p=b_2$, $a=b_N$ and applying the derivatives with respect
to $\frac{\partial^{N-2}}{\partial J_{b_2b_3}\dots 
\partial J_{b_{N-1}b_N}}$, we obtain from (\ref{eq:Ward})
\begin{align}
&(E_{b_2}-E_{b_N})\Big(
\sum_{n\in I} G_{|nb_2\dots b_N|} + 
\sum_{k=2}^{N} G_{|b_2\dots b_k|b_k\dots b_N|}+
\sum_{l=1}^{\frac{N-2}{2}} G_{|b_2\dots b_{2l+1}|}G_{|b_{2l+1}\dots b_N|}\Big)
\nonumber
\\
& = G_{|b_Nb_3\dots b_{N-1}|} - G_{|b_{2}b_3\dots b_{N-1}|}  \;.
\end{align}
Inserting (\ref{Vphi}) and the rescaling (\ref{scaling-GN:B:g}) 
we see that the functions with $B=2$ cycles increase the genus by
$1$. The restriction to the planar sector $g=0$ thus leads with 
(\ref{under-a-to-a}) to the
equation
\begin{align}
&Z(1+\mathcal{Y})(b_2-b_N) \Big(
(1+\mathcal{Y})^2 \int_0^{\Lambda^2} \!\!\! p\,dp
\; G_{pb_2\dots b_{N}} + 
\sum_{l=1}^{\frac{N-2}{2}} G_{b_2\dots b_{2l+1}}G_{b_{2l+1}\dots b_N}\Big)
\nonumber
\\
&= G_{n_Nb_3\dots b_{N-1}}- G_{b_{2}b_3\dots b_{N-1}}  \;.
\label{Ward-N}
\end{align}
It is interesting to verify (\ref{Ward-N}) for $N=4$. From the explicit
formula (\ref{GN-algebraic}) we obtain 
\begin{align}
Z(1+\mathcal{Y})\lambda\pi 
\big(G_{bc}\mathcal{H}_c^{\!\Lambda}[\bullet G_{\bullet d}]
-G_{dc} \mathcal{H}_c^{\!\Lambda}[\bullet G_{\bullet b}]
\big) + Z (1+\mathcal{Y})(b{-}d) G_{bc}G_{cd}= G_{dc} - G_{bc} \;,
\end{align}
in agreement with (\ref{ident-Z}).  

\item One may speculate that for large indices $a,b\gg 1$ there is a scaling 
relation $G_{sa,sb} =s^\eta G_{ab}$. Such a relation would by
  (\ref{GN-algebraic}) result in $G_{sb_0,\dots,sb_{N-1}}=
  s^{\frac{N-4}{2} \eta} G_{b_0\dots b_{N-1}}$.

\item Taking the difference between (\ref{ZG-inteq:diffeq}) and the
  same equation at $b\mapsto c$ we have
\begin{align}               
(b-c)-\frac{1}{Z(1+\mathcal{Y})}\Big(\frac{1}{G_{ab}}-\frac{1}{G_{ac}}\Big)
&= \lambda 
\int_0^{\Lambda^2} p\,dp \; 
\frac{\frac{G_{pb}}{G_{ab}}-\frac{G_{pc}}{G_{ac}}}{p-a}\;.
\end{align}
We divide by $b-c$ and go to the limit $b=c$. In terms of 
$G'_{ac}:=\lim_{b\to c} \frac{G_{ab}-G_{ac}}{b-c}$ we have with (\ref{ZZ2})
\begin{align*}               
\Big(\frac{1
+ \lambda\pi \mathcal{H}_a^{\!\Lambda} \big[\bullet G_{\bullet b}\big] 
-\lambda \pi \mathcal{H}_0^{\!\Lambda} \big[\bullet G_{\bullet 0}\big]}{G_{ab}}
\Big) G'_{ab}
- \lambda \pi \mathcal{H}^{\!\Lambda}_a \big[ \bullet G'_{\bullet b}\big]
= - G_{ab}\;.
\label{Gab-Gac}
\end{align*}
We insert (\ref{Gab-vartheta}) and obtain with (\ref{Tricomi-18}):
\begin{align*}               
\lambda \pi \cot \vartheta_b(a)
\big(a G'_{ab}\big)
- \lambda \pi \mathcal{H}_a^{\!\Lambda} \big[ \bullet G'_{\bullet b}\big]
= - G_{ab}\;.
\end{align*}
Recalling $\lambda \pi a \cot \vartheta_b(a)=b+\frac{1+\lambda\pi a
  \mathcal{H}^{\!\Lambda}_a[G_{\bullet 0}]}{G_{a0}}$ we have $\frac{d}{db} 
(\lambda \pi a \cot \vartheta_b(a))=1$. This allows us to write down
the Carleman equation for the $n$-th derivative $G^{(n)}_{ab}:=
\frac{d^n}{db^n} G_{ab}$
\begin{align}               
\lambda \pi \cot \vartheta_b(a)
\big(a G^{(n)}_{ab}\big)
- \lambda \pi \mathcal{H}_a^{\!\Lambda} \big[ \bullet G^{(n)}_{\bullet b}\big]
= - n G_{ab}^{(n-1)} + \frac{\delta_{n0}}{Z(1+\mathcal{Y})} \;.
\end{align}
The case $n=0$ follows directly from 
(\ref{Gab-vartheta}),  (\ref{Tricomi-18}) and (\ref{eq:Z}). 

\end{enumerate}

\section{Conclusion and outlook}

\label{sec:Conclusion}

This paper was originally intended to achieve the solution of
$\phi^4$-quantum field theory on four-dimensional Moyal space with
harmonic propagation along the lines we proposed in
\cite{Grosse:2009pa}. In developing the necessary tools we realised
that the mathematical structures extend to general quartic matrix
models with action $S=V \mathrm{tr}(E
\phi^2+\frac{\lambda}{4}\phi^4)$ for a real matrix-valued field
$\phi=\phi^*=(\phi_{pq})_{p,q\in I}$, where the external matrix $E$
encodes the dynamics and the number $V>0$ represents the volume. We
proved that in a scaling limit $V\to \infty$ and
$\frac{1}{V} \sum_{p\in I}$ finite, all such matrix models have a
2-point function satisfying a closed non-linear equation, and that all
higher correlation functions are obtained from a universal algebraic
recursion formula in terms of the 2-point function and the eigenvalues
of $E$. The remarkable and completely unexpected conclusion is that
\emph{all} renormalisable quartic matrix models have a vanishing
$\beta$-function, i.e.\ they are almost scale-invariant --- 
a fact that was perturbatively established in
\cite{Disertori:2006nq} for the noncommutative $\phi^4_4$-model.

We feel that getting thus far was only possible because there is a
deep mathematical structure behind. Our observations could be another
facet of the close connection between integrability and scale
invariance. This point deserves further investigation, in particular
in view relations to other classes of integrable models. It would also
be interesting to know whether there is more than a formal similarity
with the cubic matrix model of Kontsevich \cite{Kontsevich:1992ti}.

\medskip

This paper achieves the non-perturbative solution of a toy model of
\emph{four-dimensional} quantum field theory --- of the
$\phi^4_4$-model on noncommutative Moyal space in the limit $\theta\to
\infty$. The solution of the correlation functions can be viewed as
summation of infinitely many renormalised Feynman graphs (see also
Appendix~\ref{app:perturbative}). The main tools for these
achievements are Ward identities in field-theoretical matrix models,
the resulting Schwinger-Dyson equations and the solution theory of the
Carleman singular integral equation. The 2-point function plays a
central r\^ole in our approach. It is expressed in terms of a single
function $G$ of one variable given as solution (that we proved to
exist) of a non-linear integral equation.

We are left with a few problems:
\begin{itemize} \itemsep 0pt

\item We have ignored so far the possible non-trivial solutions of the
  homogeneous Carleman equation. The numerical treatment in
  \cite{GW-numerical} justifies this for coupling constants $0 <
  \lambda \leq \frac{1}{\pi}$.  This needs a rigorous confirmation, which is
  only possible if the whole space of solutions of the Carleman
  equation is taken into account. The investigation of the consistency
  requirements should then select the right point in the solution
  space. In this way it should be possible to continue the solution of
  the model to $\lambda > \frac{1}{\pi}$. It seems plausible that the true
  solution changes its sign for $\lambda > \frac{1}{\pi}$. This could be
  interpreted as the impossibility of an infinite correlation length
  so that $\lambda = \frac{1}{\pi}$ might be the end point of the
  $\lambda$-family of critical models.

  It makes little sense to prove the (important!) uniqueness of the
  solution of the integral equation for the two-point function before
  clarifying the freedom from the homogeneous Carleman equation.

\item Clarify the combinatorics of the weight factors in the formulae
  expressing the $N$-point function in terms of the $2$-point
  function. We already know that the non-crossing chord diagrams
  arise which are counted by the Catalan numbers.  Identifying the
  combinatorics might suggest links to other models.  

\item Reformulate the infinite volume limit $\theta\to \infty$ in
  position space. This is the subject of a paper in preparation 
\cite{GW-xspace}. 
  We can prove that the limit $\theta\to \infty$ restores the full 
  Euclidean symmetry group and that only the diagonal 
$(N_1{+}\dots{+}N_B)$-point functions $G_{\underbrace{a_1\dots a_1}_{N_1}|
\dots | \underbrace{a_B\dots a_B}_{N_B}}$ with all $N_i$ even contribute 
to position space correlation functions. These functions describe a theory 
with interaction but without exchange of momenta. In contrast to a free 
theory, clustering is violated, which corresponds to the presence of 
different topological sectors. 

\item In position space reformulation the main question
concerns the analytic continuation in the Euclidean time variable to 
a possible Minkowskian theory. We will show in \cite{GW-xspace} that
Osterwalder-Schrader reflection positivity of the two-point function 
is connected to the question whether $a\mapsto G_{aa}$ is a 
Stieltjes function. Settling this question requires more knowledge on 
the fixed point solution $G_{a0}$ of our master equation. 

\item The scaling limit in which both the size $\mathcal{N}$ of the 
matrices and the noncommutativity parameter $\theta$ are sent to infinity 
with their ratio (\ref{limit}) kept fixed restricts the 
matrix model to its planar sector. The reason is that any full 
correlation function satisfies exactly the same equation as its 
restriction to the planar sector. It is an interesting and definitely 
much harder problem to construct the model for $\mathcal{N}\to \infty$ 
but finite $\theta$. The main difficulty is to control the planar 
2-point function which now solves the non-linear discrete equation 
(\ref{ZG-norm}). There is next a hierarchy of linear equations for 
higher $(N_1{+}\dots{+}N_B)$-point functions of fixed genus, 
with explicitly known solution if one $N_i\geq 3$. Assuming that the
other basic functions with all $N_i\leq 2$ can be controlled there 
remains the resummation of the genus expansion, a problem of constructive 
physics. It would be very interesting to establish the limit 
$\theta\to \infty$ via this construction. That it 
agrees with the procedure based on (\ref{limit}) 
is plausible but by no means obvious.

\end{itemize}

\begin{appendix}

\section{Correlation functions with two boundary components}

\label{appendix:B=2}

\subsection{Two cycles of odd length: Schwinger-Dyson equations}

The case of $B=2$ cycles needs distinction of several cases. For the
($1{+}1$)-point function we use $\frac{\partial(-S_{int})}{\partial
  \phi_{aa}}=(-V \lambda_4)\sum_{p,n\in I}
\phi_{ap}\phi_{pn}\phi_{na}$ to derive
\begin{align}
G_{|a|c|}
&= \frac{1}{(E_a+E_a)\mathcal{Z}[0]} 
\Big\{ \Big(\phi_{cc}\frac{\partial(-S_{int})}{\partial
  \phi_{aa}}\Big)\Big[\frac{1}{V}\frac{\partial}{\partial J}\Big]\Big\}
\mathcal{Z}[J]\Big|_{J=0}
\nonumber
\\
&= \frac{(-\lambda_4)}{V^3 (E_a+E_a)\mathcal{Z}[0]} 
\Big\{\frac{\partial^2\big((V^2 W_a^1[J]+V W_a^2[J])\mathcal{Z}[J]\big)}{
\partial J_{aa}\partial J_{cc}  }
\nonumber
\\
&\qquad\qquad\qquad+\sum_{p,n \in I} 
\frac{V}{E_p-E_a} 
\frac{\partial^2}{\partial J_{pa} \partial J_{cc}}
\Big(J_{pn}\frac{\partial \mathcal{Z}}{\partial J_{an}}-
J_{na}\frac{\partial \mathcal{Z}}{\partial J_{np}}\Big)\Big\}_{J=0}\;.
\end{align}
The result after genus expansion, obtained by similar considerations
as in Sec.~\ref{sec:genusexpansion}, is
\begin{align}
G_{|a|c|}^{(g)}&=\frac{(-\lambda_4)}{(E_a+E_a)} 
\Big\{
\frac{1}{V} \sum_{n\in I}\sum_{g'+g''=g} G_{|a|c|}^{(g')} G_{|an|}^{(g'')}
+\frac{3}{V} \sum_{g'+g''+1=g}  G_{|a|c|}^{(g')} G^{(g'')}_{|a|a|}
\nonumber
\\*
&\qquad\qquad\qquad
+ \frac{1}{V^2}\Big( G^{(g-2)}_{|a|a|a|c|}+\sum_{n\in I} G^{(g-1)}_{|a|c|an|}
+ G^{(g-1)}_{|c|aaa|}+G^{(g-1)}_{|a|cac|}
\Big)
\nonumber
\\
&\qquad\qquad\qquad
+\frac{1}{V} \sum_{p \in I} \frac{G^{(g)}_{|a|c|}- G^{(g)}_{|p|c|}}{E_p-E_a} 
+\frac{1}{V} \frac{G^{(g)}_{|ac|}- G^{(g)}_{|cc|}}{E_c-E_a} 
\Big\}\;. 
\label{G2-irregular}
\end{align}
In the general case of two
cycles which both have odd length we have
\begin{align*}
G_{|a b_1\dots b_{2l}|c_1\dots c_{N-2l-1}|}
&= \frac{(-\lambda_4)}{V^3 (E_a+E_b)\mathcal{Z}[0]} 
\Big\{\frac{\partial^N\big((V^2 W_a^1[J]+V W_a^2[J])\mathcal{Z}[J]\big)}{
\partial J^{2l+1}_{ab_1\dots b_{2l}} \partial J^{N-2l-1}_{c_1\dots
  c_{N-2l-1}}}
\\*
&  
+\sum_{p,n \in I} 
\frac{V}{E_p-E_a} 
\frac{\partial^N \Big(J_{pn}\dfrac{\partial \mathcal{Z}}{\partial J_{an}}-
J_{na}\dfrac{\partial \mathcal{Z}}{\partial J_{np}}\Big)}{
\partial J_{pb_1} \partial J_{b_1b_2}\cdots 
\partial J_{b_{2l-1}b_{2l}} \partial J_{b_{2l}a} \partial J^{N-2l-1}_{c_1\dots
  c_{N-2l-1}}}
\Big\}_{J=0}\;,
\end{align*}
where 
$\frac{\partial^k}{\partial J^k_{p_1\dots p_k}}:= 
\frac{\partial^k}{\partial J_{p_1p_2}\dots J_{p_{k-1}p_k}\partial
  J_{p_kp_1}}$. For $l=0$ we have to put $\frac{1}{E_a+E_{b_1}}\mapsto 
\frac{1}{E_a+E_{a}}$. 
The derivatives are evaluated according to the discussion 
preceding (\ref{GN-regular}): 
\begin{align}
&G_{|a b_1\dots b_{2l}|c_1\dots c_{N-2l-1}|}
\nonumber
\\
&= \frac{(-\lambda_4)}{E_a+E_{b_1}}\Big\{
\frac{1}{V}
\Big(G_{|a b_1\dots b_{2l}|c_1\dots c_{N-2l-1}|} G_{|a|a|} 
+  \sum_{n\in I} 
G_{|a b_1\dots b_{2l}|c_1\dots c_{N-2l-1}|} G_{|an|} \Big)
\nonumber
\\
&
+ \frac{1}{V^2}\Big(G_{|a|a|a b_1\dots b_{2l}|c_1\dots c_{N-2l-1}|}
+ G_{|aaab_1\dots b_{2l}|c_1\dots c_{N-2l-1}|}
+ \sum_{n\in I} G_{|an|a b_1\dots b_{2l}|c_1\dots c_{N-2l-1}|}
\nonumber
\\[-1ex]
&\qquad +\sum_{k=1}^{2l} 
G_{|b_1\dots b_kab_k\dots b_{2l}a |c_1\dots c_{N-2l-1}|}
+\sum_{k=1}^{N-2l-1} 
G_{|c_1\dots c_kac_k \dots c_{N-2l-1}|a b_1\dots b_{2l}|}\Big)
\nonumber
\\
&
+\frac{2}{V} 
G_{|a|ab_1\dots b_{2l}|}G^{(g'')}_{|a|c_1\dots c_{N-2l-1}|}
- \frac{1}{V} \sum_{p\in I}
\frac{G_{|pb_1\dots b_{2l}|c_1\dots c_{N-2l-1}|}
-G_{|ab_1\dots b_{2l}|c_1\dots c_{N-2l-1}|}}{E_p-E_a} 
\nonumber
\\
&-\frac{1}{V} \sum_{k=1}^{2l}
\frac{G_{|b_1\dots b_k|b_{k+1}\dots b_{2l}b_k|c_1\dots c_{N-2l-1}|}-
G_{|b_1\dots b_k|b_{k+1}\dots b_{2l}a|c_1\dots c_{N-2l-1}|}}{E_{b_k}-E_a}
\nonumber
\\*
&
-\sum_{j=1}^{l} G_{|b_1\dots b_{2j-1}|c_1\dots c_{N-2l-1}|}
\frac{G_{|b_{2j}\dots  b_{2l}b_{2j-1}|}-G_{|b_{2j}\dots b_{2l}a|}
}{E_{b_{2j-1}}-E_a}
\nonumber
\\*
&
- \sum_{j=1}^{l} G_{|b_1\dots b_{2j}|}
\frac{G_{|b_{2j+1}\dots  b_{2l}b_{2j-1}|c_1\dots c_{N-2l-1}|}
-G_{|b_{2j+1}\dots b_{2l}a|c_1\dots c_{N-2l-1}|} }{E_{b_{2j}}-E_a}
\nonumber
\\*
&
-\frac{1}{V}\sum_{k=1}^{N-2l-1} 
\frac{G_{|c_1\dots c_kb_1\dots b_{2l}c_k c_{k+1}\dots c_{N-2l-1}|}
-G_{|c_1\dots c_kb_1\dots b_{2l}a c_{k+1}\dots c_{N-2l-1}|}
}{E_{c_k}-E_a}
\Big\}\;.
\label{Gabc-B=2-odd}
\end{align}
In case of a real field theory we can proceed as in
section~\ref{sec:recursion} to obtain an algebraic recursion
formula. For $l\geq 1$ we multiply (\ref{Gabc-B=2-odd}) by
$E_a+E_{b_1}$ and subtract from the resulting equation the equation
with renamed indices $b_k\leftrightarrow b_{2l+1-k}$ and 
$c_k\leftrightarrow c_{N-2l-k}$. Taking (\ref{GN-reversal}) into
account we arrive (after genus expansion) at
\begin{align}
& G^{(g)}_{|b_0 b_1\dots b_{2l}|c_1\dots c_{N-2l-1}|}
\nonumber
\\
&= 
\frac{(-\lambda_4)}{V} \sum_{k=1}^{N-2l-1} 
\frac{G^{(g)}_{|c_1\dots c_{k-1} b_0 b_1\dots b_{2l}
c_k c_{k+1}\dots c_{N-2l-1}|}
-G^{(g)}_{|c_1\dots c_{k-1} c_k b_1\dots b_{2l} b_0 
c_{k+1}\dots c_{N-2l-1}|}
}{(E_{b_1}-E_{b_{2l}})(E_{b_0}-E_{c_k})}
\nonumber
\\
&
+(-\lambda_4) \!\!\! \sum_{g'+g''=g}\sum_{j=1}^{l} 
\frac{G^{(g')}_{|b_0b_1\dots b_{2j-2}|c_1\dots c_{N-2l-1}|}
G^{(g'')}_{|b_{2j-1}b_{2j}\dots  b_{2l}|}
-G^{(g')}_{|b_{2j-1}b_1\dots b_{2j-2}|c_1\dots c_{N-2l-1}|} 
G^{(g'')}_{|b_0b_{2j}\dots b_{2l}|}
}{(E_{b_1}-E_{b_{2l}}) (E_{b_0}-E_{b_{2j-1}})}
\nonumber
\\
&
+(-\lambda_4) \!\!\! \sum_{g'+g''=g}\sum_{j=1}^{l} 
\frac{G^{(g')}_{|b_0b_1\dots b_{2j-1}|} 
G^{(g'')}_{|b_{2j}b_{2j+1}\dots  b_{2l}|c_1\dots c_{N-2l-1}|}
-G^{(g')}_{|b_{2j}b_1\dots b_{2j-1}|} 
G^{(g'')}_{|b_0b_{2j+1}\dots b_{2l}|c_1\dots  c_{N-2l-1}|} }{
(E_{b_1}-E_{b_{2l}})(E_{b_0}-E_{b_{2j}})}
\nonumber
\\
&
+
\frac{(-\lambda_4)}{V} \sum_{k=1}^{2l}
\frac{G^{(g-1)}_{|b_0b_1\dots b_{k-1}|b_kb_{k+1}\dots b_{2l}|
c_1\dots c_{N-2l-1}|}-
G^{(g-1)}_{|b_kb_1\dots b_{k-1}|b_0 b_{k+1}\dots b_{2l}|c_1\dots
  c_{N-2l-1}|}}{(E_{b_1}-E_{b_{2l}})(E_{b_0}-E_{b_k})}
\;.
\label{GN-recursion-B=2-odd}
\end{align}
The $l=0$ case is obtained from the symmetry $G_{|b_0|c_1\dots
  c_{N-1}|} =G_{|c_1\dots c_{N-1}|b_0|}$. The last line of 
(\ref{GN-recursion-B=2-odd}) increases
the genus by $1$ because the external vertex connects two cycles of
the same function, from $B=3$ to $B=2$.  
The restriction to the planar ($g=0$) sector, exact
in the limit $(V\to \infty, \frac{1}{V}\sum_{p\in I}\,\text{finite})$,
gives:%
\begin{Proposition}
\label{prop:planar-B=2-odd}
  Given a real $\phi^4$-matrix model with injective external matrix
  $E$, the (unrenormalised) planar connected $N$-point functions
  with two cycles of odd length satisfy together with the single-cycle
  functions (\ref{SD-G}) and (\ref{SD-GN}) the system of equations
\begin{align}
&G^{(0)}_{|a |c|}
= \frac{\lambda_4}{V}
\sum_{p\in I}
\frac{G^{(0)}_{|p|c|} G^{(0)}_{|aa|}
-G^{(0)}_{|a|c|}G^{(0)}_{|pa|}}{E_p-E_a} 
+\frac{\lambda_4}{V} 
G^{(0)}_{|aa|} 
\frac{G^{(0)}_{|cc|}
{-}G^{(0)}_{|ca|}}{E_{c}-E_a}\,,
\label{G1+1-planar}
\\
& G^{(0)}_{|b_0 b_1\dots b_{2l}|c_1\dots c_{N-2l-1}|}
\nonumber
\\*
&= 
\frac{(-\lambda_4)}{V} \sum_{k=1}^{N-2l-1} 
\frac{G^{(0)}_{|c_1\dots c_{k-1} b_0 b_1\dots b_{2l}c_k c_{k+1}\dots c_{N-2l-1}|}
-G^{(0)}_{|c_1\dots c_{k-1} c_k b_1\dots b_{2l} b_0 c_{k+1}\dots c_{N-2l-1}|}
}{(E_{b_1}-E_{b_{2l}})(E_{b_0}-E_{c_k})}
\nonumber
\\
&
+(-\lambda_4) \sum_{j=1}^{l} 
\frac{G^{(0)}_{|b_0b_1\dots b_{2j-2}|c_1\dots c_{N-2l-1}|}
G^{(0)}_{|b_{2j-1}b_{2j}\dots  b_{2l}|}
-G^{(0)}_{|b_{2j-1}b_1\dots b_{2j-2}|c_1\dots c_{N-2l-1}|} 
G^{(0)}_{|b_0b_{2j}\dots b_{2l}|}
}{(E_{b_1}-E_{b_{2l}}) (E_{b_0}-E_{b_{2j-1}})}
\nonumber
\\
&
+(-\lambda_4)
\sum_{j=1}^{l} 
\frac{G^{(0)}_{|b_0b_1\dots b_{2j-1}|} 
G^{(0)}_{|b_{2j}b_{2j+1}\dots  b_{2l}|c_1\dots c_{N-2l-1}|}
-G^{(0)}_{|b_{2j}b_1\dots b_{2j-1}|} G^{(0)}_{|b_0b_{2j+1}\dots b_{2l}|c_1\dots
  c_{N-2l-1}|} }{(E_{b_1}-E_{b_{2l}})(E_{b_0}-E_{b_{2j}})}
\;.
\label{GN-recursion-B=2-odd-planar}
\raisetag{2ex}
\end{align}
In the limit $(V\to
\infty,\frac{1}{V}\sum_{p\in I}\,\text{finite})$, the full function
and the genus-0 function satisfy the same equation: 
$\lim_{\di{V\to \infty}{\frac{1}{V}\sum_{p\in I}\,\text{finite}}} 
G^{(0)}_{|a b_1\dots b_{2l}|c_1\dots c_{N-2l-1}|}
=\lim_{\di{V\to \infty}{\frac{1}{V}\sum_{p\in I}\,\text{finite}}} 
G_{|a b_1\dots b_{2l}|c_1\dots c_{N-2l-1}|}$. 
\end{Proposition}
\emph{Proof}. Equation (\ref{G1+1-planar}) follows from the ($g=0$)-case of
(\ref{G2-irregular}) by elimination of
$\sum_{n\in I} G^{(0)}_{|an|}$ via (\ref{SD-G}). \hspace*{\fill}$\square$%

\subsection{Two cycles of even length: Schwinger-Dyson equations}

The situation is different in the case of two cycles which
both have even length because the action of the two $J$-cycles on
$\mathcal{Z}$ decomposes into its action on $\log \mathcal{Z}$ and the
separate action of each one cycle on one of the factors $\log \mathcal{Z}$ in
$\frac{1}{2}(\log \mathcal{Z})^2$. This means
\begin{align*}
&V G_{|a b_1\dots b_{2l-1}|c_1\dots c_{N-2l}|}
+ V^2 G_{|a b_1\dots b_{2l-1}|} G_{|c_1\dots c_{N-2l}|}
\nonumber
\\
&= V \frac{(-\lambda_4)}{V^3 (E_a+E_b)\mathcal{Z}[0]} 
\Big\{\frac{\partial^N\big((V^2W_a^1[J]+VW_a^2[J])\mathcal{Z}[J]\big)}{
\partial J^{2l}_{ab_1\dots b_{2l-1}} \partial J^{N-2l}_{c_1\dots
  c_{N-2l}}}
\\*
&  
\qquad +V \sum_{p,n \in I} 
\frac{V}{E_p-E_a} 
\frac{\partial^N \Big(J_{pn}\dfrac{\partial \mathcal{Z}}{\partial J_{an}}-
J_{na}\dfrac{\partial \mathcal{Z}}{\partial J_{np}}\Big)}{
\partial J_{pb_1} \partial J_{b_1b_2}\cdots 
\partial J_{b_{2l-2}b_{2l-1}} \partial J_{b_{2l-1}a} 
\partial J^{N-2l}_{c_1\dots  c_{N-2l}}}
\Big\}_{J=0}\;.
\end{align*}
The global prefactor $V$ arises from $\frac{\partial}{\partial J_{ab}}
\exp(\frac{V}{2}\langle J,J\rangle_E)= \frac{VJ_{ba}}{E_a+E_b}
\exp(\frac{V}{2}\langle
J,J\rangle_E)$. In addition to the discussion of (\ref{GN-regular})
there is now the possibility that the derivatives with respect to
$J^{2l}_{ab_1\dots b_{2l-1}}$ and $J^{N-2l}_{c_1\dots c_{N-2l}}$ act
separately on one of the functions $V^2W_a^1[J]+VW_a^2[J]$ and
$\mathcal{Z}[J]$. The analogous consideration applies to the last
line. In this way all terms which constitute $G_{|c_1\dots c_{N-2l}|}$
times the rhs of equation (\ref{GN-regular}) for $G_{|a b_1\dots
  b_{2l-1}|}$ are generated, and these cancel with the lhs. For 
$(N=4,l=1)$ we thus obtain after genus expansion
\begin{align}
G^{(g)}_{|a b|cd|}
&= \frac{(-\lambda_4)}{E_a+E_b}\Big\{
\frac{1}{V} 
\sum_{g'+g''+1=g} G^{(g')}_{|a b|cd|} G^{(g'')}_{|a|a|} 
+\frac{1}{V} \sum_{g'+g''=g}\sum_{n\in I}  G^{(g')}_{|a b|cd|} 
G^{(g'')}_{|an|}
\nonumber
\\
&
+ \frac{1}{V^2} G^{(g-2)}_{|a|a|a b|c d|}
+ \frac{1}{V} \sum_{g'+g''+1=g} G^{(g')}_{|a|a|cd|} G^{(g'')}_{|ab|}
+ \frac{1}{V^2} \sum_{n\in I} G^{(g-1)}_{|an|a b|c d|}
\nonumber
\\
& 
+\frac{1}{V} \!\!\sum_{g'+g''=g} \sum_{n\in I} G^{(g')}_{|an|c d|}
G^{(g'')}_{|a b|}
+ \frac{1}{V^2} \big( G^{(g-1)}_{|aaab|cd|}
+ G^{(g-1)}_{|baba|cd|}
+ G^{(g-1)}_{|cacd|a b|}
+G^{(g-1)}_{|cdad|a b|}\big)
\nonumber
\\
&
+\frac{1}{V} \sum_{g'+g''=g} 
\big(G^{(g')}_{|cacd|}G^{(g'')}_{|a b|}
+G^{(g')}_{|cdad|}G^{(g'')}_{|a b|}\big)
- \frac{1}{V} \sum_{p\in I}
\frac{G^{(g)}_{|pb|cd|}-G^{(g)}_{|ab|cd|}}{E_p-E_a} 
\nonumber
\\
&
-\frac{1}{V} \frac{
G^{(g-1)}_{|b|b|cd|}-G^{(g-1)}_{|b|a|cd|}}{E_b-E_a}
-\frac{1}{V}
\frac{G^{(g)}_{|cbcd|}-G^{(g)}_{|cbad|}}{E_{c}-E_a}
-\frac{1}{V}\frac{G^{(g)}_{|cdbd|}-G^{(g)}_{|cdba|}}{E_{d}-E_a}
\Big\}\;.
\label{G2+2}
\end{align}
For $N\geq 6$ we can achieve $l\geq 2$ by symmetry, and then the
invariance of the real theory under orientation reversal
(\ref{GN-reversal}) allows us to derive the purely algebraic solution
\begin{align}
&G^{(g)}_{|a b_1\dots b_{2l-1}|c_1\dots c_{N-2l}|}
\nonumber
\\
&= (-\lambda_4)
\sum_{j=1}^{l-1} 
\frac{
G^{(g)}_{|b_1\dots b_{2j-1}a|c_1\dots c_{N-2l}|}
G^{(g)}_{|b_{2j}b_{2j+1}\dots b_{2l-1}|}
-
G^{(g)}_{|b_1\dots b_{2j-1}b_{2j}|c_1\dots c_{N-2l}|}
G^{(g)}_{|ab_{2j+1}\dots b_{2l-1}|}
}{(E_{b_1}-E_{b_{2l-1}})(E_a-E_{b_{2j}})}
\nonumber
\\
&
+ (-\lambda_4)
\sum_{j=1}^{l-1} 
\frac{
G^{(g)}_{|b_1\dots b_{2j-1}a|}
G^{(g)}_{|b_{2j}b_{2j+1}\dots b_{2l-1}|c_1\dots c_{N-2l}|}
-
G^{(g)}_{|b_1\dots b_{2j-1}b_{2j}|}
G^{(g)}_{|ab_{2j+1}\dots b_{2l-1}|c_1\dots c_{N-2l}|}
}{(E_{b_1}-E_{b_{2l-1}})(E_a-E_{b_{2j}})}
\nonumber
\\
&
+\frac{(-\lambda_4)}{V}\sum_{k=1}^{N-2l} 
\frac{
G^{(g)}_{|c_1\dots c_{k-1}ab_1\dots b_{2l-1}c_k c_{k+1}\dots c_{N-2l}|}
-G^{(g)}_{|c_1\dots c_{k-1}c_kb_1\dots b_{2l-1}a c_{k+1}\dots c_{N-2l}|}
}{(E_{b_1}-E_{b_{2l-1}})(E_a-E_{c_k})}
\nonumber
\\
&
+\frac{(-\lambda_4)}{V} \sum_{k=1}^{2l-1}\frac{
G^{(g-1)}_{|b_1\dots b_{k-1}a|b_kb_{k+1}\dots b_{2l-1}|c_1\dots c_{N-2l}|}-
G^{(g-1)}_{|b_1\dots b_{k-1}b_k|ab_{k+1}\dots b_{2l-1}|c_1\dots c_{N-2l}|}
}{(E_{b_1}-E_{b_{2l-1}})(E_a-E_{b_k})}
\Big\}\;.
\label{Gabc-B=2-even}
\end{align}

The restriction to the planar ($g=0$) sector gives:
\begin{Proposition}
  Given a real $\phi^4$-matrix model with injective external matrix
  $E$, the (unrenormalised) planar connected $N$-point functions
  with two cycles of even length satisfy together with the single-cycle
  functions (\ref{SD-G}) and (\ref{SD-GN}) the system of equations
\begin{align}
&G^{(0)}_{|a b|cd|}
= 
\frac{(-\lambda_4)}{V} 
G^{(0)}_{|a b|}G^{(0)}_{|a b|}\Big(\sum_{n\in I} G^{(0)}_{|an|c d|}
+
G^{(0)}_{|cacd|}
+G^{(0)}_{|cdad|}\Big) 
 \nonumber
\\*
&
\qquad\quad +\frac{\lambda_4}{V} \sum_{p\in I}
\frac{G^{(g)}_{|pb|cd|}G^{(0)}_{|ab|}-G^{(g)}_{|ab|cd|}G^{(0)}_{|pb|}}{E_p-E_a} 
+\frac{\lambda_4}{V}
\frac{G^{(0)}_{|cbcd|}-G^{(0)}_{|cbad|}}{E_{c}-E_a}
+\frac{\lambda_4}{V}\frac{G^{(0)}_{|cdbd|}-G^{(0)}_{|cdba|}}{E_{d}-E_a}
\Big\}\;,
\label{G2+2-planar}
\\
&G^{(0)}_{|a b_1\dots b_{2l-1}|c_1\dots c_{N-2l}|}
\nonumber
\\*
&
 (-\lambda_4)
\sum_{j=1}^{l-1} 
\frac{
G^{(0)}_{|b_1\dots b_{2j-1}a|c_1\dots c_{N-2l}|}
G^{(0)}_{|b_{2j}b_{2j+1}\dots b_{2l-1}|}
-
G^{(0)}_{|b_1\dots b_{2j-1}b_{2j}|c_1\dots c_{N-2l}|}
G^{(0)}_{|ab_{2j+1}\dots b_{2l-1}|}
}{(E_{b_1}-E_{b_{2l-1}})(E_a-E_{b_{2j}})}
\nonumber
\\*
&
+ (-\lambda_4)
\sum_{j=1}^{l-1} 
\frac{
G^{(0)}_{|b_1\dots b_{2j-1}a|}
G^{(0)}_{|b_{2j}b_{2j+1}\dots b_{2l-1}|c_1\dots c_{N-2l}|}
-
G^{(0)}_{|b_1\dots b_{2j-1}b_{2j}|}
G^{(0)}_{|ab_{2j+1}\dots b_{2l-1}|c_1\dots c_{N-2l}|}
}{(E_{b_1}-E_{b_{2l-1}})(E_a-E_{b_{2j}})}
\nonumber
\\*
&
+\frac{(-\lambda_4)}{V}\sum_{k=1}^{N-2l} 
\frac{
G^{(0)}_{|c_1\dots c_{k-1}ab_1\dots b_{2l-1}c_k c_{k+1}\dots c_{N-2l}|}
-G^{(0)}_{|c_1\dots c_{k-1}c_kb_1\dots b_{2l-1}a c_{k+1}\dots c_{N-2l}|}
}{(E_{b_1}-E_{b_{2l-1}})(E_a-E_{c_k})}\;.
\label{GN-recursion-B=2-even-planar}
\end{align}
We have 
$\lim_{\di{V\to \infty}{\frac{1}{V}\sum_{p\in I}\,\text{finite}}} 
G^{(0)}_{|a b_1\dots b_{2l-1}|c_1\dots c_{N-2l}|}
=\lim_{\di{V\to \infty}{\frac{1}{V}\sum_{p\in I}\,\text{finite}}} 
G_{|a b_1\dots b_{2l-1}|c_1\dots c_{N-2l}|}$
in the scaling limit. 
\label{prop:planar-B=2-even}
\end{Proposition}
\emph{Proof}. Equation (\ref{G2+2-planar}) follows from the ($g=0$)-case of
(\ref{G2+2}) by elimination of
$\sum_{n\in I} G^{(0)}_{|an|}$ via (\ref{SD-G}). \hspace*{\fill}$\square$%

\subsection{Remarks on functions with more than two cycles}

Although we are not going to work out the details, it is clear how to
write down the Schwinger-Dyson equation for any planar $B$-cycle
$N$-point function and how to achieve its solution. By analogy with
the $(1{+}1)$-point function and the $(2{+}2)$-point function we
expect that the solution for the $(N_1{+}\dots{+}N_B)$-point functions
with all $N_i\leq2$ requires a case by case Carleman solution for
known input data.  If at least one $N_j\geq 3$, then the invariance
under orientation reversal leads 
to an explicit algebraic formula for the
$(N_1{+}\dots{+}N_B)$-point function.

\subsection{The $B=2$ sector of noncommutative $\phi^4_4$-theory}

We solve the basic equations (\ref{G1+1-planar}) for $G^{(0)}_{|a|c|}$
and (\ref{G2+2-planar}) for $G^{(0)}_{|ab|cd|}$ for the parameters
(\ref{Vphi}) of the noncommutative $\phi^4_4$-model in matrix
representation.  Higher correlation functions are algebraic.
The equations only depend on the $1$-norms
(\ref{under-a-to-a}) of the indices so that $\sum_{\under{p} \in
  \mathbb{N}^2_{\mathcal{N}}} \mapsto\sum_{|\under{p}|=0}^{\mathcal{N}}
(|p|+1)$.  We absorb the mass dimension and the volume factor 
\begin{equation}
G_{ab_1\dots b_{k}|c_1\dots
  c_{N-k-1}}:= \mu^{3N-4}(V\mu^4) 
G^{(0)}_{|\under{a}\under{b}_1\dots \under{b}_{k}|\under{c}_1\dots
  \under{c}_{N-k-1}|}\;.
\label{scaling-Gab:B=2}
\end{equation}
Conversely, (\ref{scaling-Gab:B=2}) suggests that if $G_{ab_1\dots
  b_{k}|c_1\dots c_{N-k}}$ has a limit for $V\to \infty$ (and that is
the case as shown in the sequel), then
$G^{(0)}_{|\under{a}\under{b}_1\dots \under{b}_{k}|\under{c}_1\dots
  \under{c}_{N-k-1}|}$ is scaled to zero. We give in
section~\ref{sec:misc} some arguments why the sector $B\geq 2$ is
nonetheless interesting.

Recalling $G_{aa}:=\mu^2 G_{|\under{a}\under{a}|}^{(0)}$,
$V=(\frac{\theta}{4})^2$, $\lambda_4=Z^2\lambda$ and $E_{\under{m}}$
from (\ref{Vphi}), multiplication of (\ref{G1+1-planar}) by
$\frac{\mu^6V}{Z (1+\mathcal{Y}) G_{aa}}$ thus leads to
\begin{align}
\frac{Z^{-1}}{(1+\mathcal{Y})}  \frac{G_{a|c}}{G_{aa}} 
&=\lambda
\Big(\frac{4}{\theta \mu^2(1+\mathcal{Y})}\Big)^2
 \sum_{|\under{p}|=0}^{\mathcal{N}} (|\under{p}|+1) \frac{G_{p|c} 
- \frac{G_{a|c}}{G_{aa}} G_{pa}}{p-a}
+\frac{\lambda}{(1+\mathcal{Y})^2} \frac{G_{cc}- G_{ac}}{c-a} \;,
\raisetag{2ex}
\label{SD-2a:B=2}
\end{align}
where $p$ and $|\under{p}|$ are related by (\ref{under-a-to-a}).  Now
the thermodynamic limit (\ref{limit}) to continuous variables
$a,c,p\in [0,\Lambda^2]$ exists. Under the assumption that
$G_{a|c}$ is H\"older-continuous, we can move $\frac{G_{a|c}}{G_{aa}}
G_{pa}$ to the lhs of (\ref{SD-2a:B=2}), and taking (\ref{ident-Z}) at
$b_1=a$ into account we arrive at the following Carleman equation for
the planar $(1{+}1)$-point function:
\begin{align}
\Big(\frac{a}{a}+
\frac{1+\lambda\pi a
\mathcal{H}_a^{\!\Lambda}[G_{\bullet 0}]}{a G_{a0}}\Big) 
\big(a G_{a|c}\big)
-  \lambda\pi \mathcal{H}_a^{\!\Lambda}
\big[\bullet G_{\bullet|c}\big]
&= \frac{\lambda}{(1+\mathcal{Y})^2}
\frac{G_{cc}-G_{ac}}{c-a}\;.
\label{G2-Carleman:B=2}
\end{align}
We let $\vartheta(a):=\vartheta_a(a)$ and stress that the (non-)presence of
the subscript at $\vartheta$ will be important in the following
calculation.  The Carleman formula (\ref{Solution:Carleman}) and
rational fraction expansion give the solution of
(\ref{G2-Carleman:B=2}) as
\begin{align}
G_{a|c}
&= \frac{\sin \vartheta(a)}{(1+\mathcal{Y})^2 \pi a}
\Big(
\frac{ G_{cc}-G_{ac}}{c-a}\cos (\vartheta(a))
+ e^{\mathcal{H}_a^{\!\Lambda}[\vartheta]}
\mathcal{H}_a^{\!\Lambda}\Big[
e^{-\mathcal{H}_\bullet^{\!\Lambda}[\vartheta]} \sin (\vartheta(\bullet))
\frac{G_{cc}-G_{\bullet c}}{c-\bullet}
\Big]\Big)
\nonumber
\\
&
= \frac{\sin \vartheta(a)}{(1{+}\mathcal{Y})^2 \pi a (c{-}a)}
\Big(
\big(G_{cc}{-}G_{ac}\big) \cos (\vartheta(a))
+ e^{\mathcal{H}_a^{\!\Lambda}[\vartheta]}
\mathcal{H}_a^{\!\Lambda}\Big[
e^{-\mathcal{H}_\bullet^{\!\Lambda}[\vartheta]} \sin (\vartheta(\bullet))
\big(G_{cc}{-}G_{\bullet c}\big) 
\Big]
\nonumber
\\*
&\qquad\qquad\qquad\qquad-
e^{\mathcal{H}_a^{\!\Lambda}[\vartheta]}
\mathcal{H}_c^{\!\Lambda}\Big[
e^{-\mathcal{H}_\bullet^{\!\Lambda}[\vartheta]} \sin (\vartheta(\bullet))
\big(G_{cc}{-}G_{\bullet c}\big) 
\Big]\Big)
\nonumber
\\
&
= \frac{\sin \vartheta(a)e^{\mathcal{H}_a^{\!\Lambda}[\vartheta]}
}{ (1+\mathcal{Y})^2 \pi a (c-a)}
\Big(
G_{cc}\cos (\vartheta(c)) e^{-\mathcal{H}_c^{\!\Lambda}[\vartheta]}
-G_{ac} \cos (\vartheta(a)) e^{-\mathcal{H}_a^{\!\Lambda}[\vartheta]}
\nonumber
\\*
&\qquad\qquad\qquad\qquad
-\mathcal{H}_a^{\!\Lambda}\big[
e^{-\mathcal{H}_\bullet^{\!\Lambda}[\vartheta]} \sin (\vartheta(\bullet))
G_{\bullet c}\big]
+
\mathcal{H}_c^{\!\Lambda}\big[
e^{-\mathcal{H}_\bullet^{\!\Lambda}[\vartheta]} \sin (\vartheta(\bullet))
G_{\bullet c} \big]\Big)\;.
\label{G2-planar-B=2}
\end{align}
The last identity follows from (\ref{Tricomi-28}). 
We see no possibility at the moment to further simplify this
expression. A non-trivial consistency check is the symmetry
$G_{a|c}=G_{c|a}$. In perturbation theory up to 
$\mathcal{O}(\lambda^2)$ we confirm the symmetry and the agreement
with the Feynman graph expansion, see 
Appendix~\ref{app:perturbative}.

Higher $(N_1{+}N_2)$-point functions with $N_i$ odd are obtained from
(\ref{GN-recursion-B=2-odd-planar}) for 
$E_a-E_b\mapsto  Z^2\mu^2(1+\mathcal{Y})(a-b)$ and
$\lambda_4=Z^2\lambda$. In the simplest case ($N=4,l=1$) the solution
in dimensionless functions (\ref{scaling-Gab:B=2}) reads 
\begin{align}
G_{a_1 a_2 a_3|c}
&= 
\frac{\lambda}{(1+\mathcal{Y})^2} 
\frac{G_{a_1|c} G_{a_3a_2}-G_{a_2|c} G_{a_3a_1}}{(a_2-a_3) (a_2-a_1)}
+\frac{\lambda}{(1+\mathcal{Y})^2} 
\frac{G_{a_2a_1} G_{a_3|c}-G_{a_2 a_3} G_{a_1|c} }{(a_2-a_3)(a_3-a_1)}
\nonumber
\\*
& +\frac{\lambda}{(1+\mathcal{Y})^2} 
\frac{G_{a_1 a_2 a_3 c} -G_{c a_2 a_3 a_1} }{(a_2-a_3)(c-a_1)}\;.
\label{Gccca-bad}
\end{align}

\medskip

The basic function for $B=2$ cycles of even length is
$G_{|ab|cd|}^{(0)}$ determined by the solution of
(\ref{G2+2-planar}). In the parametrisation~(\ref{Vphi}) and after
taking the limit (\ref{limit}) to continuous matrix indices
(\ref{under-a-to-a}), this equation reads in terms of
(\ref{scaling-Gab:B=2}) 
\begin{subequations}
 \begin{align}
\Big(\frac{b}{a}+
\frac{1+\lambda\pi a
\mathcal{H}_a^{\!\Lambda}[G_{\bullet 0}]}{a G_{a0}}\Big) 
& \big(a G_{ab|cd}\big)
-  \lambda\pi \mathcal{H}_a^{\!\Lambda}
\big[\bullet G_{\bullet b|cd}\big]
\nonumber
\\
&= - Z(1+\mathcal{Y}) G_{ab} \Big(  \lambda I_{a|cd}
+ \frac{\lambda}{(1+\mathcal{Y})^2} (G_{acdc} + G_{adcd}) \Big)
\nonumber
\\*
& +\frac{\lambda}{(1+\mathcal{Y})^2}\frac{G_{dcbc} -G_{dcba}}{c-a}
+\frac{\lambda}{(1+\mathcal{Y})^2}\frac{G_{cdbd} -G_{cdba}}{d-a}\;,
\label{G22-a}
\\
I_{a|cd} & := \int_0^{\Lambda^2}\!\!\! q\,dq\; G_{aq|cd}\;.
\end{align}
\end{subequations}
We regard $I_{a|cd}$ as a function which is independent
of $G_{ab|cd}$ so that we can apply the Carleman formula
(\ref{Solution:Carleman}). By linearity we first treat the auxiliary
problem 
\begin{align}
\Big(\frac{b_1}{a}+\frac{1+\lambda\pi a
\mathcal{H}_a^{\!\Lambda}[G_{\bullet 0}]}{a G_{a0}} \Big) \big(a F_{a b_1|
c_1\dots c_N}\big)
& -  \lambda\pi \mathcal{H}_a^{\!\Lambda}
\big[ \bullet  F_{\bullet b_1|c_1\dots c_N}\big]
\nonumber
\\
& = \frac{\lambda}{(1+\mathcal{Y})^2} \frac{G_{c_2\dots c_{N}c_1}{-}
G_{c_2\dots c_{N}a}}{c_1-a}\;.
\label{GN-Carleman-B=2-F}
\end{align}
Comparing with (\ref{GN-Carleman}) we notice that both sides of
$\displaystyle G_{ab_1c_1\dots c_N}= \sum_{l=1}^{\frac{N}{2}}
G_{b_1c_1\dots c_{2l-1}} F_{ab_1 | c_{2l-1}\dots c_N}$
satisfy the same Carleman equation. We can thus compute the functions
$F$ by iteration:
\begin{align}
F_{ab_1|c_1c_2}& = \frac{G_{ab_1c_1c_2}}{G_{b_1c_1}} \;,
\nonumber
\\
F_{ab_1|c_1c_2c_3c_4}& = \frac{
G_{ab_1c_1c_2c_3c_4}- G_{b_1c_1c_2c_3} F_{ab_1c_3c_4}}{G_{b_1c_1}}
= \frac{G_{ab_1c_1c_2c_3c_4}G_{b_1c_3} -
G_{b_1c_1c_2c_3}G_{ab_1c_3c_4}}{G_{b_1c_1} G_{b_1c_3}}\;.
\raisetag{1ex}
\label{F6}
\end{align}
Since $G_{ab}$ is differentiable, these functions extend to coinciding
indices $c_i$ so that the inhomogeneity
$\frac{\lambda}{(1+\mathcal{Y})^2} \frac{G_{dcbc} -G_{dcba}}{c-a}
+\frac{\lambda}{(1+\mathcal{Y})^2}\frac{G_{cdbd} -G_{cdba}}{d-a}$ in
(\ref{G22-a}) gives rise to the contribution $F_{ab|cdcb}+F_{ab|dcdb}$
in $G_{ab|cd}$. The inhomogeneity proportional to
$Z(1+\mathcal{Y})G_{ab}$ needs to be treated by the Carleman formula
(\ref{Solution:Carleman}):
\begin{align}
G_{ab|cd} &= F_{ab|cdcb}+F_{ab|dcdb}
\nonumber
\\
& - \frac{\sin\vartheta_b(a)}{\lambda \pi a} 
\Big\{ 
Z(1+\mathcal{Y}) G_{ab} \Big( \lambda I_{a|cd}
+ \frac{\lambda}{(1+\mathcal{Y})^2} 
(G_{acdc} + G_{adcd}) \Big) \cos \vartheta_b(a)
\nonumber
\\
& + e^{\mathcal{H}_a^{\!\Lambda}[\vartheta_b]}\mathcal{H}_a^{\!\Lambda}
\Big[e^{-\mathcal{H}_\bullet^{\!\Lambda}[\vartheta_b]} \sin \vartheta_b(\bullet)  
Z(1+\mathcal{Y}) G_{\bullet b} \Big(  \lambda I_{\bullet|cd}
+ \frac{\lambda}{(1+\mathcal{Y})^2} (G_{\bullet cdc} + G_{\bullet dcd}) \Big)
\Big]\Big\}
\nonumber
\\
&= F_{ab|cdcb}+F_{ab|dcdb}
\nonumber
\\*
& - \frac{\sin\vartheta_b(a)}{\lambda \pi a} \cos \vartheta_b(a) G_{ab}
Z(1+\mathcal{Y}) \Big(\lambda I_{a|cd}
+ \frac{\lambda}{(1+\mathcal{Y})^2} (G_{acdc} {+} G_{adcd}) \Big)
\nonumber
\\*
& -G_{ab}Z(1+\mathcal{Y})
\mathcal{H}_a^{\!\Lambda}\Big[ 
\frac{\sin^2 \vartheta_b(\bullet)}{\lambda \pi \bullet} 
\Big(  \lambda I_{\bullet|cd}
+ \frac{\lambda}{(1+\mathcal{Y})^2} (G_{\bullet cdc} + G_{\bullet dcd}) \Big)
\Big]\;,
\label{G4-B=2-even}
\end{align}
where (\ref{Gab-vartheta}) has been used.  We multiply by
$\lambda b$ and integrate over $b=:q$ to obtain an equation
for $I_{a|cd}:= \int_0^{\Lambda^2}\!\!\! q\,dq\; G_{aq|cd}$:
\begin{subequations}
\begin{align}
 X_{a|cd} \Big\{\frac{Z^{-1}}{1+\mathcal{Y}} &+ \frac{1}{\pi a} 
\int_0^{\Lambda^2} \!\!\!\!\! q\,dq \;\sin \vartheta_q(a)
\cos\vartheta_q(a) G_{aq} \Big\}
+ \mathcal{H}_a^{\!\Lambda}\Big[
\frac{X_{\bullet |cd}}{\pi \bullet} 
\int_0^{\Lambda^2} \!\!\!\!\! 
q\,dq \;\sin^2 \vartheta_q(\bullet) G_{aq}\Big]
\nonumber
\\*
&= \lambda \int_0^{\Lambda^2} \!\!\! q\,dq \,
(F_{aq|cdcq}+F_{aq|dcdq})
+ \frac{\lambda}{(1+\mathcal{Y})^2}(G_{acdc} {+} G_{adcd})\;,
\label{Xacd}
\\
X_{a|cd}&:=Z(1+\mathcal{Y}) 
\Big(\lambda I_{a|cd}
+ \frac{\lambda }{(1+\mathcal{Y})^2}(G_{acdc} + G_{adcd})\Big)\;.
\end{align}
Using $\frac{q}{\lambda\pi a} =\cot\vartheta_q(a)-\cot\vartheta_0(a)$ and
inserting $Z^{-1}$ from (\ref{ZZ2}), we have 
\begin{align}
&\frac{Z^{-1}}{1+\mathcal{Y}}  + \frac{1}{\pi a} 
\int_0^{\Lambda^2} \!\!\!\!\! q\,dq \;\sin \vartheta_q(a)
\cos\vartheta_q(a) G_{aq}
\nonumber
\\
&= 1+\lambda \int_0^{\Lambda^2} \!\!\!\!\!dq
\;(G_{aq}-G_{0q})
-\lambda \int_0^{\Lambda^2}  \!\!\!\!\!dq\;
\frac{ G_{aq} \sin \vartheta_q(a) \cos\big(\vartheta_q(a)-\vartheta_0(a)\big)}{\sin 
\vartheta_0(a)}\;.
\label{Xacd-Z}
\end{align}
\end{subequations}
All integrals in this equation, and also
$\displaystyle \int_0^{\Lambda^2} \!\!\!\!\! 
q\,dq \;\sin^2 \vartheta_q(b) G_{aq} $ and 
$\displaystyle 
\int_0^{\Lambda^2} \!\!\! q\,dq \,(F_{aq|cdcq}+F_{aq|dcdq})$
exist for $\Lambda \to
\infty$, even in perturbation theory. Therefore, (\ref{Xacd})
is again a Carleman singular integral equation (\ref{Carleman}) for
$X_{\bullet| cd}$ where the input functions $h$ and $f$ 
only depend on the known data
$G_{ab}$ and $\vartheta_b(a)$ and exist in the limit $\Lambda\to
\infty$. Its solution is given by
(\ref{Solution:Carleman}). Inserted
into (\ref{G4-B=2-even}) we thus get the (in principle explicit)
solution 
\begin{align}
G_{ab|cd} 
&= F_{ab|cdcb}+F_{ab|dcdb}
- \frac{\sin\vartheta_b(a)}{\lambda\pi a} \cos \vartheta_b(a)
G_{ab} X_{a|cd}
-G_{ab} 
\mathcal{H}_a^{\!\Lambda}\Big[ \frac{\sin^2 \vartheta_b(\bullet)}{
\lambda\pi \bullet}   X_{\bullet| cd}
\Big]\;.
\label{G4-B=2-even-final}
\end{align}

Higher $(N_1{+}N_2)$-point functions with $N_i$ even are obtained from
(\ref{GN-recursion-B=2-even-planar}) for 
$E_a-E_b\mapsto  Z^2\mu^2(1+\mathcal{Y})(a-b)$ and
$\lambda_4=Z^2\lambda$. In the simplest case ($N=6,l=2$) the solution
in dimensionless functions (\ref{scaling-Gab:B=2}) reads 
\begin{align}
G_{ab_1b_2b_3|c_1c_2}
&  = \frac{\lambda}{(1+\mathcal{Y})^2}  
\frac{
G_{b_1a} G_{b_2b_3|c_1c_2} -G_{b_1b_2} G_{ab_3|c_1 c_2}
+
G_{b_1a|c_1c_2} G_{b_2b_3} -G_{b_1b_2|c_1c_2} G_{ab_3}}{
(b_2-a)(b_1-b_3)}
\nonumber
\\*
&
+\frac{\lambda}{(1+\mathcal{Y})^2} 
\Big(
\frac{G_{a b_1b_2b_3 c_1c_2}-G_{c_1 b_1b_2b_3 ac_2}}{(c_1-a)(b_1-b_3)}
+
\frac{G_{c_1 a b_1b_2b_3 c_2}-G_{c_1 c_2 b_1b_2b_3 a}}{(c_2-a)(b_1-b_3)}\Big)\;.
\end{align}

\section{Perturbative expansion}

\label{app:perturbative}

It is interesting to expand our solutions for the correlation
functions into power series in $\lambda$. These series should
reproduce the expansion of the partition function into ribbon graphs,
and indeed this agreement was for us a non-trivial consistency check of our
equations. The Feynman rules are:
\begin{itemize} \itemsep 0pt
\item 
$\frac{1}{1+(1+\mathcal{Y})(a+b)}$ for a (fat) line 
separating faces with indices $a,b$

\item $(-Z^2\lambda)$ for a (ribbon) vertex with four outgoing ribbons
  and index conservation at every corner,

\item $\displaystyle (1+\mathcal{Y})^2 
\int_0^{\Lambda^2} p dp$ for
  every closed face with index $p$.
\end{itemize}
These rules lead to a (for $\Lambda\to \infty$) divergent one-particle
irreducible (1PI) two-point function $\Gamma_{ab}$ which therefore
needs renormalisation through subtraction of the constant and linear
Taylor terms. There is no subtraction for the four-point function: As
result of the vanishing $\beta$-function, the divergences in the
ribbon graphs of the 1PI four-point function for $\Lambda\to \infty$
cancel exactly with the divergence of the wavefunction
renormalisation $Z$.

All of the following results are given for the limit
$\Lambda\to \infty$.  Starting point is the master equation
(\ref{G-beta0-infty}) for $G_{b0}$ which yields 
$G_{a0}=\frac{1}{a+1}+\mathcal{O}(\lambda)$. 
To next order we thus have
\begin{align*}
G_{a0} &= \frac{1}{1+a}-\lambda \frac{\log(1+a)}{(1+a)}
+ \mathcal{O}(\lambda^2)\;.
\end{align*}
For the Hilbert transform we have
\begin{align*}
\lambda\pi \mathcal{H}_a^{\!\infty}[G_{\bullet 0}]
&= -\lambda \frac{\log(a)}{1+a} +
\mathcal{O}(\lambda^2)\;.
\end{align*}
This yields for (\ref{arctan-0}) the expansions
\begin{align}
&\vartheta_b(a) = \lambda \frac{a \pi}{1+a+b}
-\lambda^2 \frac{a\pi\big(
(1+a)\log(1+a)-a\log a\big)}{(1+a+b)^2}+\mathcal{O}(\lambda^3)\;,
\label{varthetaba}
\\
&\mathcal{H}_a^{\!\infty}[\vartheta_b]{-}\mathcal{H}_0^{\!\infty}[\vartheta_0]
= \lambda \frac{-(1+b)\log(1+b)-a\log a}{1+a+b}
+ \mathcal{O}(\lambda^2)\;,
\nonumber
\end{align}
which inserted into (\ref{Gab-vartheta}) give 
\begin{align}
G_{ab} &= \frac{1}{1+a+b}
-\lambda
\frac{(1+a)\log(1+a)
+(1+b)\log(1+b) }{(1+a+b)^2}+\mathcal{O}(\lambda^2)
\end{align}
for the 2-point function. This result is reproduced by renormalised
one-loop ribbon graphs if we take the following expansion of
$\mathcal{Y}$ (see (\ref{Y-int})) and $Z(1+\mathcal{Y})=
e^{\mathcal{H}_0^{\!\Lambda}[\vartheta_0]}$ (see (\ref{eq:Z})) into
account (strictly speaking, $Z$ is only defined for finite $\Lambda$):
\begin{align}
\mathcal{Y} &= \lambda +
\mathcal{O}(\lambda^2)\;,\qquad 
Z(1+\mathcal{Y})=1+\mathcal{O}(\lambda)\;.
\end{align}

It is interesting to compare this with the perturbative solution of the
true master equation~(\ref{master-Ta}). This leads to
$\mathcal{T}_a=1+a+\mathcal{O}(\lambda)$ and then in next order 
\begin{align}
\mathcal{T}_a=1+a+ \lambda\Big((1+a)\log(1+a)+a \log
a\Big)+\mathcal{O}(\lambda^2)= \frac{\lambda\pi a}{
 \tan \vartheta_0(a)}\;,
\end{align}
in agreement with (\ref{varthetaba}). We see this as good indication that
in the limit $\Lambda\to \infty$ the master equations
(\ref{master-Ta}) and (\ref{G-beta0-infty}) are equivalent.

From (\ref{G4-new}) we then obtain the 4-point function 
$G_{abcd}=: -G_{ab}G_{bc}G_{cd}G_{da}\Gamma_{abcd}$ up to one
loop, with the 1PI contribution
\begin{align}
\Gamma_{abcd}=\lambda\Big(1
&-\lambda
\frac{a- (1+a)\log(1+a)-c+(1+c)\log(1+c) }{a-c}
\nonumber
\\
&-\lambda
\frac{b- (1+b)\log(1+b)-d+(1+d)\log(1+d) }{b-d}
\Big) + \mathcal{O}(\lambda^3)
\end{align}
that agrees with the ribbon graph calculation. For the 6-point
function, the result of (\ref{N=6}) can be arranged as
\begin{align}
G_{abcdef}
&= G_{ab}G_{bc}G_{cd}G_{de}G_{ef}G_{fa}\Big(
\Gamma_{abcd} G_{ad} \Gamma_{defa}
{+}\Gamma_{bcde} G_{be} \Gamma_{efab}
{+}\Gamma_{cdef} G_{cf} \Gamma_{fabc}
- \Gamma_{abcdef}\Big)\;,
\nonumber
\\
\Gamma_{abcdef}
&= (-\lambda)^3\frac{(a{-}c)(1{+}e)\log(1{+}e)
+(c{-}e)(1{+}a)\log(1{+}a)
+(e{-}a)(1{+}c)\log(1{+}c)}{(a-c)(c-e)(e-a)}
\nonumber
\\
&+(-\lambda)^3
\frac{(b{-}d)(1{+}f)\log(1{+}f)
+(d{-}f)(1{+}b)\log(1{+}b)
+(f{-}b)(1{+}d)\log(1{+}d)}{(b-d)(d-f)(f-b)}
\nonumber
\\
& + \mathcal{O}(\lambda^4)\;,
\label{Gamma6}
\end{align}
also in agreement with the ribbon graph calculation.

For the $(1{+}1)$-point function (\ref{G2-planar-B=2}) we need 
\begin{align*}
\mathcal{H}_a^{\!\Lambda}[\vartheta]-\mathcal{H}_c^{\!\Lambda}[\vartheta]
&= \lambda\Big(\frac{c\log c+c\log 2}{1+2c}
-\frac{a\log a+a\log 2}{1+2a}
\Big)+ \mathcal{O}(\lambda^2)\;,
\\
\mathcal{H}_b^{\!\Lambda}\big[e^{\mathcal{H}_a^{\!\Lambda}[\vartheta]
-\mathcal{H}_\bullet^{\!\Lambda}[\vartheta]}
\sin \vartheta(\bullet) G_{\bullet c}\big]
&= \lambda\Big(
\frac{(1+c)\log (1+c)}{(1+2c)(1+b+c)}
+\frac{\log 2}{(1+2c)(1+2b)}
\\
& \qquad\qquad 
-\frac{b\log b }{(1+b+c)(1+2b)}\Big)+ \mathcal{O}(\lambda^2)\;.
\end{align*}
Now (\ref{G2-planar-B=2}) gives after straightforward but lengthy
computation 
\begin{align}
G_{a|c}&= -G_{aa}G_{cc}\big( \Gamma_{aacc}G_{ac}-\Gamma_{a|c}\big)\;,
\nonumber
\\
\Gamma_{a|c} &= \lambda^2
\Big(
\frac{\frac{(1+c)\log(1+c)}{1+c+c}-
  \frac{(1+a)\log(1+a)}{1+a+a}}{(c-a)}
- 
\frac{\log(2)}{(1+a+a)(1+c+c)}\Big)+\mathcal{O}(\lambda^3)\;,
\end{align}
where $\Gamma_{a|c}$ agrees with the Feynman graph computation 
\[
\Gamma_{a|c}=\int_0^\infty  p\,dp \;
\frac{(-\lambda)^2 }{(p+a+1)(p+p+1)(p+c+1)}+\mathcal{O}(\lambda^3)\;.
\]

For the $(2{+}2)$-point function (\ref{G4-B=2-even-final}) we first
have to provide some intermediate results. From (\ref{F6}) and 
(\ref{Gamma6}) we have
\begin{subequations}
\begin{align*}
F_{ab|cdcb} 
&=G_{ab}G_{cd}G_{dc}G_{ba}\Big(
\Gamma_{abcd} G_{ad} G_{bc}\Gamma_{dcba}
+\Gamma_{bcdc} G_{bc}G_{bc} \Gamma_{cbab}
-G_{bc} \Gamma_{abcdcb}
\Big)\;.
\end{align*}
Next we compute $X_{a|cd}$
from (\ref{Xacd}). Since $\mathcal{H}_a^{\!\Lambda}\Big[ \frac{X_{\bullet
    |cd}}{\pi \bullet} \int_0^{\Lambda^2} \!\!  q\,dq \;\sin^2
\vartheta_q(\bullet) G_{aq}\Big]=\mathcal{O}(\lambda^4)$, the equations
(\ref{Xacd}+\ref{Xacd-Z}) are to $\mathcal{O}(\lambda^3)$ purely
algebraic with solution
\begin{align}
X_{a|cd}&= G_{cd}^2\Big\{
-G_{ac}^2\Gamma_{acdc}
\Big(\lambda-\lambda^2(1-\log(1+a))\Big)
-\Gamma_{adcd} G_{ad}^2 \Big(\lambda-\lambda^2(1-\log(1+a))\Big)
\nonumber
\\*
& \qquad\qquad - G_{ad} J_{aac}-G_{ac} J_{aad}-J_{aadd}-J_{aacc}\Big\}
+ \mathcal{O}(\lambda^4)\;,
\label{Xacd-3}
\\
J_{aac} &:= \int_0^\infty p\,dp  \;
\frac{(-\lambda^3)}{(1{+}a{+}p)^2(1{+}c{+}p)} 
=(-\lambda)^3
\frac{a-c+ (1+c)\big(\log(1{+}c)-\log(1{+}a)\big)}{(a-c)^2} \;,
\\
J_{aacc} &:= \int_0^\infty p\,dp  \;
\frac{(-\lambda^3)}{(1+a+p)^2(1+c+p)^2} 
\nonumber
\\
& =(-\lambda)^3
\frac{2c-2a+ (2+a+c)\log(1+a)-(2+a+c)\log(1+c)}{(a-c)^3} \;.
\end{align}
From (\ref{Xacd-3}) we obtain after some rearrangements
\begin{align}
&\mathcal{H}_a^{\!\infty}\Big[ 
\frac{\sin^2 \vartheta_b(\bullet)}{\lambda\pi \bullet}  X_{\bullet| cd}
\Big]
\nonumber
\\
&= (-\lambda)^3 G_{cd}^2 G_{ab}^2
\Big( 
-\frac{(1+c) \log(1+c)+ a \log(a)}{(1+a+c)^2} 
+\frac{(1+b)\log(1+b) - (1+c) \log(1+c)}{(1+a+c) (b-c)}
\nonumber
\\
& 
\qquad\qquad\qquad\quad -\frac{(1+d) \log(1+d)+ a \log(a)}{(1+a+d)^2} 
+\frac{(1+b)\log(1+b) - (1+d) \log(1+d)}{(1+a+d) (b-d)}
\nonumber
\\
& 
\qquad -G_{cd}^2 G_{ab} \Big( J_{ccb} G_{ca} +J_{ddb} G_{da}
+J_{bbcc} +J_{bbdd} 
\Big)
+\mathcal{O}(\lambda^4)\;.
\end{align}
\end{subequations}
These results and $\Gamma_{abcdcb}= -J_{cca}-J_{bbd}$ give for 
(\ref{G4-B=2-even-final})
\begin{align}
G_{ab|cd}
&= G_{ab}G_{cd}G_{dc}G_{ba}\Big(
\Gamma_{abcd} G_{ad} G_{bc}\Gamma_{dcba}
+\Gamma_{abdc} G_{ac} G_{bd}\Gamma_{cdba}
+\Gamma_{bcdc} G_{bc}G_{bc} \Gamma_{cbab}
\nonumber
\\
&\qquad\qquad 
+\Gamma_{bdcd} G_{bd}G_{bd} \Gamma_{dbab}
+\Gamma_{acdc} G_{ac}G_{ac} \Gamma_{caba}
+\Gamma_{adcd} G_{ad}G_{ad} \Gamma_{daba}
+ \Gamma_{ab|cd}
\Big)\;,
\nonumber
\\
 \Gamma_{ab|cd}&= 
G_{ad} J_{aac}+ G_{bd} J_{bbc}+G_{ac} J_{aad}+G_{bc} J_{bbd}
+G_{cb} J_{cca}+ G_{ca} J_{ccb}+G_{db} J_{dda}+G_{da} J_{ddb}
\nonumber
\\
&+ J_{aacc}+ J_{aadd}+ J_{bbcc}+ J_{bbdd}+\mathcal{O}(\lambda^4)\;.
\end{align}

\end{appendix}

\section*{Acknowledgements}
\addcontentsline{toc}{section}{Acknowledgements}

We would like to thank the Erwin-Schr\"odinger-Institute for
Mathematical Physics in Vienna and the Collaborative Research Centre
``Groups, Geometry and Actions'' (SFB 878) in M\"unster for financing
several mutual visits. We also thank Alan Carey, Christian Krattenthaler 
and Vincent Rivasseau for helpful discussions. We are grateful to the 
referee for suggestions of improvements and most importantly for the 
remarks concerning the type of summability of the genus expansion.

\end{document}